\documentclass[pdflatex,sn-mathphys-num]{sn-jnl}

\usepackage{graphicx}%
\usepackage{multirow}%
\usepackage{amsmath,amssymb,amsfonts}%
\usepackage{amsthm}%
\usepackage{mathrsfs}%
\usepackage[title]{appendix}%
\usepackage{xcolor}%
\usepackage{textcomp}%
\usepackage{manyfoot}%
\usepackage{booktabs}%
\usepackage{algorithm}%
\usepackage{algorithmicx}%
\usepackage{algpseudocode}%
\usepackage{listings}%
\usepackage{graphicx}
\usepackage{caption}
\usepackage{subcaption}
\usepackage{tikz}
\usepackage{float}
\usepackage[utf8]{inputenc} 
\usepackage[T1]{fontenc}    
\usepackage{hyperref}       
\usepackage{url}            
\usepackage{booktabs}       
\usepackage{amsfonts}       
\usepackage{nicefrac}       
\usepackage{microtype}      
\usepackage{xcolor}         
\usepackage{graphicx}
\usepackage{subcaption}
\usepackage{multirow}
\usepackage{array}
\usepackage{rotating}
\usepackage{wrapfig}
\usepackage{amsmath}
\usepackage{amssymb}
\usepackage{fdsymbol}
\usepackage{comment}

\theoremstyle{thmstyleone}%
\theoremstyle{thmstyletwo}%

\theoremstyle{thmstylethree}%

\begin{document}


\title{%
  \includegraphics[height=1cm]{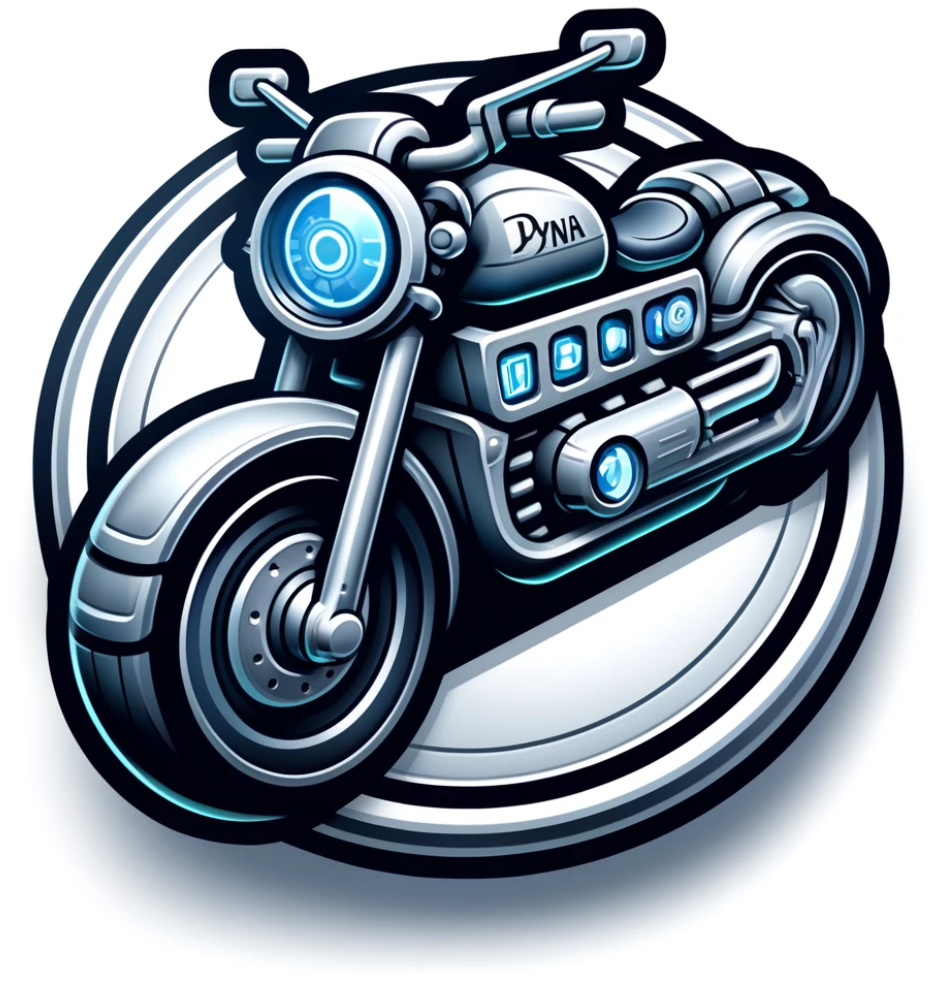}
  \textsc{dyna}: Disease-Specific Language Model for Variant Pathogenicity%
}








\author[1]{\fnm{Huixin} \sur{Zhan}}\email{huixin.zhan@cshs.org}

\author*[1,2]{\fnm{Zijun} \sur{Zhang}}\email{zijun.zhang@cshs.org}

\affil[1]{\orgdiv{Division of Artificial Intelligence in Medicine}, \orgname{Cedars-Sinai Medical Center}, \orgaddress{\city{Los Angeles}, \postcode{90048}, \state{CA}, \country{USA}}}

\affil[2]{\orgdiv{Department of Computational Biomedicine}, \orgname{Cedars-Sinai Medical Center}, \orgaddress{\city{Los Angeles}, \postcode{90048}, \state{CA}, \country{USA}}}


\abstract{

Clinical variant classification of pathogenic versus benign genetic variants remains a challenge in clinical genetics. Recently, the proposition of genomic foundation models has improved the generic variant effect prediction (VEP) accuracy via weakly-supervised or unsupervised training. However, these VEPs are not disease-specific, limiting their adaptation at the point of care. To address this problem, we propose DYNA: Disease-specificity fine-tuning via a Siamese neural network broadly applicable to all genomic foundation models for more effective variant effect predictions in disease-specific contexts. We evaluate DYNA in two distinct disease-relevant tasks. For coding VEPs, we focus on various cardiovascular diseases, where gene-disease relationships of loss-of-function vs. gain-of-function dictate disease-specific VEP. For non-coding VEPs, we apply DYNA to an essential post-transcriptional regulatory axis of RNA splicing, the most common non-coding pathogenic mechanism in established clinical VEP guidelines. In both cases, DYNA fine-tunes various pre-trained genomic foundation models on small, rare variant sets. The DYNA fine-tuned models show superior performance in the held-out rare variant testing set and are further replicated in large, clinically-relevant variant annotations in ClinVAR. Thus, DYNA offers a potent disease-specific variant effect prediction method, excelling in intra-gene generalization and generalization to unseen genetic variants, making it particularly valuable for disease associations and clinical applicability.

}




\maketitle
\section{Main}\label{sec-main}

Clinical variant interpretation is transforming precision medicine, yet limitations exist that prevent its further adaptations and utilities~\citep{katsanis2013molecular}.
Following a disease diagnosis, the identification and classification of pathogenic vs benign genetic variant has important clinical implications.
The outcome of clinical variant interpretation provides a basis for clinical screening~\citep{cocchi2020clinical,xie2023genomic} and genetic testing of first-degree family members~\citep{ni2023screening}, and may serve as a prognostic marker for the affected patient~\citep{lee2019cag,musunuru2020genetic}. Currently, the utility of genetic testing is limited by the fact that a substantial proportion (30-50\%) of yielded variants are classified as variant of uncertain significance (VUS) according to the ACMG guidelines~\citep{richards2015standards}. The presence of VUSs complicates the genetic counseling and patient management, while they can not be used in clinical decision making. Given the large number of genetic variations classified as VUS, both common and rare, \textit{in silico} methods are ideal avenues to aid the clinical variant interpretation, and to facilitate downstream prioritization of experimental validation and clinical testing~\citep{frazer2021disease}.

Predicting the phenotypic outcomes of genetic variations, commonly referred to as variant effect prediction (VEP), remains challenging. This is because existing variant annotations are limited in amount and biased by human curations. As such, it is highly desirable for computational VEP methods to be trained via weakly-supervised or unsupervised approaches that are independent of human bias. Conventionally, evolutionary conservation-based methods~\citep{reva2011predicting,rentzsch2019cadd} have been considered as weak evidence (PP3/BP4) for clinical variant interpretation in ACMG guidelines~\citep{richards2015standards}. Using a more sophisticated machine learning approach, a deep generative model EVE based on variational autoencoders~\citep{frazer2021disease} achieved state-of-the-art performance in classifying clinical variants and outperformed conservation-based method in ClinVar~\citep{landrum2014clinvar}. EVE is trained on multiple sequence alignment (MSA) that captures evolutionary-related sequence variations across species, with the goal of reconstructing the MSA from a latent bottleneck. More recently, the emergence of protein language models have expanded the arsenal of unsupervised and weakly-supervised VEP methods with new, powerful tools. Similar to EVE, the MSA transformer~\citep{rao2021msa} is also trained on MSA data, but employs a masked language modeling objective using self-attention mechanisms. 
The AlphaMissense model~\citep{cheng2023accurate} leverages the potential of integrating evolutionary data with protein structural modeling. It's trained on population frequency data, utilizes sequences from MSAs, and incorporates predicted structural contexts, all of which collectively augment its predictive performance. 
Strikingly, ~\citet{brandes2023genome} proposed a zero-shot workflow that adapts ESM1b for protein sequences of any length and employed it to predict potential missense variant impacts in the human genome. 
Unlike the above methods leveraging MSAs, EMS1b-zero shot does not explicitly rely on any evolutionary data.
Breaking the reliance on MSA is significant, especially for orphan genes with poor MSA coverage~\citep{chowdhury2022single,michaud2022language} and for rare variants that are underrepresented in population~\citep{manolio2009finding}. 
The zero-shot ESM-1b outperformed EVE, suggesting that large protein language models have learned and generalized over evolutionary constraints for the VEP task. 

However, unsupervised VEP methods are not disease specific, substantially limiting their utility and adoption in clinical variant interpretations at point-of-care. This is especially problematic when different variants in a single gene would lead to various closely-related, but distinct disease phenotypes, such as in cardiomyopathies~\citep{mcnally2015genetic,zhang2021disease}. For each missense genetic variant, unsupervised VEP will yield a score representing whether the variant is damaging to the disease-relevant protein function or structure, without distinguishing important disease-specific parameters underlying the gene-disease relationship. Such gene-disease relationships include the distinction between gain-of-function (GoF) vs loss-of-function (LoF), and primary vs modifier effects, etc. For instance, GoF vs LoF may lead to distinct phenotypes under a disease-specific context. Computational modeling for disease-specific variants is difficult, because pathogenic and benign variant annotations are even more sparse when restricted to a single disease condition. Failure to account for these disease-specific information will decrease the predictive power, and more importantly, lead to incorrect clinical decisions and sub-optimal patient management.  

To address this challenge, we introduce a disease-specific language model approach for variant pathogenicity, designed to more effectively capture the pseudo-log-likelihood ratio of both rare missense variants and non-coding variants in disease-specific contexts. Our \textsc{dyna} approach, i.e., \textsc{d}isease-specificit\textsc{y} fine-tu\textsc{n}ing via a Si\textsc{a}mese neural network, is an analogy to the concept of semantic textual similarity (STS)~\citep{han2013umbc_ebiquity} in natural language processing (NLP), where a score is assigned to measure the similarity between two text segments~\citep{reimers2019sentence}.
Employing a Siamese neural network, the similarity between the wild-type sequence and the mutated sequence determined through embeddings generated by two language model branches that share weights. Concurrently, the pre-trained knowledge within the genomic foundation model is fine-tuned using a limited set of disease-specific variant annotations.

\textsc{dyna} is broadly applicable to all genomic foundation models, including both protein language models and DNA foundation models.
We applied \textsc{dyna} to the coding VEP in cardiovascular diseases, and the non-coding VEP of RNA splicing regulation.
These two tasks cover a wide range of specific disease-gene relationships and disease-causing mechanisms, therefore their performance will inform the general utility of \textsc{dyna}.
In both cases, \textsc{dyna} is fine-tuned on small, rare variants and demonstrate superior VEP accuracy on large, human-annotated variant sets of ClinVar~\citep{landrum2014clinvar}.
\textsc{dyna} provides a powerful computational framework for deploying various pretrained genomic foundation models to meet the specific clinical needs.

\section{Results}\label{sec-results}

\subsection{Overview of the DYNA Framework}\label{sec:dyna_framework}

\begin{figure}[!htbp]
    \centering
     \begin{minipage}{0.2\textwidth}
     \begin{subfigure}[b]{\linewidth}
     \caption{}\label{fig:input_genomic_foundation_model}
         \includegraphics[width=\linewidth]{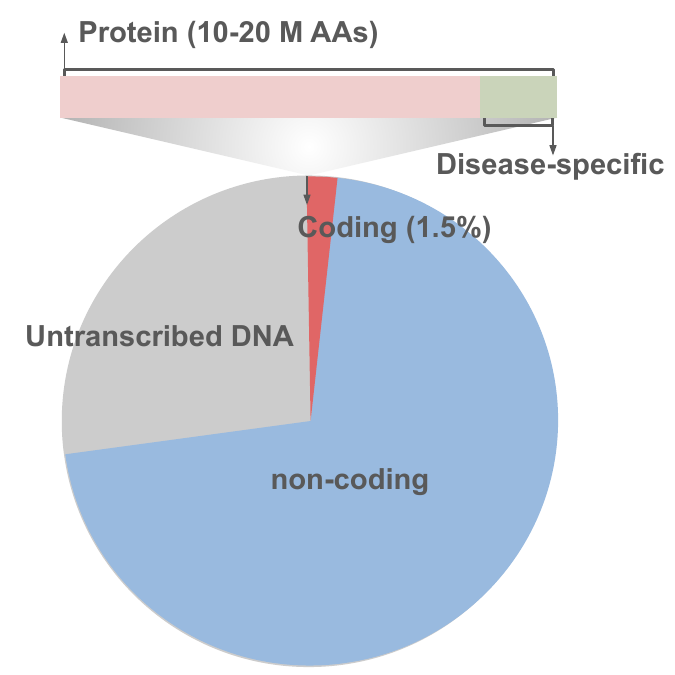}
     \end{subfigure}

             \begin{subfigure}[b]{\linewidth}
             \vspace{25pt}
                      \captionsetup{labelformat=empty}
        \includegraphics[width=\linewidth]{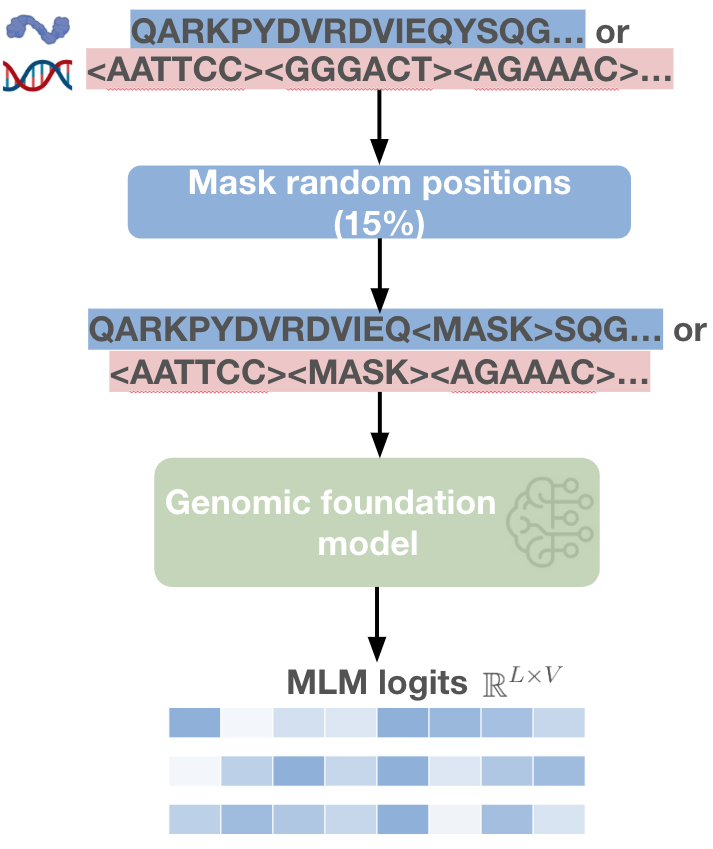}
    \end{subfigure}
     \end{minipage}
     \hfill
    \begin{minipage}{0.73\textwidth}
    \begin{subfigure}[b]{\linewidth}
    \caption{}\label{fig:dyna_framework}
        \includegraphics[width=\linewidth]{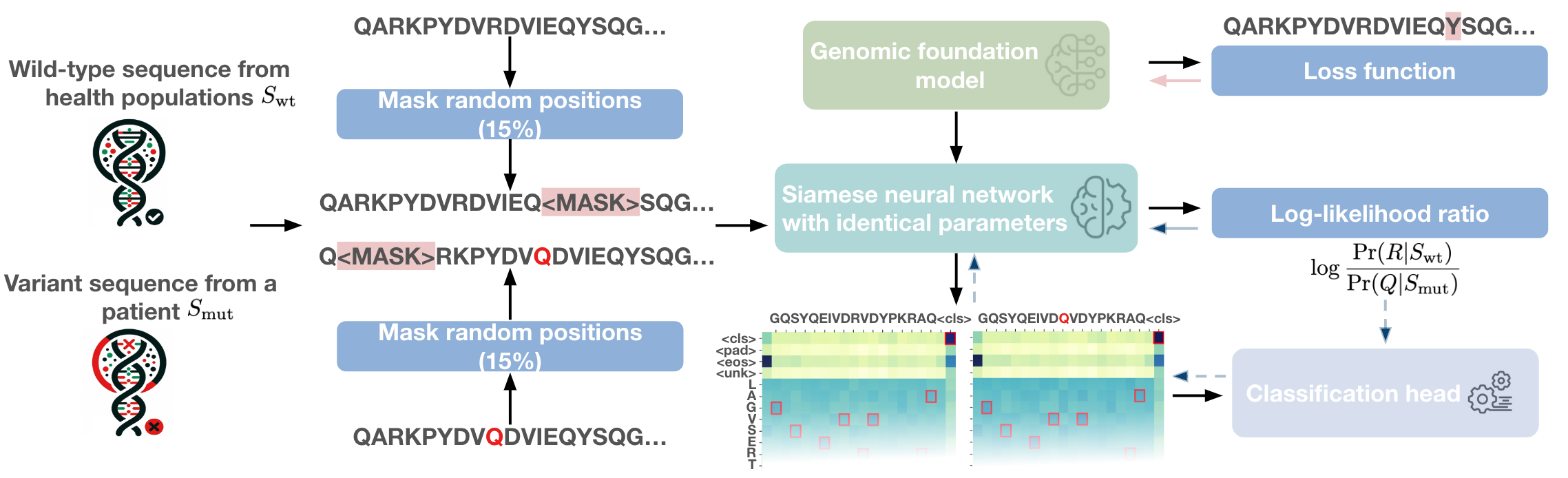}
    \end{subfigure}

        \begin{subfigure}[b]{0.32\linewidth}
              \caption{}\label{fig:PLLR_dyna}
          \includegraphics[width=\linewidth]{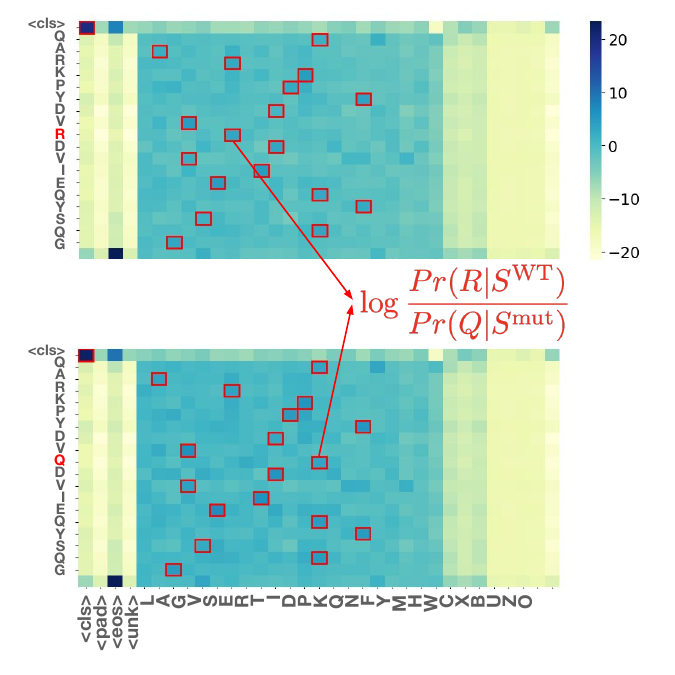}
      \end{subfigure}
     \hfill
    \begin{subfigure}[b]{0.32\linewidth}
           \caption{}\label{fig:PLLR_dyna_cm_mode}
        \includegraphics[width=\linewidth]{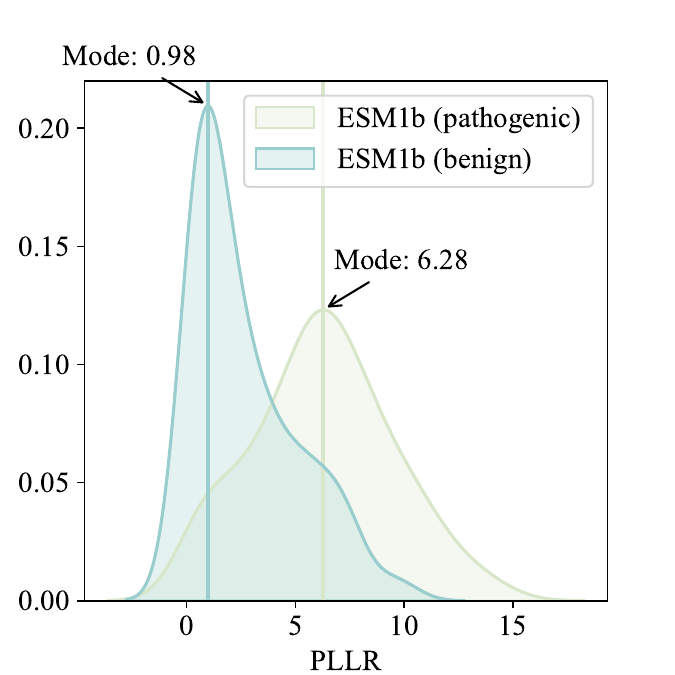}
    \end{subfigure}
    \hfill
    \begin{subfigure}[b]{0.32\linewidth}
           \captionsetup{labelformat=empty}
        \includegraphics[width=\linewidth]{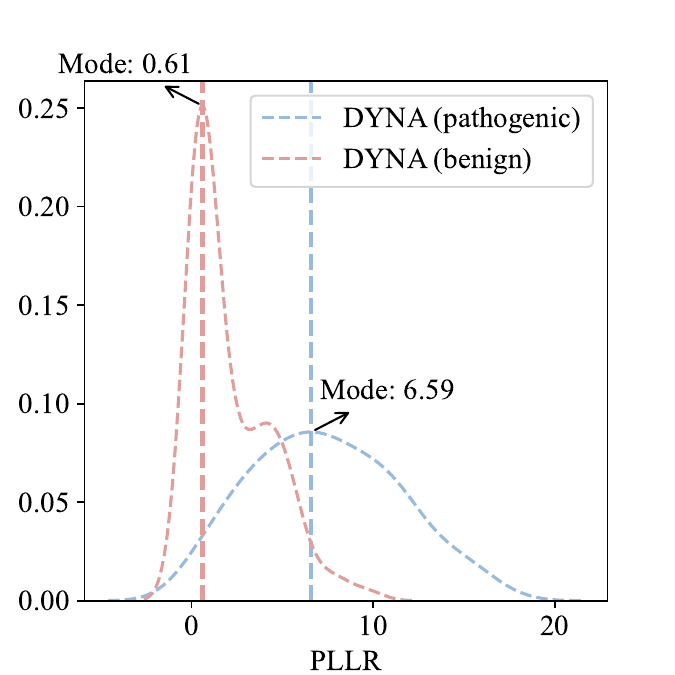}
    \end{subfigure}
    
    \end{minipage}


    \begin{minipage}{0.9\textwidth}   

\begin{subfigure}[b]{\linewidth}
        \caption{}\label{fig:heatmap}
        \begin{tikzpicture}
\def\barlength{11.72} 
\def\coverage{0.67} 
\def\coveragel{2.79} 
\def\coveragen{2.4} 
\def\barheight{0.2} 
\def\start{0.18}
\fill[color=white!60] (0,0) rectangle (\barlength,\barheight);
\fill[color=red!30] (\barlength/2.11,0) rectangle (\barlength/2.11 + \coverage,\barheight);
\fill[color=red!30] (\barlength/1.308,0) rectangle (\barlength/1.308 + \coveragel,\barheight);
\fill[color=blue!30] (\start,0) rectangle (\start+\coveragen,\barheight);


\end{tikzpicture}

        \includegraphics[width=\linewidth]{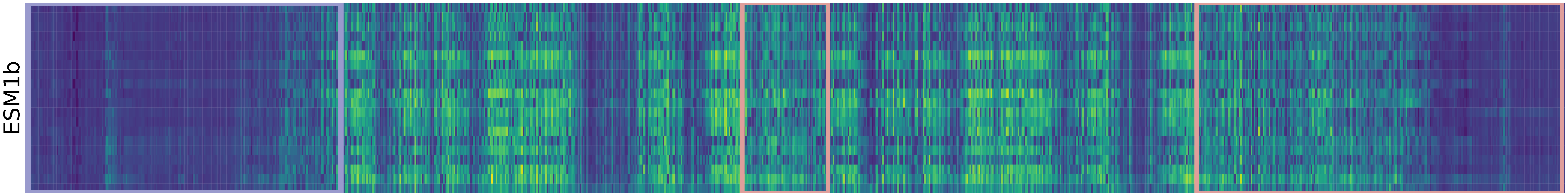}
    \end{subfigure}
    
            \begin{subfigure}[b]{\linewidth}
                    \captionsetup{labelformat=empty} 
        \includegraphics[width=\linewidth]{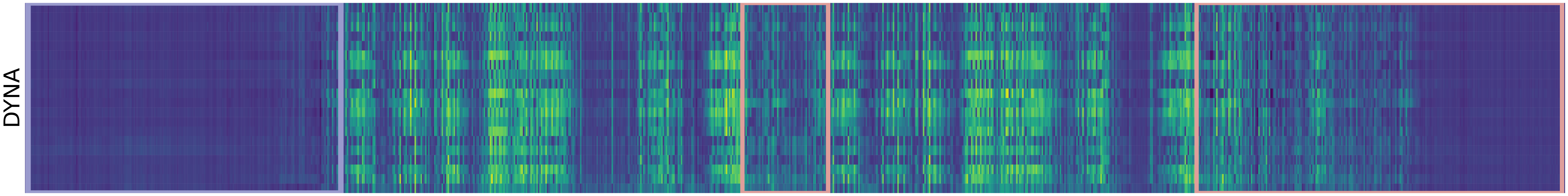}
    \end{subfigure}
    
            \begin{subfigure}[b]{\linewidth}
                    \captionsetup{labelformat=empty} 
        \includegraphics[width=\linewidth]{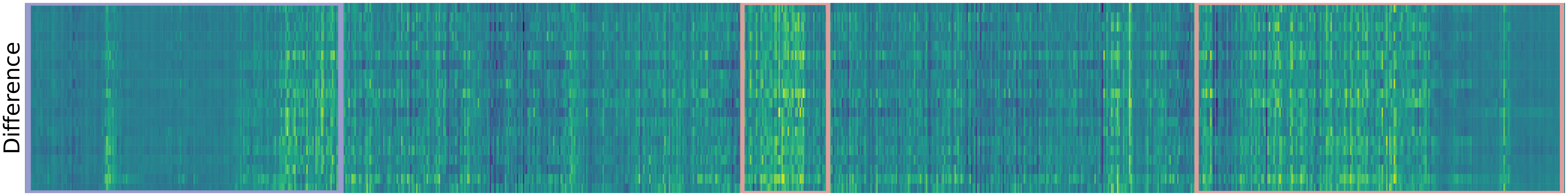}
    \end{subfigure}
    \end{minipage}
    \begin{minipage}{0.037\textwidth}
    \vspace{29pt}
    \hspace{30pt}
                \begin{subfigure}[b]{\linewidth}
                        \captionsetup{labelformat=empty} 
        \includegraphics[width=\linewidth]{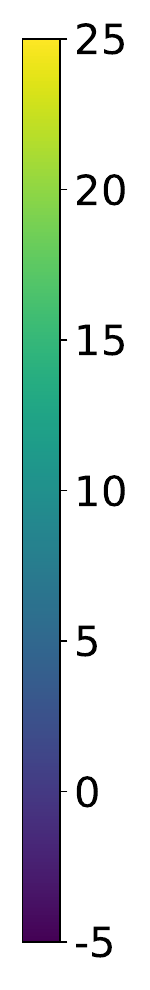}
    \end{subfigure}
                    \begin{subfigure}[b]{\linewidth}
                        \captionsetup{labelformat=empty} 
        \includegraphics[width=\linewidth]{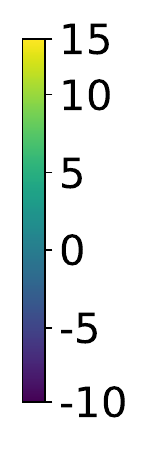}
    \end{subfigure}

\end{minipage}
\caption{
\textbf{\large{a}} In genomic foundation models, we analyze two primary types of biological inputs: protein sequences, representing the coding regions of the genome (approximately 1.5\%), and DNA sequences, corresponding to the non-coding regions. \textbf{\large{b}} \textsc{dyna} incorporates a Siamese network to enhance the analysis of genomic sequences through and similarity comparison. \textbf{\large{c}} Illustration of PLLR Computation in \textsc{dyna} for a pair of wild-type and mutated sequences. \textbf{\large{d}} The distributions of PLLR values for benign and pathogenic sequences under both the ESM1b and \textsc{dyna} models on cardiomyopathies were compared, showing variations in model performance. \textbf{\large{e}} Attribution matrix that visualizes the PLLR scores for all $V \times L$ possible missense variants, where $V$ denotes the vocabulary size and $L$ denotes the protein length. The attribution matrix sequentially displays the PLLR values from the ESM1b model, the \textsc{dyna} model, and their differences, with cytoplasmic domains marked in red and non-cytoplasmic domains in blue.}\label{fig:cm_arm}
\end{figure}
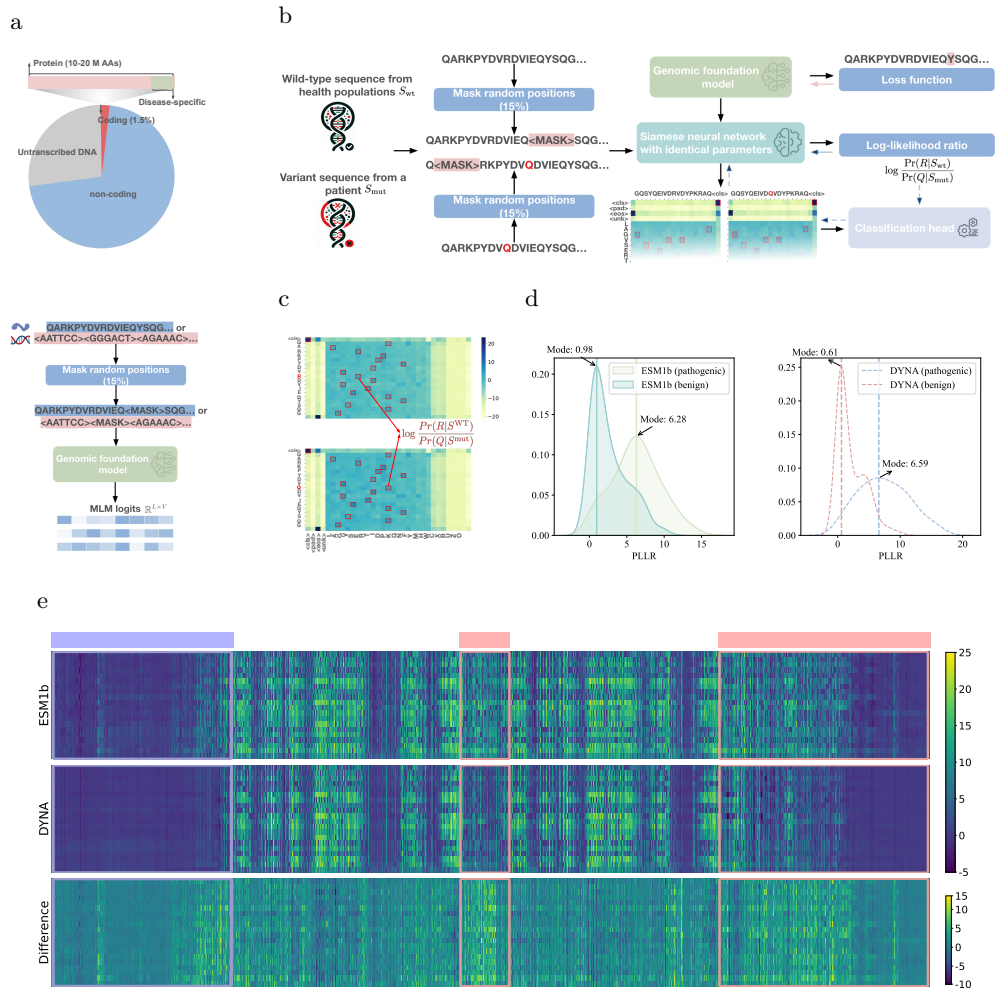

To compute the similarity between the wild-type sequence and the mutated sequence, \textsc{dyna} employs a Siamese neural network on embeddings generated by two language model branches that share weights. In the context of genomic foundation models, we analyze two distinct types of biological inputs: protein sequences, which represent the coding regions of the genome, and DNA sequences, corresponding to both coding and non-coding regions (shown in Figure~\ref{fig:input_genomic_foundation_model}). This dual approach allows us to comprehensively assess the functional impacts of genetic variations on both the protein products and the regulatory elements. The genomic foundation models employ a Masked Language Model (MLM) setting. Within this MLM paradigm, specific amino acid residues/DNA tokens in these sequences are ``masked'' or hidden, and the model is pre-trained to predict the identity of these masked residues. As the model makes these predictions, it produces raw scores or predictions for each potential amino acid or DNA letters that could replace the masked token, commonly referred to as ``MLM logits''. These are represented in the format $\mathbb{R}^{L \times V}$, where $L$ is the length of the input sequence and $V$ is the size of the vocabulary. The pre-trained model uses backpropagation based on these raw scores and the ground truth tokens, depicted as a red arrow in Figure~\ref{fig:dyna_framework}. 

\textsc{dyna} is designed with various loss functions and optimization schemes, tailored to support both protein and DNA foundation models, respectively. This is because creating semantically meaningful MLM logits poses unique challenges for protein and DNA sequences due to their disparate vocabulary sizes. Protein language models, constrained by the relatively small set of amino acids, allow clearer semantic interpretation of MLM logits. In contrast, DNA models must manage a vastly larger vocabulary, ranging from $4^k$ possible $k$-mer combinations to byte pair encoding (BPE) tokenization~\citep{zhou2023dnabert}, which complicates semantic clarity due to the extensive dictionary size. To effectively address these challenges, we apply distinct loss functions tailored to each model type: a pseudo-log-likelihood ratio (PLLR) for protein language models for coding VEPs to enhance the semantic relevance of sequence variations, and combined loss (contrastive + cross-entropy loss) for DNA foundation models for non-coding VEPs to effectively minimize prediction errors across a diverse array of tokens. The PLLR in Figure~\ref{fig:PLLR_dyna} is computed by taking the absolute difference between the pseudo-log-likelihood (PLL) of the wild-type sequence and the PLL of the mutated sequence, indicating the relative change in likelihood caused by the mutation. The PLLR are backpropagated through the siamese network as depicted by the blue bold arrow. For non-coding VEPs, the wild-type sequence and its corresponding mutation are fed into \textsc{dyna} similarly. A contrastive loss of the concatenated MLM logits is backpropagated through the siamese network as shown in the blue dotted arrows (in Figure~\ref{fig:dyna_framework}).


The superior performance of \textsc{dyna} are attributable to two key improvements: PLLR distribution (Figure ~\ref{fig:PLLR_dyna_cm_mode}) and amino acid resolution (Figure \ref{fig:heatmap}). 
Figure\ref{fig:PLLR_dyna_cm_mode} shows \textsc{dyna}'s ability to differentiate between pathogenic and benign variants in cardiomyopathies by adjusting the PLLR values more distinctly than ESM1b. Meanwhile, Figure~\ref{fig:heatmap} illustrates \textsc{dyna}'s enhanced accuracy in identifying pathogenic variants at the amino acid level, prompting us to further investigate the positional basis of disease-specific pathogenic variant intolerance in protein domains. We aimed to assess the similarities and differences in variant intolerance between the models at the resolution of single amino acids. Given that zero-shot ESM1b underperformed on disease-specific variant sets, we hypothesized that \textsc{dyna}'s enhanced performance stems from more precisely mapped pathogenic variant intolerance that is specific to cardiomyopathies and arrhythmias. To quantify these positional effects after fine-tuning, we employed In Silico Mutagenesis (ISM) to calculate the difference in PLLR between ESM1b and \textsc{dyna}. ISM involves systematically replacing each amino acid in the SCN5A protein sequence and computing the resulting PLLR differences. This analysis was conducted separately on the same protein sequence using both the ESM1b and \textsc{dyna} models, visualizing PLLR scores for all $V \times L$ possible missense variants, where $V$ represents the vocabulary size and $L$ represents the protein length. The attribution matrix displays these PLLR values sequentially for the ESM1b model, the \textsc{dyna} model, and their differences, highlighting cytoplasmic domains in red and non-cytoplasmic domains in blue. Our findings confirm that \textsc{dyna}'s ISM not only successfully removes false positives in both cytoplasmic and non-cytoplasmic domains but also shows the model's refined ability to precisely identify and differentiate pathogenic variants. 

We apply \textsc{dyna} on both coding and non-coding VEP tasks to demonstrate its intra- and inter-gene generalization capability. For coding VEPs, we evaluated \textsc{dyna}'s performance against conventional neural networks and pre-trained protein language models, utilizing clinical rare missense variant sets in cardiovascular diseases~\citep{zhang2021disease}. For non-coding VEPs, we focus on a dataset compiled from multiplexed functional assay of splicing using Sort-seq (MFASS)~\citep{chong2019multiplexed} that has demonstrated how rare genetic variants can lead to substantial splicing disruptions. For more details about the dataset and setting-up, we recommend the readers refer to Section~\ref{sp:dataset} in the supplementary material. 
In both cases, we expand the \textsc{dyna} fine-tuned models to ClinVAR, and demonstrate their generalization to large, clinically-relevant inter-gene variant annotations via leveraging small, rare variant sets and pre-trained genomic foundation models.

\subsection{Assessing DYNA's intra-gene generalization ability on Inherited CM and ARM}
To demonstrate the clinical utility of \textsc{dyna} in distinguishing disease-specific missense variants, we first focus on two common inherited cardiovascular diseases: inherited cardiomyopathies (CM) and arrhythmias (ARM).
It is known that different variants in a single cardiovascular disease-relevant gene would lead to different disease phenotypes~\citep{mcnally2015genetic}, necessitating more fine-grained disease-specific fine-tune by \textsc{dyna}.
A pre-compiled dataset of rare missense pathogenic and benign variants, defined by a cohort-based approach, in CM and ARM, respectively, were employed. 
Two-thirds were used for training and one-third reserved as a holdout test set, consistent with the previous report~\citep{zhang2021disease}. 
To further evaluate the models' generalization ability, we replicate our observation in another independent, held-out clinically annotated variant set, which includes data collected from multiple centers, by extracting missense ClinVar CM and ClinVar ARM sequences based on ClinVar disease type. 
The statistics for all datasets are shown in Table~\ref{tab:dataset-1}.
The investigation of generalization and replication in genomic foundation models is driven by the imperative to enhance the accuracy and reliability of predictions regarding the clinical consequences of genetic variants. This research is essential for deciphering the complex relationship between genotypic variability within the same gene and phenotypic manifestations, thereby refining our understanding of gene function and its implications for disease. 
By ensuring that predictive models can robustly handle various mutations within a gene, we aim to improve the utility of these models in clinical diagnostics and personalized medicine. 
Thus, in this subsection, we will show \textsc{dyna} effectively predicts missense coding variant effects for cardiovascular diseases, i.e., CM, ARM, ClinVar CM, and ClinVar ARM.

For both CM and ARM, we fine-tuned two base protein language models on \textsc{dyna} framework, i.e., esm1b\_t33\_650M\_UR50S (ESM1b)~\citep{rives2021biological} and esm2\_t33\_650M\_UR50D (ESM2)~\citep{lin2023evolutionary}. Because ESM1b consistently outperformed ESM2, we refer to the \textsc{dyna} fine-tuned ESM1b as \textsc{dyna} for simplicity in the following sections, whereas ESM2 and \textsc{dyna} fine-tuned ESM2 results can be found in Figure~\ref{fig:esm2}.
This demonstrates that a larger language model does not always perform better on specific downstream tasks, and highlights the significance of \textsc{dyna} as a universal disease-specific fine-tuning framework.

\subsubsection{DYNA effectively identifies pathogenic and benign rare missense genetic variants over ESM1b.}

We fine-tuned \textsc{dyna} on the CM training set with 238 pathogenic and 202 benign variants, then tested it on the testing set with 118 pathogenic and 100 benign variants.
Importantly, both the training and testing variant set are rare (gnomAD minor allele frequency <0.1\%), representing a challenging VEP task that will unlikely be characterized by evolutionary conservation-based methods.
In Figure~\ref{fig:PLLR_dyna_cm}, we illustrate the distribution of the testing PLLR values for benign and pathogenic sequences following fine-tuning with \textsc{dyna} on CM. A one-sided p-value test was conducted between the pathogenic and benign sequences, yielding a p-value of $7.085e-22$, indicating that DYNA can effectively discriminate between pathogenic and benign sequences. In contrast, 
The ESM1b model exhibits a smaller Kullback-Leibler (KL) divergence between benign and pathogenic sequences, specifically $24.2143$ (Figure~\ref{fig:PLLR_dyna_ESM1b_cm}). Post fine-tuning with \textsc{dyna}, the PLLR distribution for benign data shifts closer to zero, whereas the PLLR values for pathogenic sequences are notably higher compared to ESM1b.

This pattern of PLLR distribution shift is consistent with ARM, where we fine-tuned with 168 pathogenic and 158 benign variants, then tested on 84 pathogenic and 79 benign variants. On the ARM test set, the pathogenic sequences have significantly higher PLLRs than benign sequences ($6.174e-20$, one-sided paired t-test; Figure~\ref{fig:PLLR_dyna_arm}) and demonstrates a larger KL divergence than ESM1b (Figure~\ref{fig:PLLR_dyna_ESM1b_arm}).

These findings underscore \textsc{dyna}'s effectiveness in learning disease-specific information, thereby enhancing its ability to distinguish between pathogenic and benign rare missense genetic variants associated with CM and ARM.

\subsubsection{DYNA outperforms baseline methods on cardiovascular diseases.}
We further assessed the capability of \textsc{dyna} to discern disease-specific information pertinent to CM by comparing its performance against nine baseline methods. These methods encompass three tree-based models—classification and regression tree (CART)~\citep{breiman2017classification}, random forest (RF)~\citep{breiman2001random}, and Bayesian additive regression trees (BART)~\citep{chipman2010bart}; three boosting algorithms—XGBoost~\citep{chen2016xgboost}, AdaBoost~\citep{ying2013advance}, and M-CAP~\citep{jagadeesh2016m}; one ensemble model, REVEL~\citep{ioannidis2016revel}; and two language models, ESM1b and ESM2. 
The comparative results are illustrated in Figure~\ref{fig:auc_cm_error} and Figure~\ref{fig:aupr_cm}. The first five comparison methods and \textsc{DYNA} are disease-specific models, while others are non-disease-specific. 
The best performances in terms of AUC and AUPR are shown in bold in Table~\ref{tab:cm-baselines}. Our \textsc{dyna}, utilizing ESM1b as its foundational model, outperforms all nine baselines with an AUPR of $0.91$ in Figure~\ref{fig:aupr_cm}. 
Because pathogenic variants are usually substantially fewer than benign variants (Table ~\ref{tab:dataset-1}), and clinically, the precision-recall of accurately calling pathogenic variants accurately is of interests and better characterized by AUPR.
Nevertheless, \textsc{dyna} remains highly competitive when evaluated by AUC. A bootstrapping analysis, performed to evaluate the robustness of the AUC measurement due to the relatively small test set, shows that the AUC of \textsc{dyna} is non-inferior (Figure~\ref{fig:auc_cm_error}).



Finally, we compare \textsc{dyna} performances in ARM with the nine benchmark methods shown in Figure~\ref{fig:auc_arm_error} and~\ref{fig:aupr_arm}, detailed in Table~\ref{tab:arm-baselines}, where the best AUC and AUPR are highlighted in bold. With ESM1b as its base, our \textsc{dyna} surpasses all nie baselines, recording an AUC of $0.94$ and an AUPR of $0.95$. This performance marks a $3\%$ AUPR improvement over the second-best performer, BART (indicated in italic). Such finding further shows \textsc{dyna} sets a new benchmark by obtaining the highest statistics across all baselines in ARM dataset. The detailed 1000 bootstrap results are shown in Figure~\ref{fig:auc_arm_error} and~\ref{fig:aupr_arm_error}.

Together, the superior and consistent performance of \textsc{dyna} on the rare variant testing sets of CM and ARM reaffirms the capability of \textsc{dyna} as a robust tool for disease-specific VEP in discerning disease associations.

\subsubsection{Replication of DYNA in ClinVar.}

\begin{figure}[!htbp]
    \centering
    
        \begin{subfigure}[b]{0.24\linewidth}
            \caption{}\label{fig:PLLR_dyna_cm}
        \includegraphics[width=\linewidth]{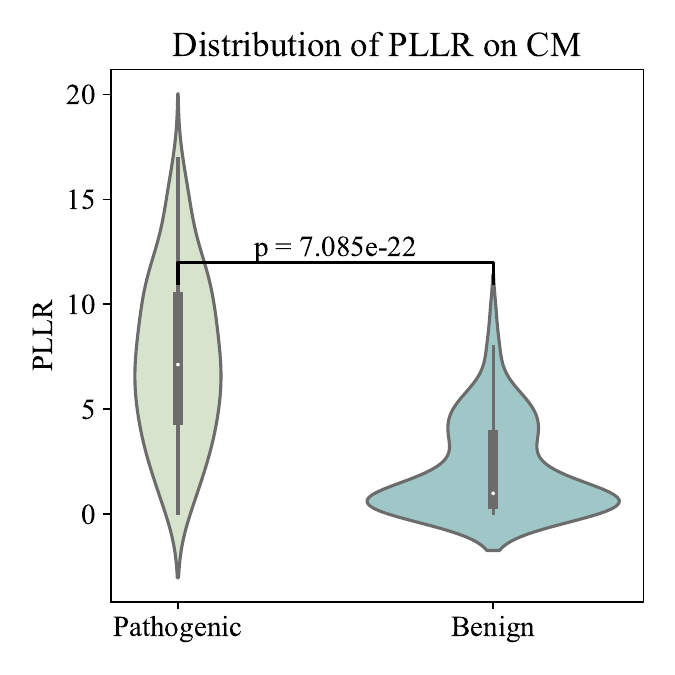}
    \end{subfigure}
    \hfill
    \begin{subfigure}[b]{0.24\linewidth}
            \caption{}\label{fig:PLLR_dyna_arm}
        \includegraphics[width=\linewidth]{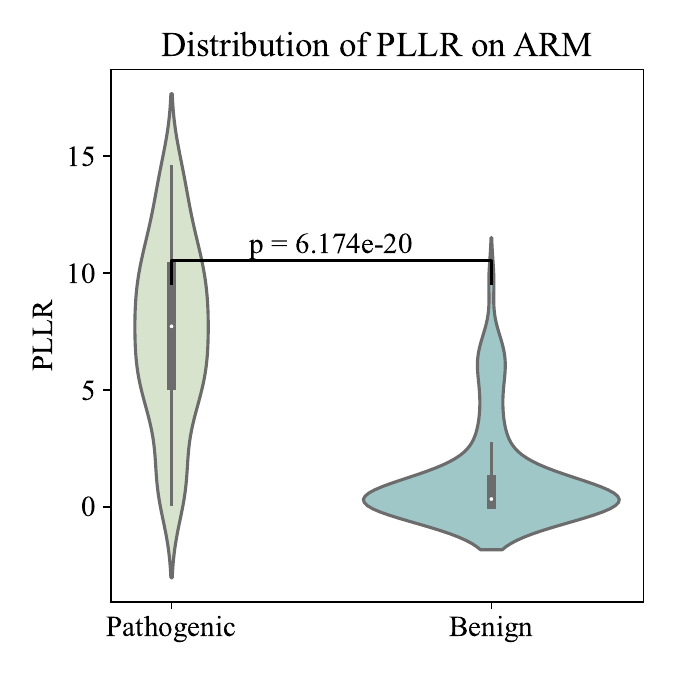}
    \end{subfigure}
    \hfill
    \begin{subfigure}[b]{0.24\linewidth}
           \caption{}\label{fig:PLLR_dyna_ESM1b_cm}
        \includegraphics[width=\linewidth]{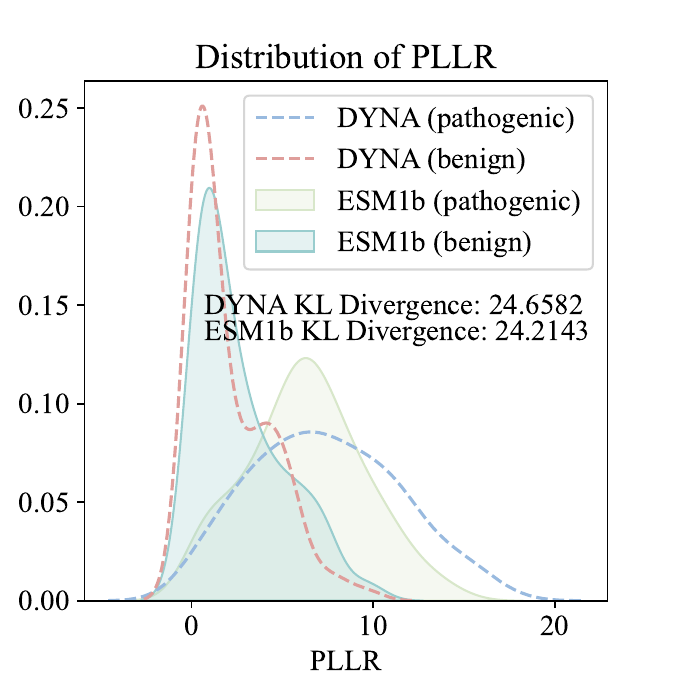}
    \end{subfigure}
    \hfill
    \begin{subfigure}[b]{0.24\linewidth}
            \caption{}\label{fig:PLLR_dyna_ESM1b_arm}
        \includegraphics[width=\linewidth]{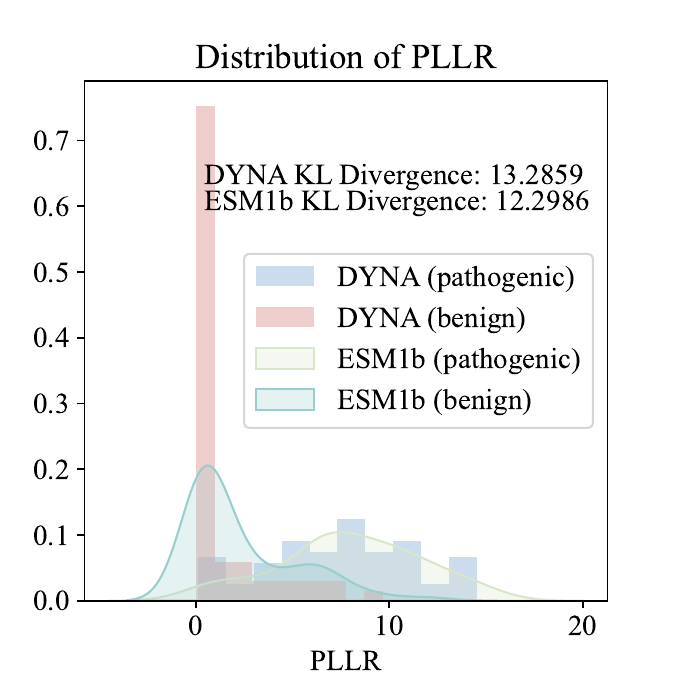}
    \end{subfigure}

       \begin{subfigure}[b]{0.48\linewidth}
            \caption{}\label{fig:aupr_cm}
        \includegraphics[width=\linewidth]{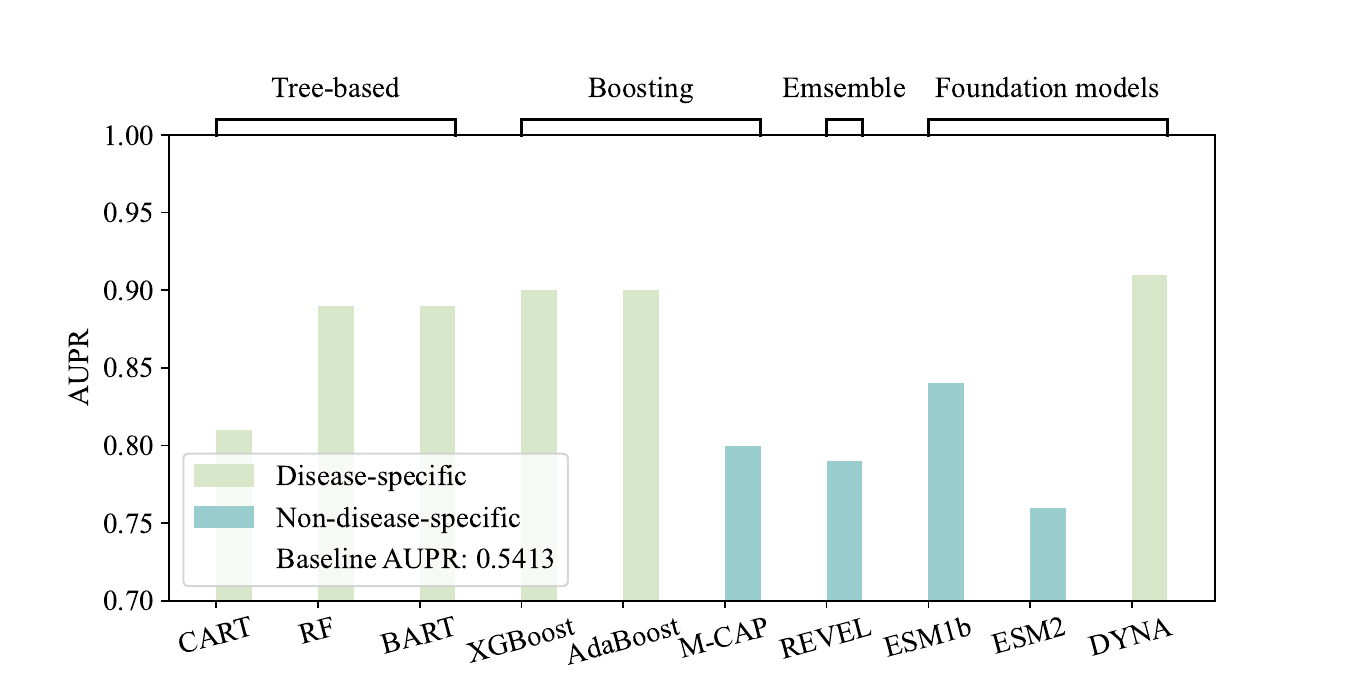}
    \end{subfigure}
    \hfill
       \begin{subfigure}[b]{0.48\linewidth}
            \caption{}\label{fig:aupr_arm}
        \includegraphics[width=\linewidth]{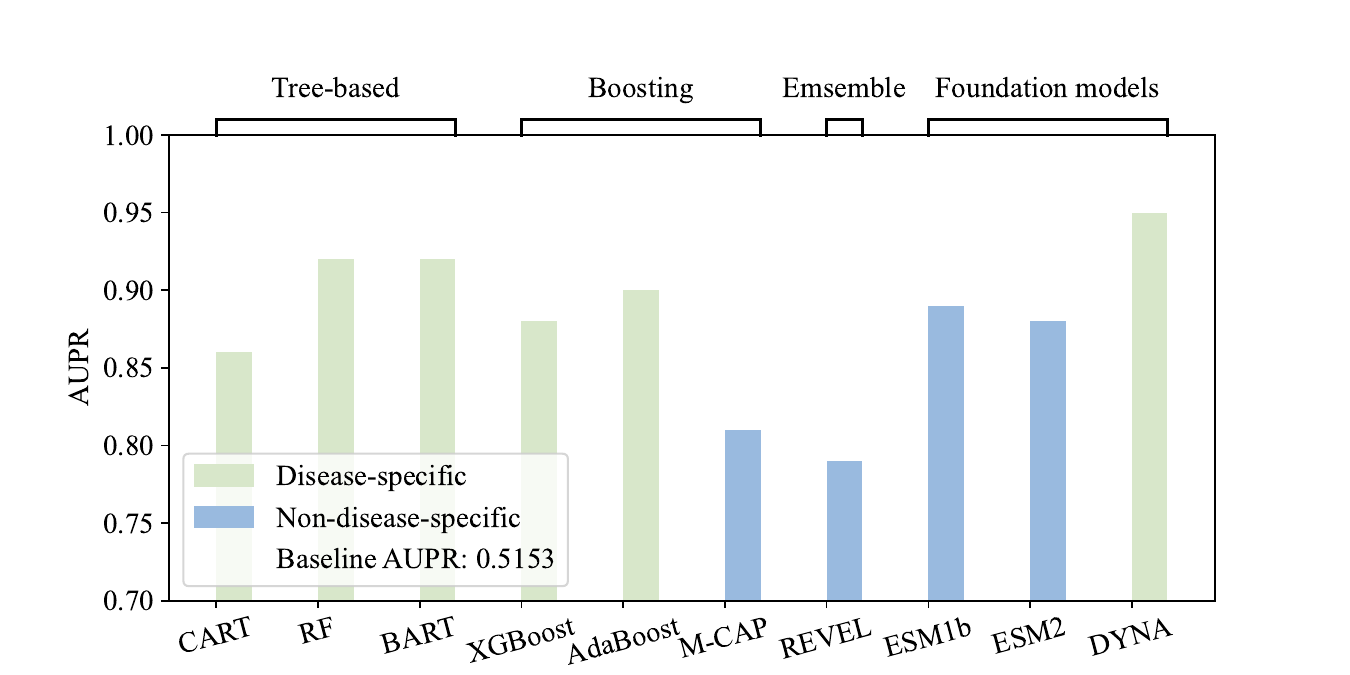}
    \end{subfigure}

        \begin{subfigure}[b]{0.24\linewidth}
            \caption{}\label{fig:venn_cm}
        \includegraphics[width=\linewidth]{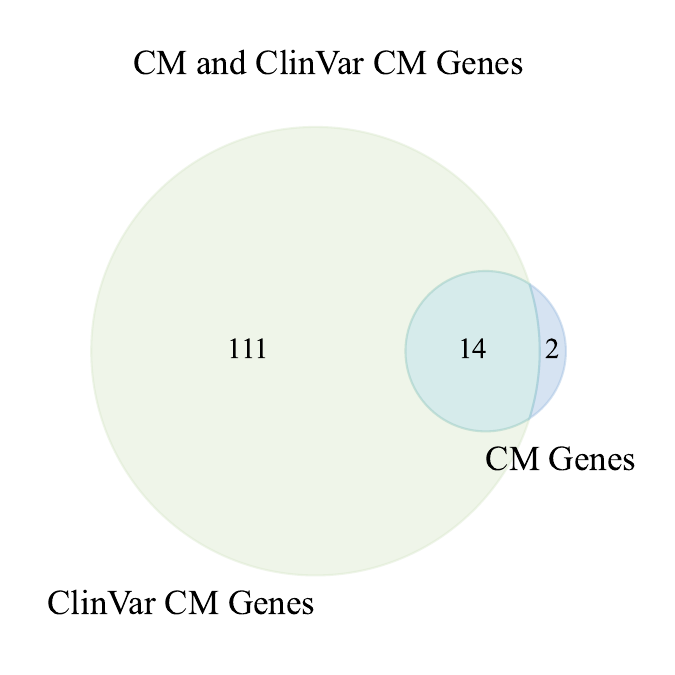}
    \end{subfigure}
\hfill
        \begin{subfigure}[b]{0.24\linewidth}
            \caption{}\label{fig:venn_arm}
        \includegraphics[width=\linewidth]{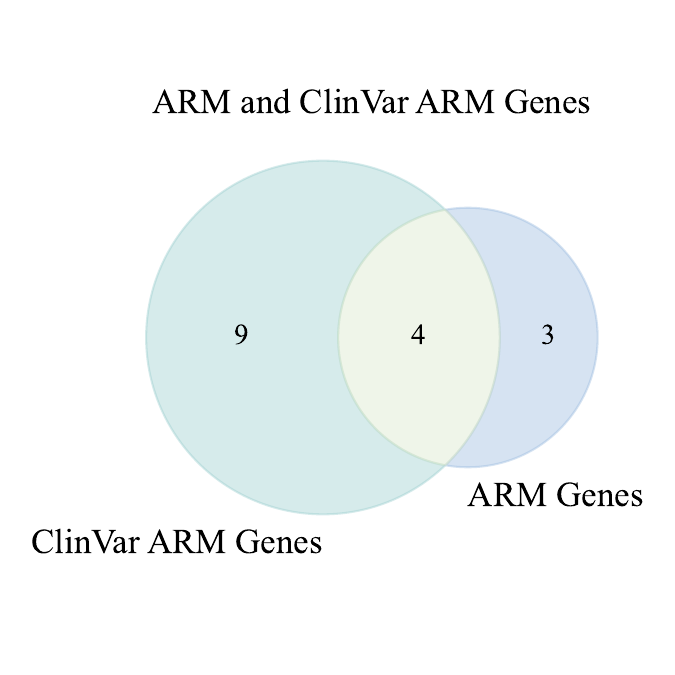}
    \end{subfigure}
\hfill
            \begin{subfigure}[b]{0.24\linewidth}
            \caption{}\label{fig:clinvar_cm_aupr}
        \includegraphics[width=\linewidth]{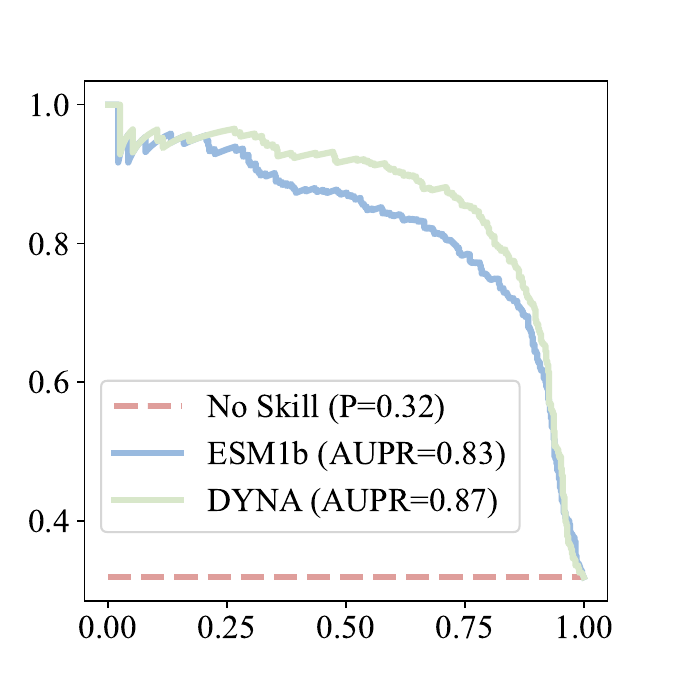}
    \end{subfigure}
\hfill
        \begin{subfigure}[b]{0.24\linewidth}
            \caption{}\label{fig:clinvar_arm_aupr}
        \includegraphics[width=\linewidth]{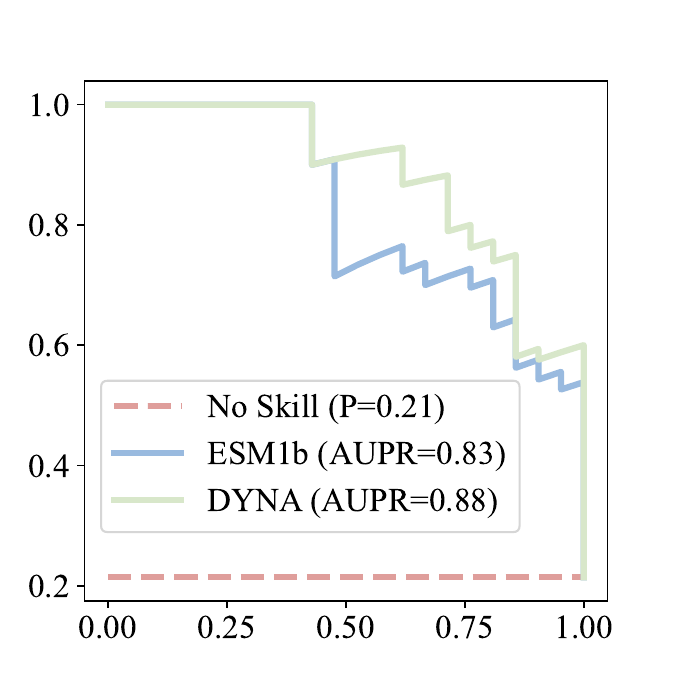}
    \end{subfigure}

    \begin{subfigure}[b]{\linewidth}
            \caption{}\label{fig:cm_gmm}
        \includegraphics[width=\linewidth,height=3cm]{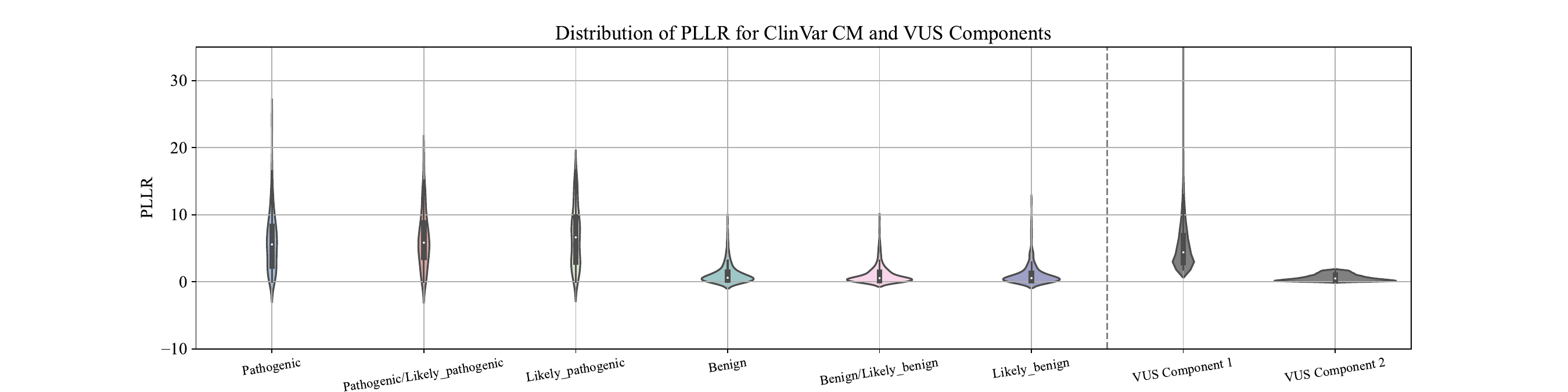}
    \end{subfigure}

\caption{  \textbf{\large{a}} The distribution of PLLR values for benign and pathogenic sequences on CM, where a one-sided p-value test confirmed significant differentiation between the pathogenic and benign sequences. \textbf{\large{b}} Similarly, we evaluated the PLLR distribution for benign and pathogenic sequences within ARM. \textbf{\large{c}} The distributions of PLLR values for benign and pathogenic sequences under both the ESM1b and \textsc{dyna} models on CM were compared, showing variations in model performance. \textbf{\large{d}} For ARM, PLLR distributions under the ESM1b and \textsc{dyna} scenarios were analyzed. \textbf{\large{e}} AUPR performances on CM for \textsc{dyna} and baselines. \textbf{\large{f}} AUPR performances on ARM for \textsc{dyna} and baselines.  }

\end{figure}

To evaluate the generalization adaptability and replication of disease fine-tuned \textsc{dyna} to clinically annotated pathogenic variants, we conduct zero-shot testing on ClinVar CM and ARM, a broader dataset sampled from ClinVar based on disease name, differing from inherited CM and ARM because it's a database collected from multiple centers across the globe. 
Indeed, the ClinVar datasets exhibit divergence from the distribution of inherited CM and ARM genes. In Figure~\ref{fig:venn_cm}, there are $125$ variants from ClinVar CM genes and $16$ variants from CM genes, with $14$ overlapping genes but variants occurring at distinct positions. Figure~\ref{fig:venn_arm} illustrates $13$ variants from ClinVar ARM genes and $7$ from ARM genes, sharing $4$ overlapping genes but differing in variant positions. 

Our results demonstrate that \textsc{dyna}'s superior performance is successfully replicable, as evidenced by its performance on the ClinVar CM and ARM datasets. 
In both ClinVar CM and ARM, \textsc{dyna} outperformed its base ESM1b by a large margin, as evaluated by AUPR (4\% improvement in CM, 5\% improvement in ARM; Figure ~\ref{fig:clinvar_cm_aupr}-\ref{fig:clinvar_arm_aupr}).
The improvement is significant, because it demonstrates that \textsc{dyna} learns the CM and ARM disease-specific variant pathogenecity patterns in small (fewer than 450 training examples for CM, and fewer than 250 training examples for ARM) rare variant set, and successfully generalizes to the independent ClinVar datasets.
Thus, the generalization ability and replication are rigorously demonstrated through zero-shot testing on the ClinVar CM and ARM datasets, which are broader and encompass a more varied set of genetic conditions compared to the initial training datasets. 

To further understand the effect of \textsc{dyna}'s disease-specific fine-tuning and the source of its improved performance over ESM1b, we characterize \textsc{dyna}'s disease-specific prediction on ClinVar CM and ARM, and compare them to generic predictions from the base model ESM1b. 
For benign sequences, \textsc{dyna} demonstrates superior discriminative ability, with 95\% of its predicted PLLR values falling below 3.2, in contrast to a higher median PLLR value of 4.4 observed with ESM1b (Figure~\ref{fig:cm_benign_box}).
This pattern is highly consistent between ClinVar CM and ARM (Figure ~\ref{fig:arm_benign_box}).
Conversely, for pathogenic sequences, \textsc{dyna} consistently generates comparable PLLR values to ESM1b in both ClinVar CM and ARM. 
By comparing the distributions of benign and pathogenic sequences from ClinVar CM (Figure~\ref{fig:tpr_cm} and ClinVar ARM (Figure~\ref{fig:tpr_arm}), \textsc{dyna} achieved high true positive rates of 0.79 and 0.95 while controlling for false positive rate less than 0.15 for ClinVar CM and ARM, respectively, using threshold values derived from the Youden's J statistic.
These results suggest that \textsc{dyna}'s improved performance is driven by pruning false positive predictions, i.e., variants not pathogenic to the target disease but predicted as pathogenic (see Figure ~\ref{fig:heatmap}), explaining the improvement as measured by AUPR.

Next, we employ \textsc{dyna} to further analyze Variants of Uncertain Significance (VUS) using a Gaussian Mixture Model (GMM) to categorize them into pathogenic versus benign classes. This analysis is driven by a fundamental hypothesis in VEP: do VUS exhibit intermediate phenotypic effects between pathogenic and benign variants, or are their effects similar to known categories but obscured by our limited understanding? Figures~\ref{fig:cm_gmm} depict that the PLLR distribution for VUS closely aligns with those of true pathogenic and benign sequences. Specifically, the component labeled VUS\_component\_1 mirrors the pathogenic distribution, while VUS\_component\_2 aligns with the benign distribution, matching the PLLR distributions for ClinVar CM and ARM in their respective common class labels. We also conducted one-sided paired t tests between all pathogenic and benign group pairs across both datasets, consistently revealing p-values smaller than 0.05, thus statistically affirming the significant distinction between these groups (detailed in Figures~\ref{fig:cm_violin_p_value} and~\ref{fig:arm_violin_p_value}).

\subsection{DYNA generalizes to unseen disease-relevant genes.}
Critically, expanding \textsc{dyna} to ClinVar CM and ARM enables a more fine-grained evaluation to the unseen disease-relevant genes (see the number of ClinVar-only genes in the Venn diagrams in Figure~\ref{fig:venn_cm}-~\ref{fig:venn_arm}). Generalization to unseen disease-relevant genes is critical for the applicability of genomic models in the evolving field of genomic medicine, where new disease-relevant genes are continuously being discovered through clinical trials and research~\citep{ata2021recent, chatzianastasis2023explainable}.
Therefore, the ability to generalize disease-specific VEP patterns to unseen disease-relevant genes is highly desirable.

\begin{figure}[!htbp]
    \centering
                    \begin{subfigure}[b]{0.24\linewidth}
                \caption{}\label{fig:auc_radar}
        \includegraphics[width=\linewidth]{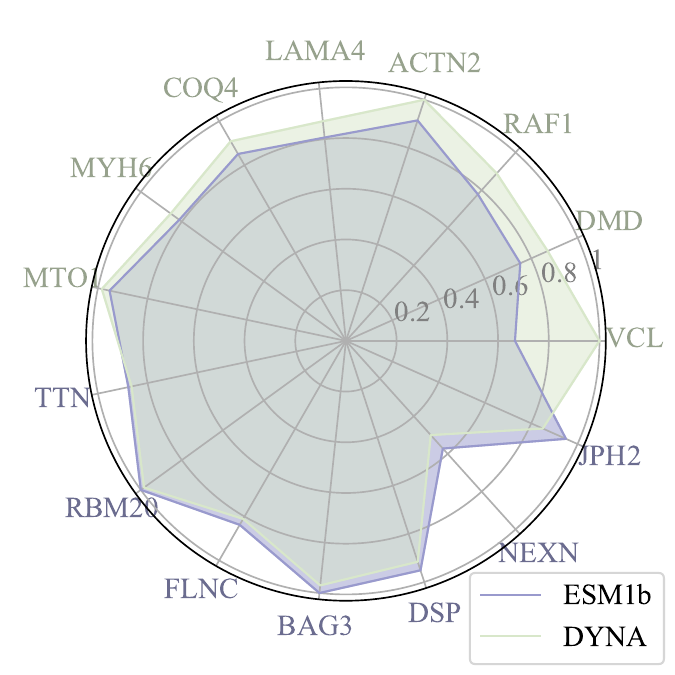}
    \end{subfigure}
    \hfill
    \begin{subfigure}[b]{0.24\linewidth}
            \caption{}\label{fig:aupr_radar}
        \includegraphics[width=\linewidth]{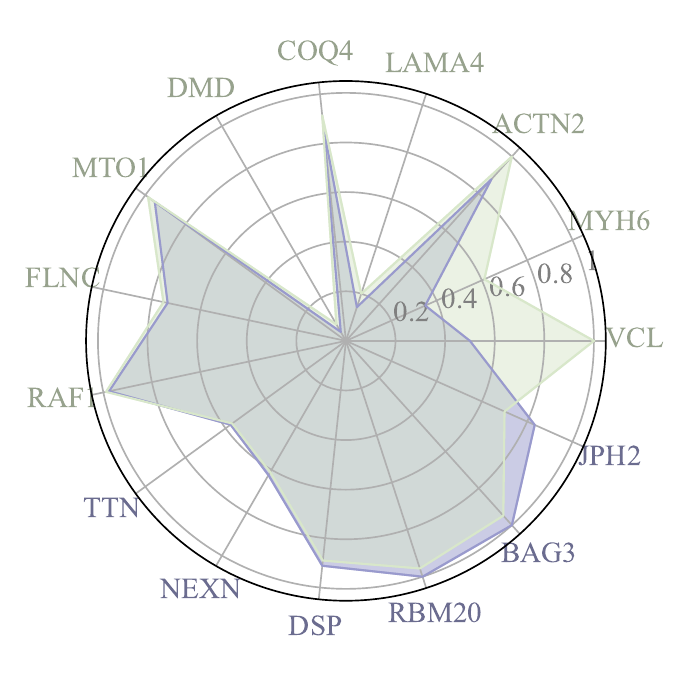}
    \end{subfigure}
\hfill
        \begin{subfigure}[b]{0.48\linewidth}
            \caption{}\label{fig:kde}
        \includegraphics[width=\linewidth]{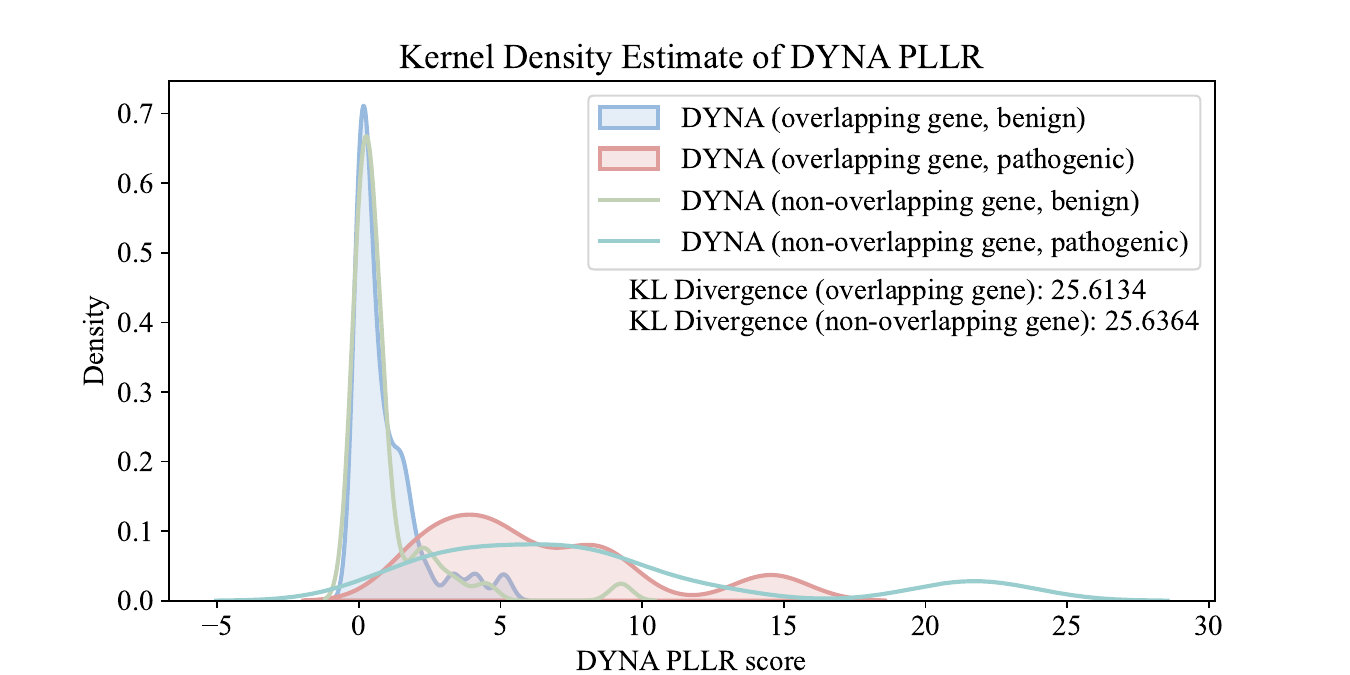}
    \end{subfigure}

    \begin{subfigure}[b]{0.24\linewidth}
            \caption{}\label{fig:auc_clinvar_cm}
        \includegraphics[width=\linewidth]{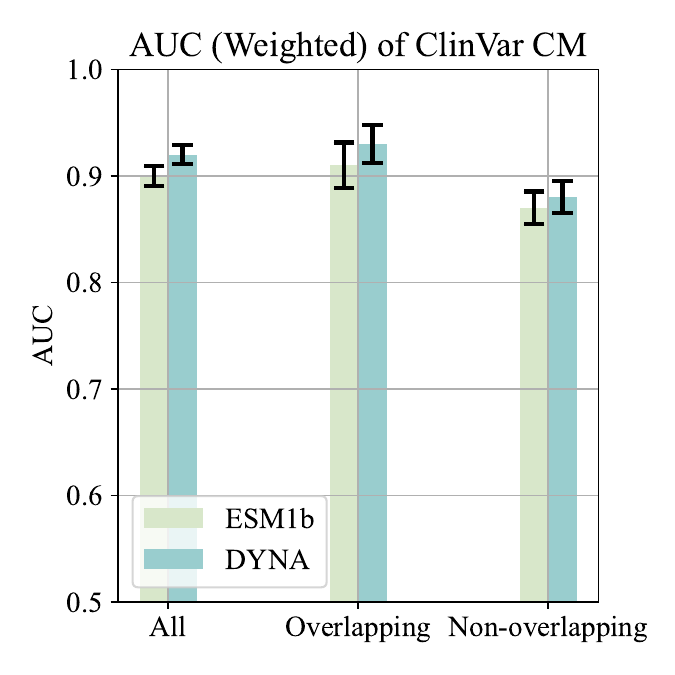}
    \end{subfigure}
        \hfill
    \begin{subfigure}[b]{0.24\linewidth}
            \captionsetup{labelformat=empty}
        \includegraphics[width=\linewidth]{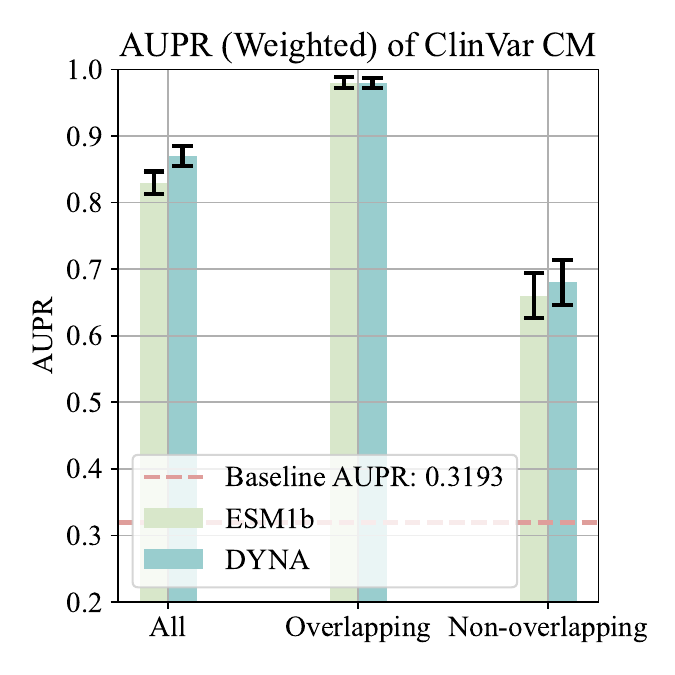}
    \end{subfigure}
                \hfill
        \begin{subfigure}[b]{0.24\linewidth}
               \captionsetup{labelformat=empty}
        \includegraphics[width=\linewidth]{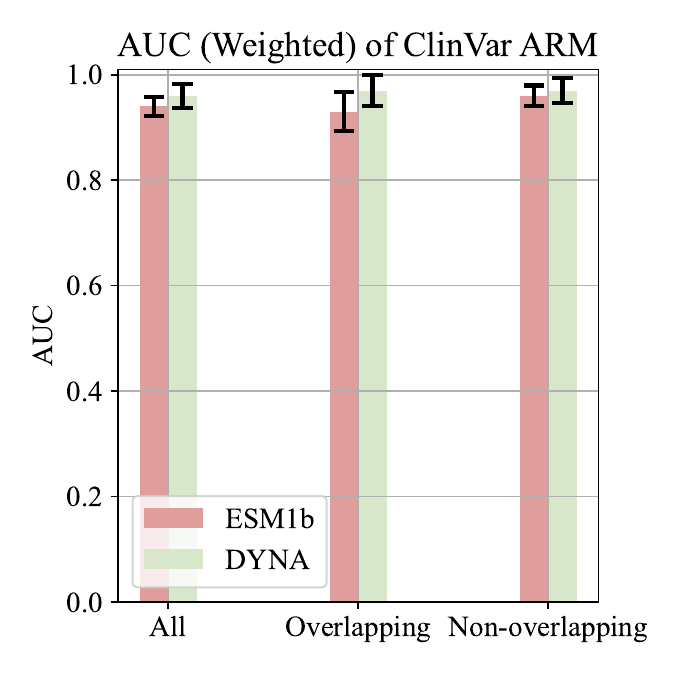}
    \end{subfigure}
        \hfill
    \begin{subfigure}[b]{0.24\linewidth}
           \captionsetup{labelformat=empty}
        \includegraphics[width=\linewidth]{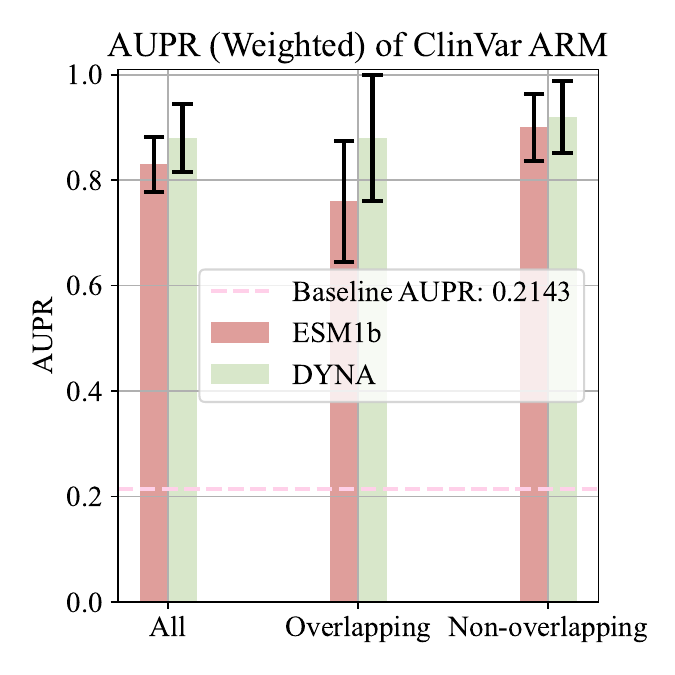}
        
    \end{subfigure}
        \caption{\textbf{\large{a}} Comparison of AUC scores between \textsc{dyna} and ESM1b for non-overlapping genes with diverse mutation positions in the ClinVar CM dataset. The radar chart illustrates \textsc{dyna}'s significantly higher AUC, achieving scores nine times greater than those of ESM1b, along with a detailed representation of the range where \textsc{dyna} shows a sevenfold improvement in the lower AUC range. 
\textbf{\large{b}} Performance comparison using AUPR for \textsc{dyna} and ESM1b, highlighting the results on non-overlapping genes within the ClinVar CM dataset. 
\textbf{\large{c}} Kernel Density Estimate (KDE) plots of the Probabilistic Log-Likelihood Ratio PLLR for overlapping and non-overlapping gene mutations in the ClinVar ARM dataset. 
\textbf{\large{d}}
Graphical representation of AUC and AUPR scores (with 1000 bootstrap) for \textsc{dyna} compared to ESM1b across the ClinVar CM and ARM datasets. These figures detail the performance metrics for weighted PLLR evaluation, demonstrating \textsc{dyna}'s superior generalization capabilities across all groups. Enhanced performance is particularly noted in non-overlapping genes, validating the model's strength in generalization to unseen disease-relevant genes compared to intra-gene generalization.
}
\end{figure}

For robust gene-specific evaluation of \textsc{dyna}'s performance, we first selected a set of non-overlapping genes, each containing more than five positive and negative variants. Figure~\ref{fig:auc_radar} illustrates the performance disparity between \textsc{dyna} and ESM1b on these genes with varying mutation positions. 
Remarkably, \textsc{dyna} achieves an AUC that is higher than that of ESM1b for nine out of 15 genes. Similarly, Figure~\ref{fig:aupr_radar} highlights \textsc{dyna}'s superior performance by demonstrating that the AUPR is higher for ten genes over ESM1b for the same gene set on ClinVar CM, underscoring significant enhancements in generalization to unseen disease-relevant genes. Both radar plots show only the genes where there was either improvement or a decrease in AUC scores, omitting genes without changes. 

To compare \textsc{dyna}'s performance on the unseen disease-relevant genes, i.e. non-overlapping genes in ClinVar ARM, to the overlapping genes between ClinVar ARM and ARM datasets, we analyze the Kernel Density Estimate (KDE) of PLLR for overlapping and non-overlapping gene mutations in ClinVar ARM (Figure~\ref{fig:kde}). The KDE plots reveal that the mean PLLR of benign sequences in non-overlapping genes approaches zero, whereas the pathogenic sequences exhibit higher values. The KL-divergence between pathogenic and benign sequences in non-overlapping genes is 25.6364, comparable and even marginally higher than that observed in overlapping genes, indicating a more pronounced distinction and better generalization to unseen disease-relevant genes by \textsc{dyna}.

Finally, we show that \textsc{dyna} outperforms ESM1b on variants in both overlapping and non-overlapping genes consistently across ClinVar CM and ARM, demonstrating that the disease-specific patterns learned by \textsc{dyna} are generalized to unseen disease-relevant genes. 
Figures~\ref{fig:auc_clinvar_cm} depicts the AUC and AUPR for both weighted and non-weighted PLLR evaluations (in Figure~\ref{fig:auc_aupr_non_weighted}) in ClinVar CM and ARM datasets. These results reveal two key insights: firstly, \textsc{dyna} exhibits superior generalization to unseen disease-relevant genes capabilities compared to ESM1b across all examined groups. Secondly, enhanced performance is evident in non-overlapping genes versus overlapping genes within the ARM dataset, reaffirming the model’s robustness in generalization to unseen disease-relevant genes over intra-gene generalization. Detailed AUC and AUPR curves are further presented in Figures~\ref{fig:clinvar_auc_esm1b} and~\ref{fig:clinvar_aupr_esm1b}.

\subsection{Assessing DYNA's Generalization Ability for Non-Coding VEPs on MFASS}
\subsubsection{DYNA outperforms other genomic foundation models.}

\textsc{dyna} is a broadly applicable to fine-tuning both protein and DNA foundation models to learn VEP patterns for specific disease and regulatory mechanisms underlying disease. Thus far, we have demonstrated the superior performance of \textsc{dyna} fine-tuned protein language models for missense VEPs; yet, as the vast genome is non-coding (Figure ~\ref{fig:input_genomic_foundation_model}), predicting the regulatory effect of non-coding variants and linking such VEPs to clinical utility presents a unique challenge. In this section, we present compelling evidence that \textsc{dyna} outperforms alternative genomic foundation models in predicting the effects of non-coding variants. Our analysis specifically targets non-coding VEPs on their regulatory impacts of exon recognition and RNA splicing.
We focus our analysis on RNA splicing because the variant impact for RNA splicing disruption is included as a key factor in the ACMG guidelines for the interpretation of sequence variants, underscoring its clinical importance for various diseases and variant interpretation. We apply \textsc{dyna} on a dataset compiled from multiplexed functional assay of splicing using Sort-seq (MFASS)~\citep{chong2019multiplexed} that demonstrated how rare genetic variants can lead to substantial splicing disruptions. 
For more details about the dataset and setting-up, we recommend the readers refer to the dataset statistics in Table~\ref{tab:dataset-2} and Section~\ref{sp:dataset} in the supplementary material.

Our results highlight the flexibility and capability of \textsc{dyna} in enhancing the predictive accuracy of non-coding VEPs beyond conventional approaches (Figure~\ref{fig:mfass_performances}). 
\textsc{dyna} is used to fine-tune four base DNA foundation models, i.e., Nucleotide-Transformer-2.5B-1000G~\citep{dalla2023nucleotide}, GPN-Brassicales~\citep{benegas2023dna}, DNABERT-2~\citep{zhou2023dnabert}, and SpliceBERT~\citep{chen2023self}. 
Note that SpliceBERT is specifically tailored for non-coding sequences. Additionally, we compare these models using three different loss constructions: combined loss with a classification head ($\clubsuit$), contrastive loss with a classification head ($\diamondsuit$), and cosine similarity loss with a classification head ($\spadesuit$). We further compare these model variants against three conventional metrics: PhastCons, PhyloP, and distance measurement from variants to splice sites. Our findings demonstrate that all four genomic foundation models leveraging \textsc{dyna} significantly outperform the conventional metrics. For example, NT+$\clubsuit$~achieves an AUPR of 0.1246, markedly surpassing PhyloP (100 way) which scores only 0.0816. 

The varied testing performances observed in different genomic foundation models after applying the same \textsc{dyna} fine-tuning process on the MFASS data (Figure~\ref{fig:mfass_performances}) underscore their differing representation abilities for learning specific downstream tasks. The flexibility and broad applicability of \textsc{dyna} will allow users to easily fine-tune foundation models for VEP tasks, allowing them to select the most effective genomic foundation model tailored to their specific needs.

Furthermore, incorporation of Splice-Disrupting Variants (SDVs) in \textsc{dyna} fine-tuning is essential for an improved performance. 
As shown in Figures~\ref{fig:mfass_auc} and~\ref{fig:mfass_aupr}, \textsc{dyna} fine-tuning with sequences containing Splice-Disrupting Variants (SDVs) consistently outperform those without SDVs. Specifically, the performance of the NT with SDVs achieves an AUC of 0.6963, significantly surpassing its counterparts without SDVs, which only reach an AUC of 0.5650. This pattern underscores the enhanced predictive capability of models when analyzing sequences that include SDVs.

\subsubsection{DYNA shows generalization ability to unseen clinically-relevant splicing non-coding VEPs.}

\begin{figure}[!htbp]
    \centering
  \begin{minipage}{0.24\linewidth}
        \begin{subfigure}[b]{\linewidth}
            \caption{}\label{fig:mfass_performances}
        \includegraphics[width=\linewidth]{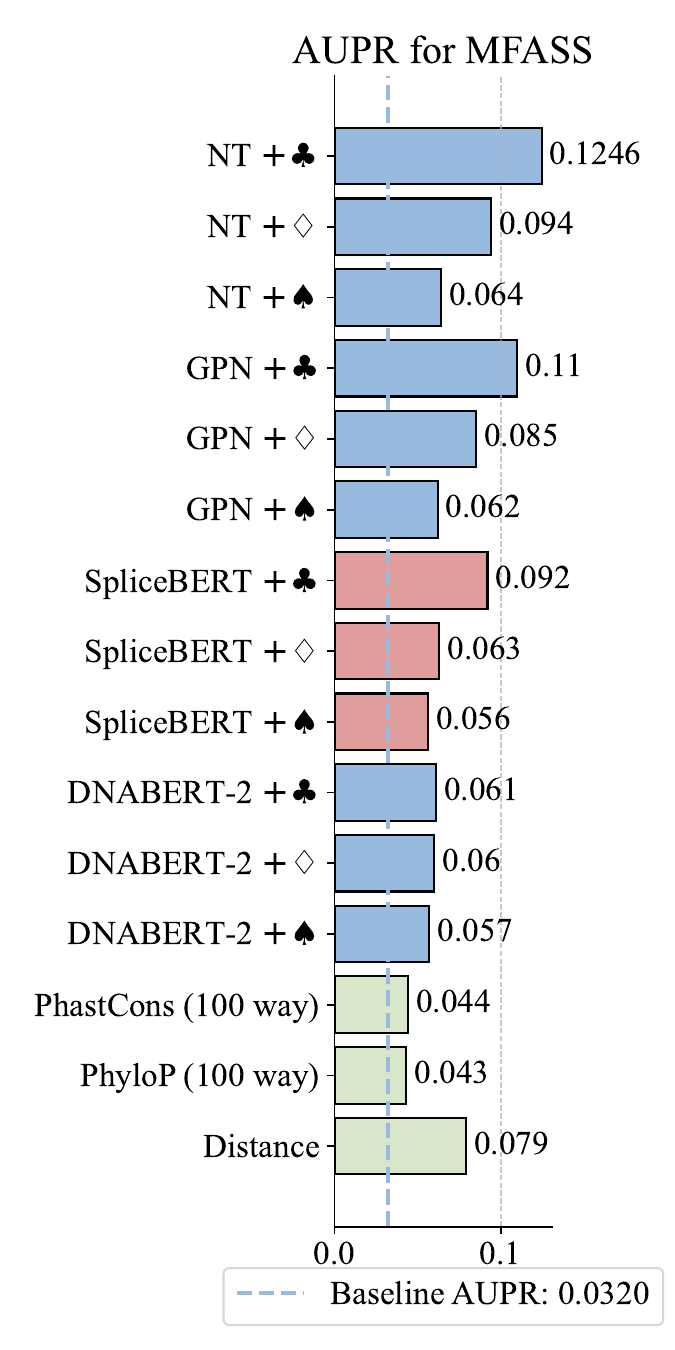}
    \end{subfigure}
      \end{minipage}
      \begin{minipage}{0.72\linewidth}
         \begin{subfigure}[b]{0.24\linewidth}
                \caption{}\label{fig:mfass_auc}
        \includegraphics[width=\linewidth]{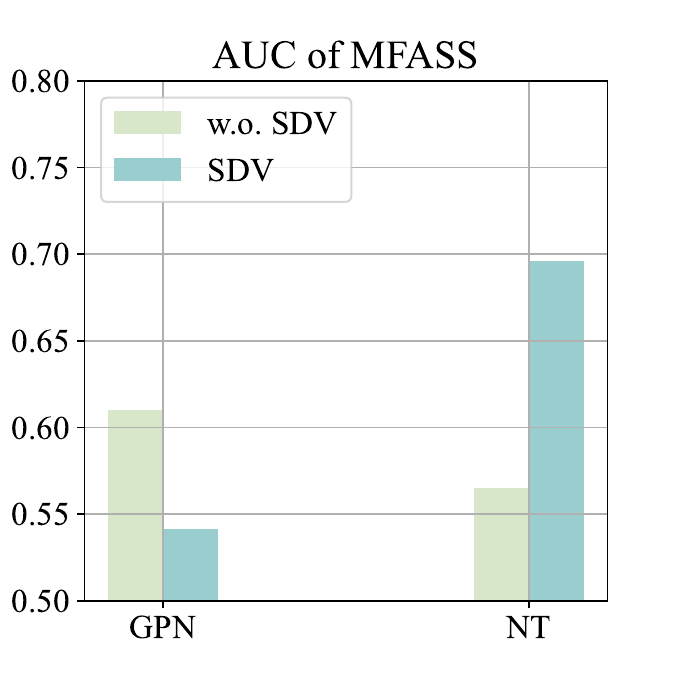}
    \end{subfigure}
    \hfill
            \begin{subfigure}[b]{0.24\linewidth}
                    \caption{}\label{fig:mfass_aupr}
            \includegraphics[width=\linewidth]{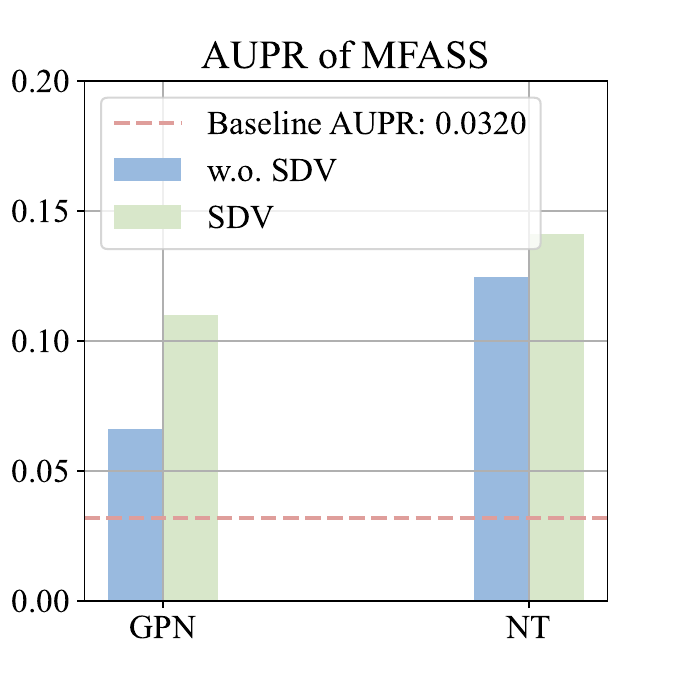}
    \end{subfigure} 
    \hfill
            \begin{subfigure}[b]{0.24\linewidth}
            \caption{}\label{fig:auc_clinvar_zero_shot}
        \includegraphics[width=\linewidth]{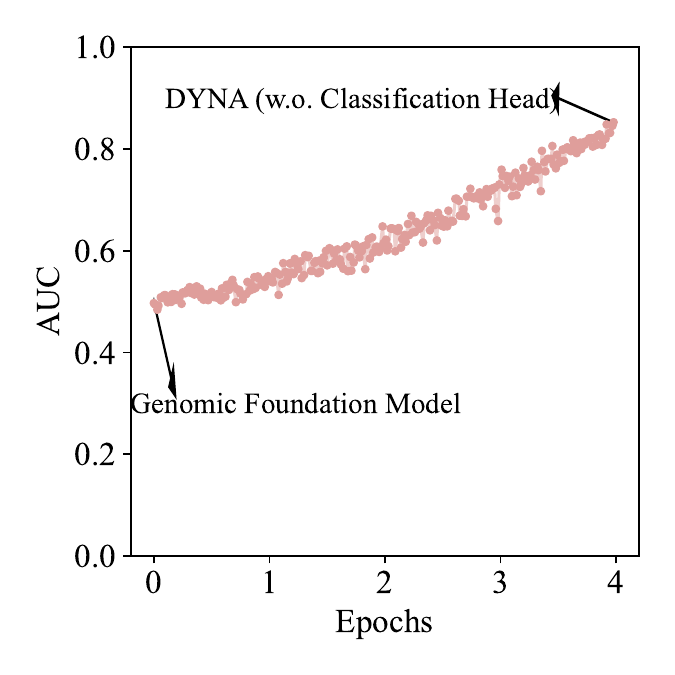}
    \end{subfigure}
    \hfill
        \begin{subfigure}[b]{0.24\linewidth}
      \caption{}\label{fig:aupr_clinvar_zero_shot}
        \includegraphics[width=\linewidth]{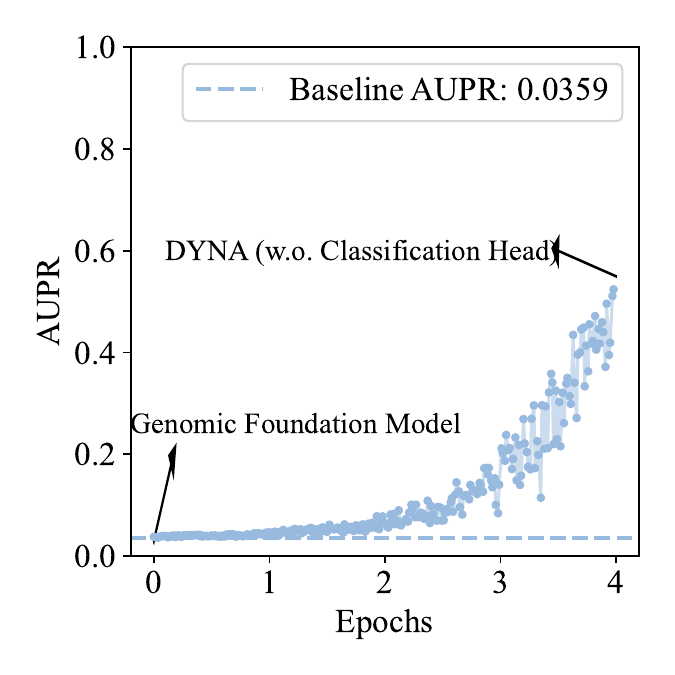}
    \end{subfigure}

            \begin{subfigure}[b]{0.24\linewidth}
          \caption{}\label{fig:auc_clinvar_five_shot}
        \includegraphics[width=\linewidth]{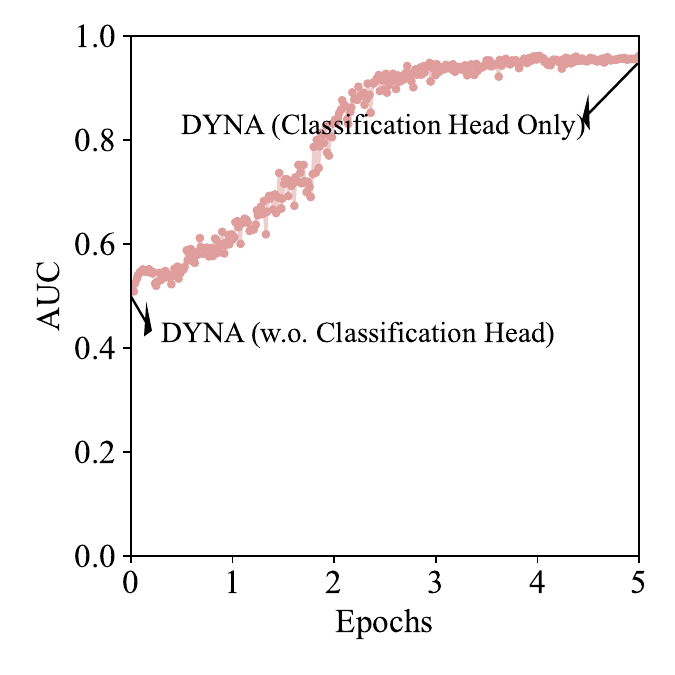}
    \end{subfigure}
    \hfill
            \begin{subfigure}[b]{0.24\linewidth}
         \caption{}\label{fig:aupr_clinvar_five_shot}
        \includegraphics[width=\linewidth]{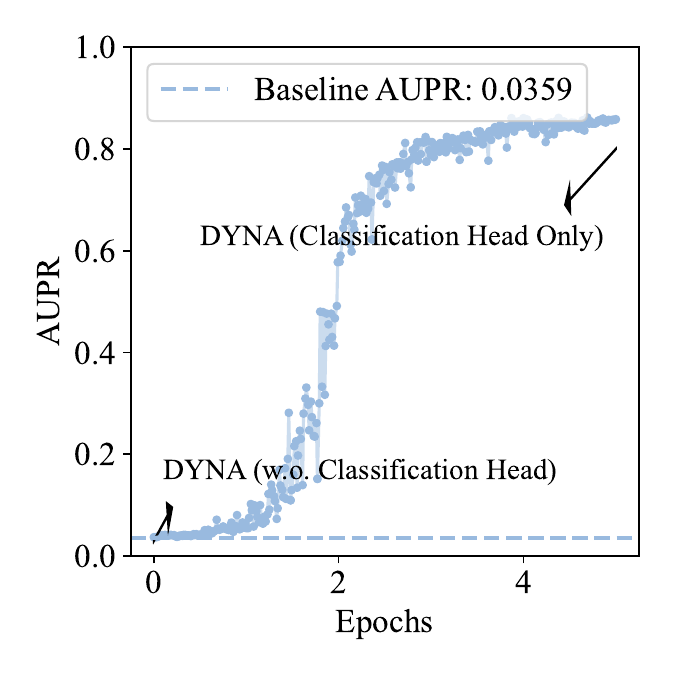}
    \end{subfigure}
    \hfill
                    \begin{subfigure}[b]{0.48\linewidth}
                    \caption{}\label{fig:clinvar_splicing_auc}
            \includegraphics[width=\linewidth]{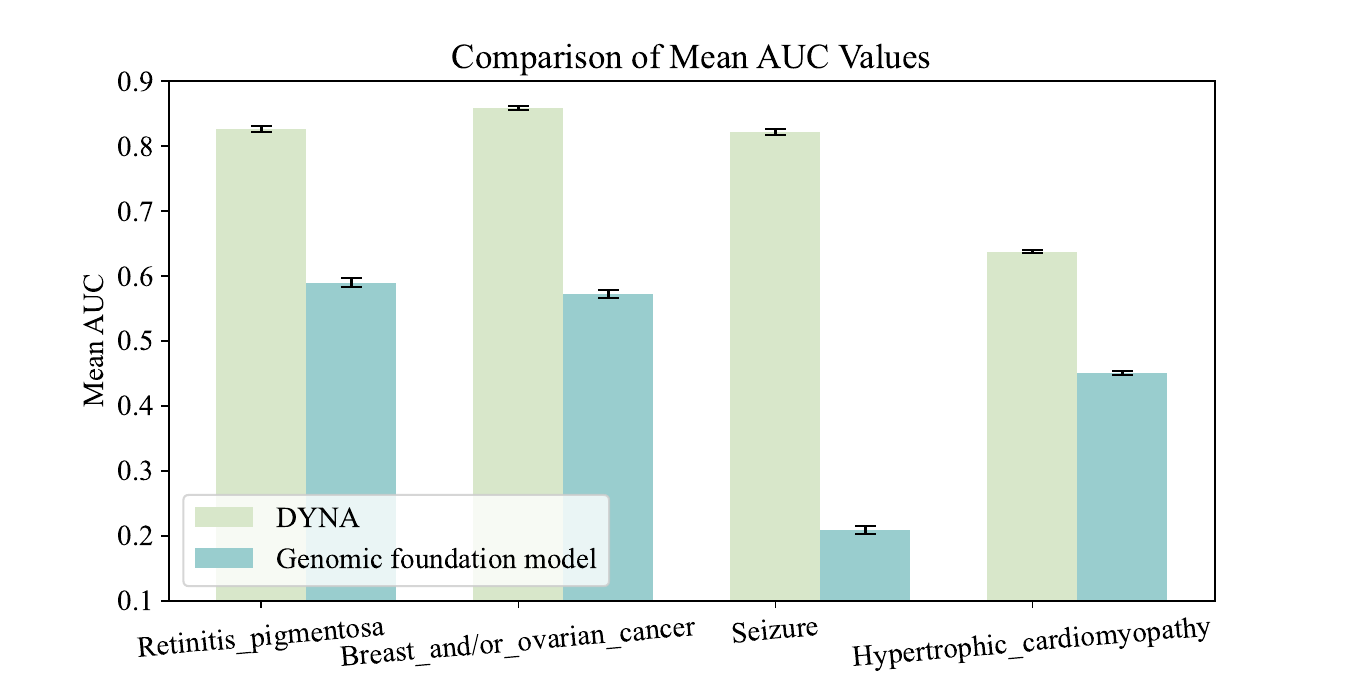}
    \end{subfigure} 
 \end{minipage}


   
        \caption{\textbf{\large{a}} Comparative analysis of MFASS AUPR performances for genomic foundation models, and conventional metrics. \textbf{\large{b}} Comparison of AUC w. and w.o. SDVs. \textbf{\large{c}} Comparison of AUPR w. and w.o. SDVs. \textbf{\large{d}} AUC Comparison between GPN Fine-tuned model with MFASS and Zero-Shot ClinVar Splicing Performance. \textbf{\large{e}} AUPR Comparison between GPN Fine-tuned with MFASS and Zero-Shot ClinVar Splicing Performance. \textbf{\large{f}} AUC results for the \textsc{dyna} model after only fine-tuning a classification head on the ClinVar Splicing dataset. This figure shows the AUC results after a five-shot fine-tuning phase exclusively on the ClinVar Splicing dataset, following the initial fine-tuning on the MFASS dataset using the GPN model. The AUC achieved is 0.95, demonstrating the model's effective use of prior non-coding VEP knowledge within the \textsc{dyna} framework and its strong out-of-distribution generalization capabilities. \textbf{\large{g}} AUPR results for \textsc{dyna} post five-shot fine-tuning on the ClinVar Splicing dataset. \textbf{\large{h}} AUC results per disease within the ClinVar Splicing dataset. This figure displays the AUC four specific disease, each with more than five positive and negative variants, highlighting the performance improvements achieved by \textsc{dyna}, fine-tuned using the GPN model, over the baseline GPN model. Retinitis Pigmentosa, Breast and/or Ovarian Cancer, Seizure Disorders, and Hypertrophic Cardiomyopathy are shown due to their known splicing-related pathologies.}
        \end{figure}

Next, we show \textsc{dyna}'s strong generalization capabilities to unseen clinically-relevant splicing non-coding VEPs. We extracted benign and splicing-related pathogenic sequences, i.e., ClinVar Splicing, based on the molecular consequences annotated in ClinVar (Table~\ref{tab:dataset-2}). 
This ClinVar Splicing dataset is over 27 times larger than the MFASS dataset, covering pathogenic variants that disrupt splicing in diverse diseases.
Initially, \textsc{dyna}, utilizing the Genomic Prediction Network (GPN) as its foundation model, begins with no specific training for non-coding VEPs, evidenced by an initial AUC of 0.5 and a baseline AUPR of 0.0359 at epoch 0, as depicted in Figures~\ref{fig:auc_clinvar_zero_shot} and~\ref{fig:aupr_clinvar_zero_shot}.

\textsc{dyna} enables the GPN model to effectively generalize to the ClinVar Splicing using the MFASS dataset. We design a two-step process. First, the GPN model undergoes a rigorous fine-tuning process on the MFASS dataset, which includes fewer than 27,000 samples. We implement early stopping by holding out a validation MFASS dataset in the fine-tuning process to prevent overfitting on MFASS effectively. Following fine-tuning on MFASS for five epoches, \textsc{dyna} is evaluated in a zero-shot testing scenario on the ClinVar Splicing dataset, where it shows an AUC of 0.85. This testing phase validates the model’s ability to generalize to completely new, clinically annotated data without prior exposure, confirming its effectiveness in acquiring and applying splicing-related non-coding VEP knowledge from a relatively small dataset. 

After the initial few-shot fine-tuning on the MFASS dataset, the genomic foundation model, integrated within the \textsc{dyna} framework and based on the GPN, is posited to have incorporated prior knowledge for non-coding VEPs. To leverage this embedded knowledge, we subsequently attach a classification head with randomly initialized weights to the fine-tuned GPN model. This classification head is then fine-tuned for an additional five epochs exclusively on the ClinVar Splicing dataset, aiming to harness and apply the acquired non-coding VEP insights directly in a targeted classification task.Figures~\ref{fig:auc_clinvar_five_shot} and~\ref{fig:aupr_clinvar_five_shot} display the outcomes of this approach, showing the AUC and AUPR metrics respectively. The results demonstrate that the fine-tuned classification head, applied to the GPN model within the \textsc{dyna} architecture, achieves an AUC of 0.95 and an AUPR of 0.87. This phase of the experiment involves pre-training the GPN model on the MFASS dataset followed by a focused training of only the classification head on the ClinVar dataset for five shots, highlighting \textsc{dyna}'s robust generalization capabilities to unseen disease-relevant genes in non-coding VEPs.

\subsubsection{DYNA improves the accuracy of non-coding VEPs on splicing-related diseases.}

To demonstrate the clinical utility of splicing VEP improvement by \textsc{dyna}, we analyze its predictive power of detecting splicing-disruptive pathogenic variants when stratified by diseases in ClinVar.
Because MFASS is not directly relevant to any disease, this test will inform how a functional genetics assay can be translated to clinical variant interpretation by \textsc{dyna} and foundation models.
Initially, we selected a robust set of diseases, each with more than five positive and negative variants, to ensure a reliable analysis. The AUC improvement achieved by \textsc{dyna}, when fine-tuned using the GPN model compared to the baseline GPN model, is detailed in Figure~\ref{fig:robust_splicing}. We focus on four specific diseases—Retinitis Pigmentosa, Breast and/or Ovarian Cancer, Seizure Disorders, and Hypertrophic Cardiomyopathy—due to their strong associations with splicing anomalies~\citep{yang2021pre,wappenschmidt2012analysis}. 
For these known diseases with a strong splicing basis, \textsc{dyna} consistently exhibits significantly improved AUC results with 1000 bootstrap iterations, compared to the baseline GPN genomic foundation model (Figure~\ref{fig:clinvar_splicing_auc}).
A similar pattern is observed when evaluated by AUPR (Figure ~\ref{fig:clinvar_splicing_aupr}).
The AUC for the top-50 best-performing diseases can be found in Figure~\ref{fig:top-50}. 
Such improvement across various diseases underscores \textsc{dyna}'s enhancement of DNA foundation model's capability to accurately capture the functional implications of genetic variants in splicing, then translate its predictions to disease contexts.

\section{Discussion}\label{sec-discussion}


The development of protein language models, such as the MSA transformer \citep{rao2021msa}, which is trained using a masked language modeling objective and self-attention mechanisms, has further enriched the VEP toolkit for unsupervised and weakly-supervised VEP methods. Additionally, the AlphaMissense model \citep{cheng2023accurate} integrates evolutionary data with protein structural modeling. It utilizes population frequency data, sequences from MSAs, and predicted structural contexts to enhance its predictive performance. Notably, \citet{brandes2023genome} introduced a zero-shot workflow employing ESM1b for predicting the impacts of potential missense variants across proteins of any length, demonstrating its utility without explicitly relying on evolutionary data. This approach is particularly vital for orphan genes with sparse MSA coverage \citep{chowdhury2022single, michaud2022language} and rare variants that are underrepresented in populations \citep{manolio2009finding}.

The success of ESM1b in a zero-shot capacity underscores the capability of large protein language models to learn and generalize evolutionary constraints, potentially reshaping the landscape of VEP by offering robust predictions independent of direct evolutionary data inputs.However, the lack of disease specificity in unsupervised VEP methods substantially limits their utility and impedes their adoption for clinical variant interpretations at the point-of-care. Therefore, our main contributions for \textsc{dyna} are as follows:
 \begin{itemize}
 \item Current deep-learning models often struggle to effectively handle both coding and non-coding variants. In contrast, our \textsc{dyna} model addresses this challenge by leveraging the flexibility of a Siamese neural network architecture to fine-tune any foundation model. This approach enhances the model's comprehensiveness, allowing for a more holistic understanding of genomic influences on disease mechanisms and phenotypic outcomes.
     \item Technically, we introduce a novel disease-specific protein language model for variant pathogenicity to more effectively capture the pseudo-log-likelihood ratio in rare missense variants using both masked language model logits through a siamese network.
     \item Clinically, we propose an efficient way to fine-tune genomic foundation models to estimate the probability of pathogenicity for rare missense variants in coding and non-coding VEPs to address a persistent challenge in clinical genetics.
     \item \textsc{dyna} exhibits OOD generalization capabilities. By fine-tuning a genomic foundation model employing a siamese framework on a dataset from a massively parallel reporter assay comprising fewer than $27,000$ samples, \textsc{dyna} attained an AUC of $0.85$ in a zero-shot learning scenario and $0.95$ in a five-shot learning context when evaluated on a ClinVar Splicing dataset exceeding $650,000$ examples.     
     \end{itemize}

For future developments, integrating \textsc{dyna} into a clinical decision support system can revolutionize clinical practices by providing healthcare professionals with advanced tools to make more informed decisions regarding diagnosis and treatment. This system would leverage genetic information to offer tailored guidelines, risk assessments, and treatment recommendations specific to individual genetic profiles, ensuring personalized medicine approaches. Additionally, undertaking cross-population validation studies is critical. By expanding these studies to include diverse populations, the robustness of \textsc{dyna}'s predictions can be confirmed across various genetic backgrounds, thereby enhancing the model's global utility and contributing to the reduction of disparities in genetic research. Furthermore, establishing collaborations with biobanks and genomic consortia can significantly enrich the datasets available to \textsc{dyna}, improving the model's accuracy and robustness. Access to a broader array of conditions and population data through these partnerships will not only refine \textsc{dyna}'s predictive capabilities but also extend its reach across a more extensive range of clinical and research applications, thereby reinforcing its position as a pivotal tool in the advancement of genomics and personalized medicine.

\section{Methods}\label{sec-methods}

Within the framework of genomic foundation models, such as protein language models and DNA foundation models, we analyze two distinct types of biological inputs: protein sequences, representing the coding regions of the genome, and DNA sequences, corresponding to the non-coding regions. For protein language models, the MLM logits are predicted by a protein language model for observing the input amino acid $s_i$ at position $i$ given the sequence $s$. Similarly, for DNA foundation models, the MLM logits are predicted by observing the DNA tokens $t_i$ at position $i$ given the input DNA sequence.    

When passed through an activation function, i.e., softmax function, these logits provide probabilities over the possible tokens, guiding the prediction process. In order to fine-tune \textsc{dyna} on pairs of wild-type and mutant sequences with labels, i,e., $\{\mathbf{s}^{\text{WT}}$, $\mathbf{s}^{\text{mut}}, \mathbf{Y}\} = \{s_k^{\text{WT}}, s_k^{\text{mut}}, y_k\}^K_{k=1} $ and $y_k \in \{0,1\}$, we create a siamese network with two weight-sharing genomic foundation model branches to update the genomic foundation model weights. In genomic foundation models, the construction of semantically meaningful MLM logits presents distinct challenges across protein and DNA sequences due to varying vocabulary complexities.  To address this challenge, we employ tailored loss functions for each model type: a pseudo-log-likelihood ratio (PLLR) for protein language models, enhancing the semantic relevance of sequence variations, and contrastive loss for DNA foundation models, effectively minimizing prediction error across a broad spectrum of tokens. Next, observing that zero-shot ESM1b underperformed on disease-specific variant sets, we proposed that \textsc{dyna}'s superior results are due to its more accurate mapping of pathogenic variant intolerance specific to cardiomyopathies and arrhythmias. To measure these positional effects post-fine-tuning, we used In Silico Mutagenesis (ISM) to determine the differences in PLLR between ESM1b and \textsc{dyna}.

\subsection{Pseudo-Log-Likelihood Ratio for Coding VEPs}

For a sequence $s = s_1, …, s_L$, the pseudo-log-likelihood is calculated as $\text{PLL}(s) = \sum_{i=1}^L \log P(x_i = s_i|s)$, where $L$ denotes the sequence length, $s_i$ represents the amino acid at position $i$, and $\log P(x_i = s_i|s)$ denotes the log-likelihood predicted by the protein language model when observing amino acid $s_i$ at position $i$ within sequence $s$. The PLLR between the $k$-th pair of $s_k^{\text{WT}}$ and $s_k^{\text{mut}}$ is then computed as:
\begin{equation}\label{eq:ab_pllr}
    \lambda_k = |\text{PLL}(s_k^{\text{WT}}) - \text{PLL}(s_k^{\text{mut}})|, 
\end{equation}
because a wild-type sequence typically tends to have a higher log-likelihood in a protein language model and a mutation disrupts the protein with a lower log-likelihood. To perform the classification of pathogenic vs benign genetic variant via the siamese network, we will utilize a binary cross entropy loss.
The non-weighted PLLR uses the absolute value of PLLR as detailed in Equation~\ref{eq:ab_pllr}, while weighted PLLR is calculated as $\tilde{\lambda} = \text{PLL}(s_k^{\text{WT}})/L_{s_k^{\text{WT}}} - \text{PLL}(s_k^{\text{mut}})/L_{s_k^{\text{mut}}}.$ 

\textbf{Binary Cross-Entropy Loss:}
In order to fine-tune the siamese network using binary cross entropy loss, we calibrate the PLLR to a probability $\hat{\sigma_k}$ between $0$ to $1$ by: $\hat{\sigma}(\lambda_k) = 2\sigma(\lambda_k) - 1,$
due to the sigmoid function $\sigma$ is between $0.5$ to $1$ for $|\text{PLLR}|$ between $0$ to $+\infty$. We then fine-tune the siamese network with the binary cross entropy loss as follows:
\begin{equation}
    \mathcal{L}(\mathcal{W}_{\text{coding}}, \mathbf{Y},\mathbf{s}^{\text{WT}},\mathbf{s}^{\text{mut}}) =  -\sum_{k=1}^K (1 - y_k) \cdot \log (\hat{\sigma}(\lambda_k)) + y_k\cdot \log (1 - \hat{\sigma}(\lambda_k)),
\end{equation}
where the protein language model is parameterized by $\mathcal{W}_{\text{coding}}$, $\hat{\sigma}(\lambda_k)$ is the sigmoid output for the $k$-th PLLR $\lambda_k$, and $y_k$ is the corresponding label. When the mutation $s_k^{\text{mut}}$ is benign, the loss function simplifies to $- \log (\hat{\sigma}(\lambda_k))$, thereby maximizing the model's confidence in the similarity of the embeddings for the pair, and conversely for pathogenic mutations.

\subsection{Contrastive Loss for Non-Coding VEPs}
Let $\mathbf{h}_{s_k^{\text{WT}}}$ and $\mathbf{h}_{s_k^{\text{mut}}}$ represent the last hidden states from the $k$-th pair input. The Euclidean distance $\mathcal{D}$ between these embeddings is defined as:
\begin{equation}
    \mathcal{D} = \| \mathbf{h}_{s_k^{\text{WT}}} - \mathbf{h}_{s_k^{\text{mut}}} \|_2.
\end{equation}

The contrastive loss $\mathcal{L}_{\text{contrast}}$ is formulated as follows, where $y_k$ denotes the binary label indicating whether the mutation $s_k^{\text{mut}}$ is benign as $s_k^{\text{WT}}$ or pathogenic:
\begin{equation}
\mathcal{L}_{\text{contrast}}(\mathcal{W}_{\text{non-coding}}, \mathbf{Y}, \mathbf{s}^{\text{WT}}, \mathbf{s}^{\text{mut}}) = \sum_{k=1}^K y_k \cdot \mathcal{D}^2 + (1 - y_k) \cdot \max(m - \mathcal{D}, 0)^2.
\end{equation}
Thus, for pairs with benign mutation, the loss aims to ensure that the distance $\mathcal{D}$ between the pairs within a proximity defined by the margin 
$m$. For pairs with pathogenic mutation, the term $\mathcal{D}^2$ is active, which penalizes small distances between the pair, encouraging the distance to be as large as possible to increase the loss.
The expectation of this loss across all pairs in the batch is then computed:
\begin{equation}
\mathcal{L}_{\text{contrast}} = \mathbb{E}[\mathcal{L}_{\text{contrast}}]
\end{equation}

\textbf{Binary Cross-Entropy Loss:}
The binary cross-entropy loss $\mathcal{L}_{\text{BCE}}$ is employed for classification, applied to the concatenated embeddings $[\mathbf{h}_{s_k^{\text{WT}}}, \mathbf{h}_{s_k^{\text{mut}}}]$ passed through a classification head $\mathcal{F}$:
\begin{equation}
\mathcal{L}_{\text{BCE}}(\mathcal{W}_{\text{non-coding}}, \mathcal{F}, \mathbf{Y}, \mathbf{s}^{\text{WT}}, \mathbf{s}^{\text{mut}}) = -\sum_{k=1}^K y_k \log(\mathcal{F}_k) + (1-y_k) \log(1-\mathcal{F}_k),
\end{equation}
where the DNA foundation model is parameterized by $\mathcal{W}_{\text{non-coding}}$, $\mathcal{F}_k$ is the sigmoid output of the classifier for the $k$-th sample.

\textbf{Combined Loss:}
The total loss $\mathcal{L}$ is a weighted sum of the contrastive and BCE losses, where $\lambda$ is a weighting factor which may adapt over training epochs:
\begin{equation}
\mathcal{L} = \mathcal{L}_{\text{contrast}} + \lambda \cdot \mathcal{L}_{\text{BCE}}
\end{equation}

\subsection{Genomic and Protein sequence extraction and annotation}

For the ClinVar CM and ARM datasets, we translate the DNA sequences into protein sequences using the human genome assembly hg38\footnote{\url{https://www.ncbi.nlm.nih.gov/grc/human}}. We employed the GFF file, MANE.GRCh38.v1.1.ensembl\_genomic.gff.gz\footnote{\url{https://www.ncbi.nlm.nih.gov/refseq/MANE/}}, to annotate coding versus non-coding regions for each gene, as only coding DNA sequences are translated into proteins. Additionally, protein domains, cataloged in the Pfam database~\citep{finn2016pfam}, are essential for the functional characterization of proteins. These domains are identified by aligning the translated sequences to known domain structures, thereby facilitating deeper insights into protein function.

\subsection{Construction of cardiovascular disease testing data from ClinVar}
The ClinVar CM and ARM datasets include all missense variants specific to CM and ARM, respectively, extracted from ClinVar according to disease types. We exclude any variants that overlap with the existing CM and ARM datasets. To assess generalization to previously unexamined disease-relevant genes, we exclusively analyze novel genes that are included in the ClinVar CM and ARM datasets but are absent from the pre-compiled datasets. For evaluating intra-gene generalization, our analysis is focused solely on variants within the same genes but located at different mutational positions.

\subsection{Gaussian mixture model for VUS analysis}
We compute the PLLR values as analyzed by \textsc{dyna} for variants of uncertain significance, categorizing them into pathogenic and benign classes using a GMM. The Expectation-Maximization algorithm~\citep{mclachlan2007algorithm} is employed to iteratively divide the variants into two clusters. This iterative process continues until the model converges, which occurs either when the change in the log-likelihood of the data under the model is less than $1e-3$, or when 500 iterations are reached.

\subsection{Construction of splice disruption testing data from ClinVar}
Similarly, the ClinVar Splicing dataset, compiled from ClinVar, includes all benign sequences and pathogenic variants pertinent to splicing based on the molecular consequences annotated in ClinVar. For fine-tuning with a five-shot approach on the ClinVar Splicing dataset, we divided the training, validation, and test sets in an 8:1:1 ratio.

\subsection{In Silico Mutagenesis for the PLLR Difference}

ISM involves systematically replacing each amino acid in a sequence and computing the resulting PLLR differences. To perform ISM in practice, consider an input sequence $s$ of length $L$. For each index $i$ (nucleotide or amino acid) in $s$, each of the three alternative nucleotides/amino acids is substituted in turn, generating an alternative sequence $s^\prime$ by altering only the $i$-th position. The mutations for $s$ and $s^\prime$ are accordingly denoted as $s^{\text{mut}}$ and $s^{\text{mut}^\prime}$. The PLLR difference, which quantifies the impact of each nucleotide/ amino acid substitution, is calculated as the difference in the model’s predictions between the alternative sequence and the original:

\begin{equation}
\text{PLLR\_diff} = \lambda(s,s^\text{mut}) - \lambda(s^\prime,s^{\text{mut}^\prime}),
\end{equation}

where $\lambda(s,s^\text{mut})$ represents the PLLR for the oringinal sequence and its mutation and $\lambda(s^\prime,s^{\text{mut}^\prime})$ represents the PLLR for the sequence  with the nucleotide substitution at position $i$ and its mutation. This process is repeated for every nucleotide, resulting in a $4 \times L$ attribution matrix, visualizable as a sequence logo in Figure~\ref{fig:heatmap}. This method has proven more effective than other attribution techniques~\citep{novakovsky2023obtaining}. While ISM typically quantifies changes in the final layer output, it is also adaptable for revealing contributions of hidden neurons.

Applying ISM at the single-nucleotide level across numerous inputs incurs substantial computational costs, on the order of $3L$ operations per sequence. To optimize, analysis may be restricted to a select subset of sequences anticipated to yield the most significant insights, such as SCN5A, due to its strong association with cardiomyopathy diseases and robust predictions.

\section{Data Availability}
For coding variant effect predictions (VEPs), our approach centers on clinical variant sets specifically related to inherited cardiomyopathies (CM) and arrhythmias (ARM). We utilize a pre-compiled dataset comprised of rare missense pathogenic and benign variants, categorized using a cohort-based approach for diseases such as cardiomyopathy and arrhythmias, as detailed in the previous report by \citet{zhang2021disease}. ClinVar CM and ARM datasets include all missense variants in CM and ARM, respectively, are extracted from ClinVar~\citep{landrum2014clinvar}. In the realm of non-coding VEPs, our focus shifts to splicing-related variants, utilizing a dataset from the multiplexed assay for exon recognition by \citet{chong2019multiplexed}, which highlights the significant impact of rare genetic variants on splicing disruptions. Similarly, the ClinVar Splicing dataset, compiled from ClinVar, encompasses all benign sequences and pathogenic variants pertinent to splicing.

\section{Code Availability}
Our \textsc{dyna} code is available on Github at \url{https://github.com/zhanglab-aim/DYNA}.

\clearpage

\noindent \textbf{Fig. 2 (cont.):} \textbf{\large{g}} Venn diagram illustrating the distribution of mutations in ClinVar CM and inherited CM genes. There are 125 mutations from ClinVar CM genes and 16 mutations from inherited CM genes, with 14 overlapping genes but mutations occurring at distinct positions. \textbf{\large{h}} Venn diagram illustrating the distribution of mutations in ClinVar ARM and inherited ARM genes. \textbf{\large{i}} Distribution of ClinVar CM benign and pathogenic sequences. \textbf{\large{j}} Distribution of ClinVar ARM benign and pathogenic sequences.  \textbf{\large{k}} Violin plots displaying the distribution of PLLR values for Variants of Uncertain Significance (VUS) categorized into pathogenic and benign classes using a Gaussian Mixture Model (GMM) as analyzed by \textsc{dyna}.

\begin{appendices}

\section{Detailed Dataset and Set-Up}\label{sp:dataset}
\paragraph{Dataset}
For coding VEPs, we focus on clinical variant sets pertaining to inherited cardiomyopathies (CM) and arrhythmias (ARM) for fine-tuning. We employed a pre-compiled dataset of rare missense pathogenic and benign variants, defined by a cohort-based approach, in cardiomyopathy and arrhythmias, respectively. The details can be found in the previous report~\citep{zhang2021disease}. The statistics for all datasets are shown in Table~\ref{tab:dataset-1}. 

\begin{table}[!htbp]
    \caption{Cardiomyopathies and arrhythmias datasets}
    \label{tab:dataset-1}
    \centering
    \begin{tabular}{l|c c c c  c c c c}
    \hline
      & \multicolumn{4}{c}{Cardiomyopathies} & \multicolumn{4}{c}{Arrhythmias}\\
     & Pathogenic & Benign & VUS & Total & Pathogenic & Benign & VUS &Total \\
    \hline
     Train    & 238 & 202 & -  & 440 & 168 & 158 & - & 326 \\
      Test    & 118  & 100 & - & 218 & 84 & 79 &- & 163\\
      Total & 356 & 302 &- & 658 & 252 & 237 &- & 489\\
      \hline
    & \multicolumn{4}{c}{ClinVar CM\textsuperscript{*}} & \multicolumn{4}{c}{ClinVar ARM\textsuperscript{*}}\\
     & Pathogenic & Benign & VUS & Total & Pathogenic & Benign & VUS & Total \\
    \hline
      Test    & 522  & 1113 &  17734 & 19369 & 21 & 77 & 885 & 983\\
      \hline
    \end{tabular}
    \textsuperscript{*}{Because the data construction on ClinVar CM and ARM missense variants is noisy, the dataset is utilized only for testing DYNA's generalization ability.}
\end{table}

We fine-tune two base protein language models on \textsc{dyna} framework, i.e., esm1b\_t33\_650M\_UR50S (ESM1b)~\citep{rives2021biological} and esm2\_t33\_650M\_UR50D (ESM2)~\citep{lin2023evolutionary}. To further test the models' generalization ability to unseen disease-relevant genes, we extract missense CM and ARM sequences on ClinVar based on ClinVar disease type. Besides pathogenic and benign mutant sequences, we extract Variants of Uncertain Significance (VUS) and employ a Gaussian mixture model to decompose these as a mixture of two Gaussian distributions. This approach captures variants predicted as more likely pathogenic or more likely benign and compare both distributions with true ClinVar CM/ARM distributions. 

For non-coding VEPs, we focus on a dataset of splicing-related variants from MFASS~\citep{chong2019multiplexed}—a multiplexed assay for exon recognition—that has demonstrated how rare genetic variants can lead to substantial splicing disruptions. We fine-tune four base DNA foundation models, i.e., Nucleotide-Transformer-2.5B-1000G~\citep{dalla2023nucleotide}, GPN-Brassicales~\citep{benegas2023dna}, DNABERT-2~\citep{zhou2023dnabert}, and SpliceBERT~\citep{chen2023self}. To evaluate the models' generalization capabilities to unseen disease-relevant genes, we extracted benign and splicing-related pathogenic sequences from ClinVar based on the molecular consequences annotated in ClinVar. The statistics for both datasets are shown in Table~\ref{tab:dataset-2}.

\begin{table}[!htbp]
    \caption{MFASS and ClinVar datasets}
    \label{tab:dataset-2}
    \centering
    \begin{tabular}{l|c c c }
    \hline
      & \multicolumn{3}{c}{MFASS} \\
     & Pathogenic & Benign  & Total  \\
    \hline
     Train    & 697 & 21316  & 22013  \\
      Test    & 88  & 2664  & 2752 \\
      Total & 785 & 23980  & 24765 \\
      \hline
    & \multicolumn{3}{c}{ClinVar Splicing\textsuperscript{*}} \\
     & Pathogenic & Benign & Total \\
    \hline
      Test    & 24347  & 653053 & 677400\\
      \hline
    \end{tabular}
        \textsuperscript{*}{Because the data construction on ClinVar Splicing gene variants is noisy, the dataset is utilized for testing DYNA's generalization ability on non-coding VEPs. We only fine-tune this dataset in the five-shot fine-tuning scenario. We divided the training, validation, and test sets in an 8:1:1 ratio.}
\end{table}

\paragraph{Set-Up}
For coding VEPs, we use a batch size of $8$; the evaluation batch size is also set to $8$. The model is trained over $10$ epochs on CM and ARM datasets for both tasks, respectively. We evaluate the validation loss every $50$ step, and if the best validation loss has not decreased for $5$ evaluations, we early stop the fine-tuning process. We leverage the pseudo-log-likelihood ratio for coding VEPs and employ the Adam optimizer, with a learning rate of $1\text{e}-5$. Additionally, we implement a warmup ratio of $0.1$ followed by linear decay. The L2 regularization weight decay is set at $0.01$. For sequences longer than 1024 that exceed the max length limit of protein language models, we truncate them to a length of 1024, centering them around the variant position.

For non-coding VEPs, we set both the training and evaluation batch sizes to $8$. The model undergoes training over $5$ epochs on the MFASS dataset. We monitor the validation loss every $500$ steps, implementing early stopping if there is no improvement in the best validation loss after $4$ consecutive evaluations. The model utilizes a combined loss function—comprising contrastive and BCE losses—and is optimized using the Adam optimizer with a learning rate of $2\text{e}-5$. The proximity threshold $m$ is established at $2.0$. The weighting factor $\beta$ for the combined loss is dynamically adjusted according to the formula $\min\left(1, \frac{\text{current\_training\_steps}}{\text{total\_training\_steps}}\right).$ Additionally, a warmup phase accounting for $10\%$ of the training duration is followed by a linear decay in the learning rate. The L2 regularization, aimed at minimizing weight decay, is set to $1e-4$. For sequences exceeding the maximum length limit of $512$, we truncate them to this length, centering the truncation around the variant position.

\section{Performance for CM and ARM Variant Pathogenicity Prediction.}
\begin{figure}[!htbp]
    \centering
      \begin{subfigure}[b]{0.5\linewidth}
            \caption{AUC on CM for ESM2}\label{fig:auc_cm_esm2}
        \includegraphics[width=\linewidth]{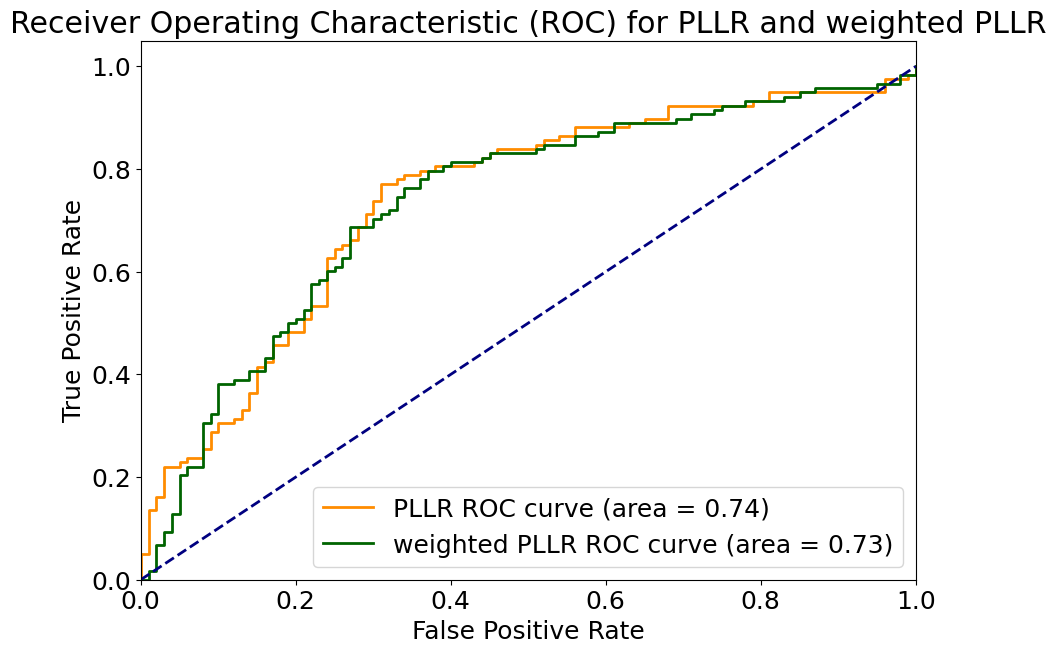}
    \end{subfigure}
    \hfill
        \begin{subfigure}[b]{0.42\linewidth}
            \caption{AUPR on CM for ESM2}\label{fig:apr_cm_esm2}
        \includegraphics[width=\linewidth]{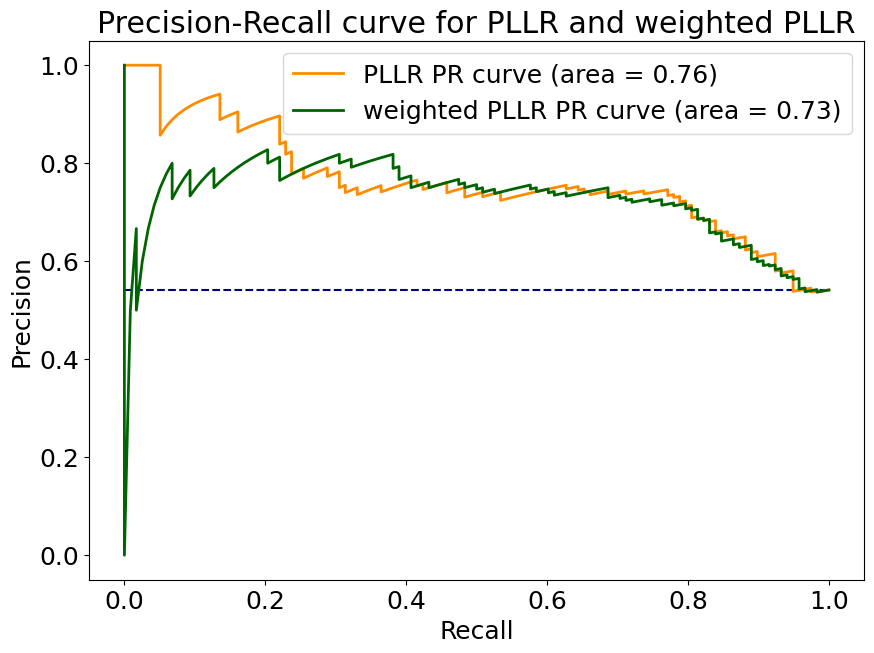}
    \end{subfigure}
       \begin{subfigure}[b]{0.5\linewidth}
            \caption{AUC on CM for \textsc{dyna} fine-tuned ESM2}\label{fig:auc_cm_esm2}
        \includegraphics[width=\linewidth]{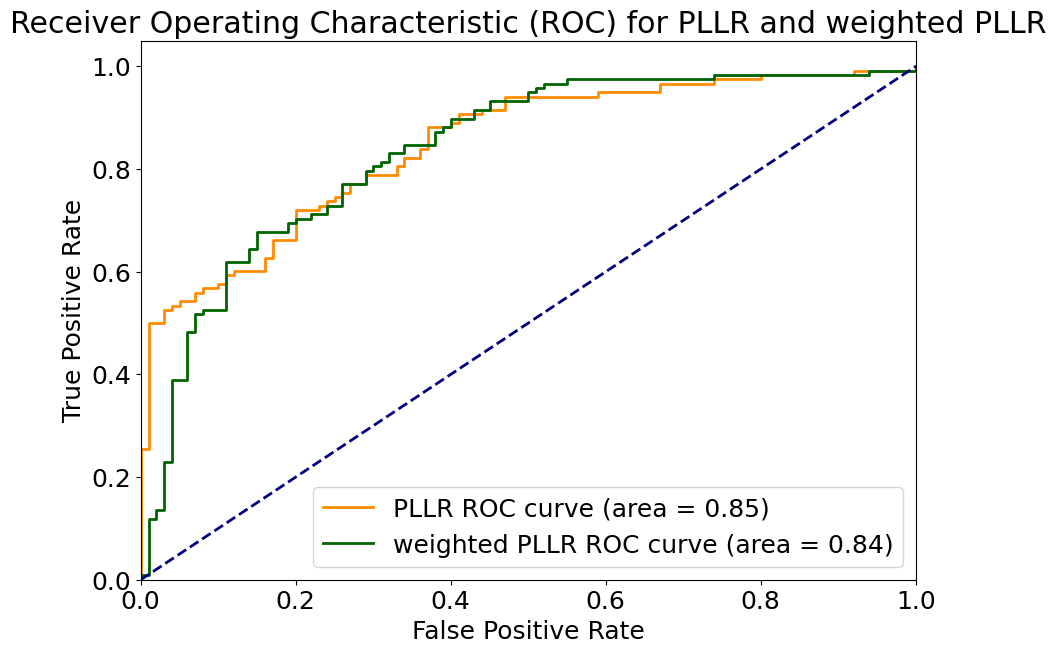}
    \end{subfigure}
    \hfill
        \begin{subfigure}[b]{0.42\linewidth}
            \caption{AUPR on CM for \textsc{dyna} fine-tuned ESM2}\label{fig:apr_cm_esm2}
        \includegraphics[width=\linewidth]{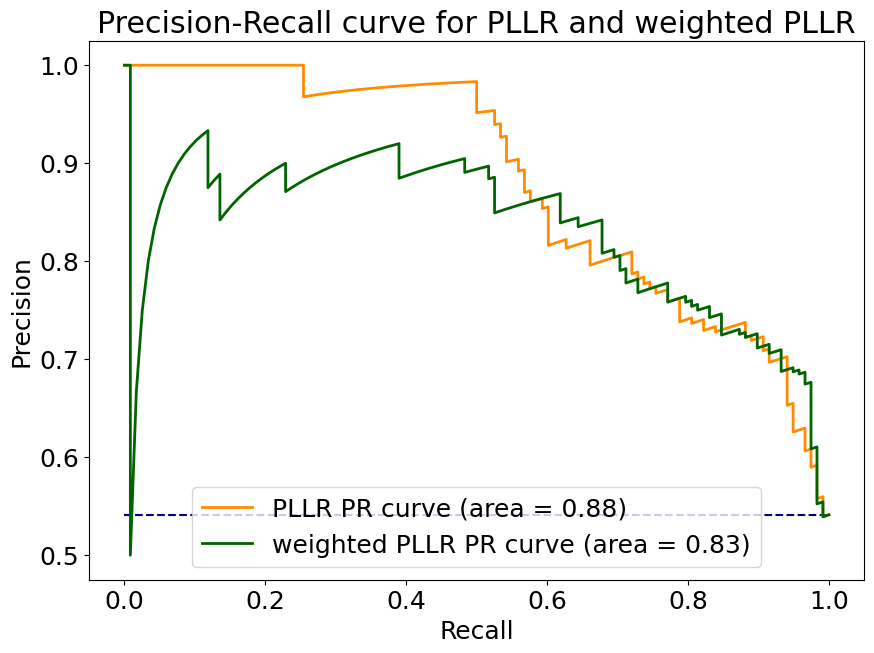}
    \end{subfigure}
    
      \begin{subfigure}[b]{0.5\linewidth}
            \caption{AUC on ARM for ESM2}\label{fig:auc_arm_esm2}
        \includegraphics[width=\linewidth]{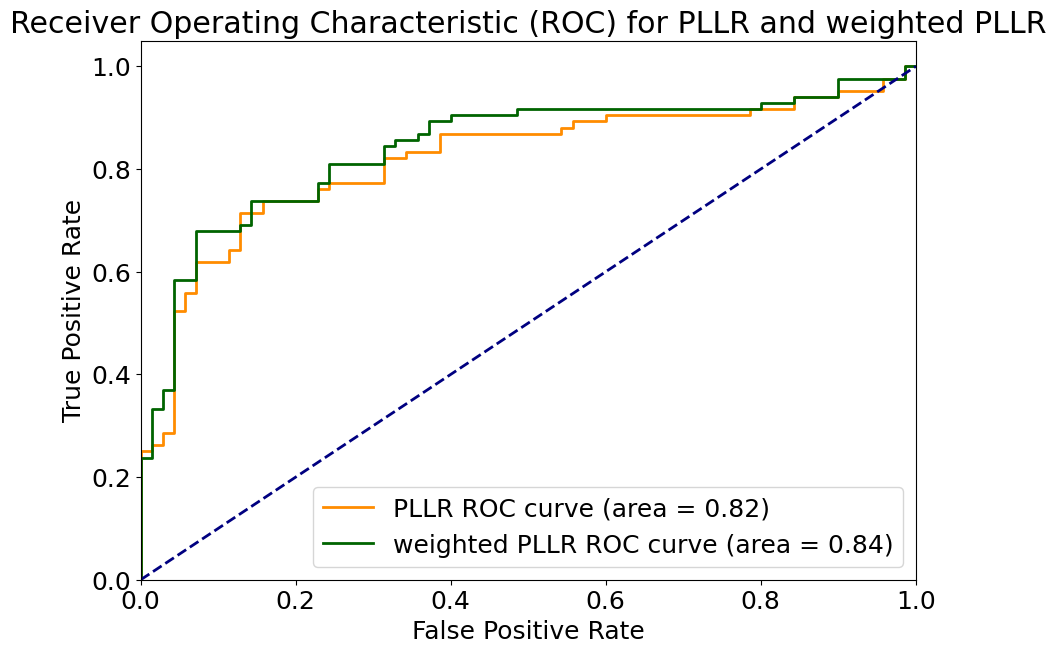}
    \end{subfigure}
    \hfill
        \begin{subfigure}[b]{0.42\linewidth}
            \caption{AUPR on ARM for ESM2}\label{fig:apr_arm_esm2}
        \includegraphics[width=\linewidth]{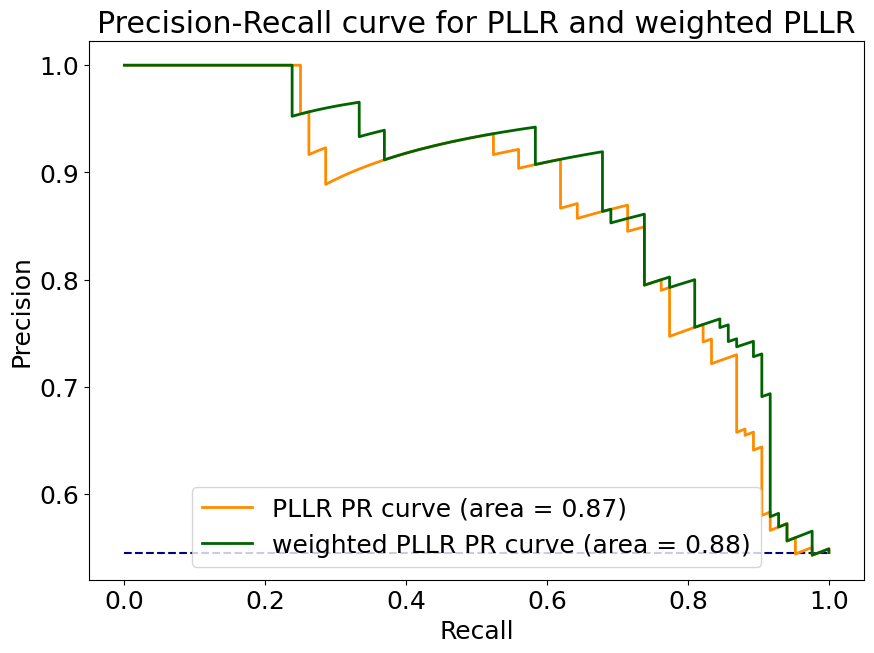}
    \end{subfigure}
       \begin{subfigure}[b]{0.5\linewidth}
            \caption{AUC on ARM for \textsc{dyna} fine-tuned ESM2}\label{fig:auc_arm_esm2}
        \includegraphics[width=\linewidth]{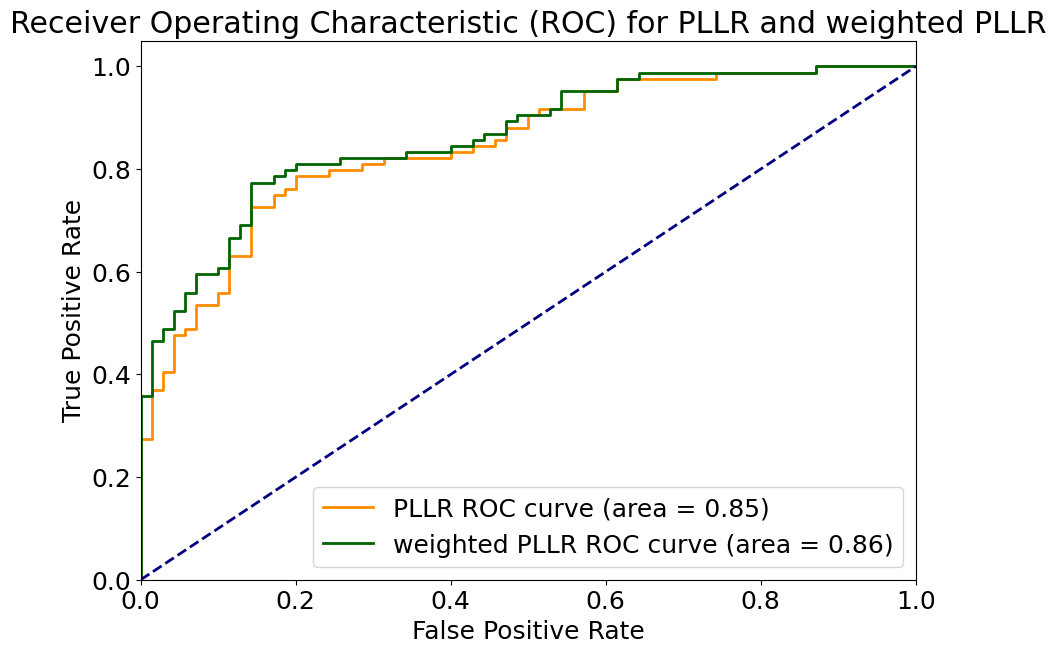}
    \end{subfigure}
    \hfill
        \begin{subfigure}[b]{0.42\linewidth}
            \caption{AUPR on ARM for \textsc{dyna} fine-tuned ESM2}\label{fig:apr_arm_esm2}
        \includegraphics[width=\linewidth]{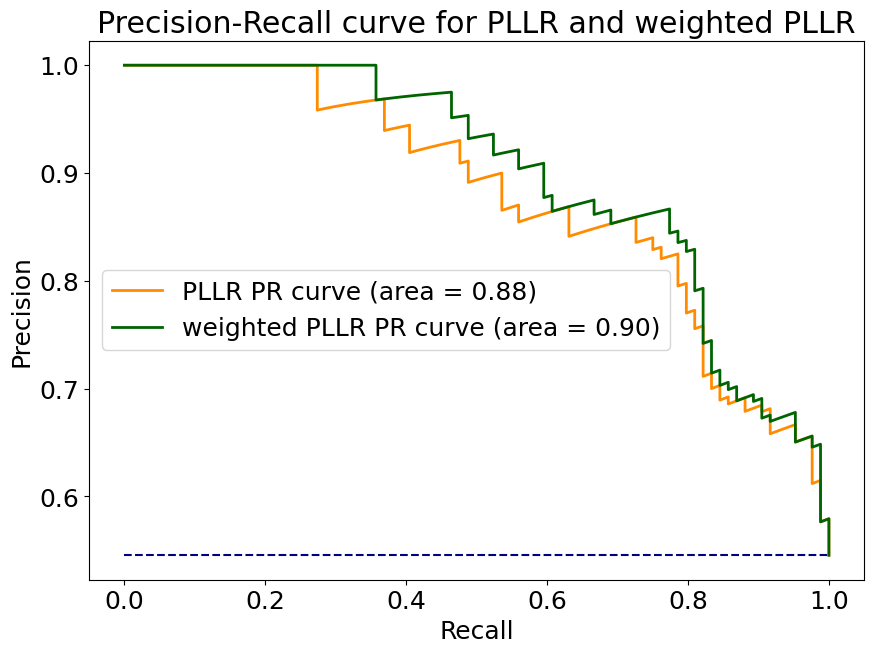}
    \end{subfigure}
\caption{ AUC and AUPR performances on CM and ARM for ESM2 and \textsc{dyna} fine-tuned ESM2.
}\label{fig:esm2}
\end{figure} 

\begin{table}[!htbp]
\caption{ Performance for cardiomyopathy variant pathogenicity prediction.}
\label{tab:cm-baselines}
\centering
 \begin{tabular}{c |>{\centering\arraybackslash}p{3cm} >{\centering\arraybackslash}p{1.5cm} >{\centering\arraybackslash}p{1.5cm} >{\centering\arraybackslash}p{1.5cm} }
    \hline
     \multirow{2}{*}{\textbf{Category}}   & \multirow{2}{*}{\textbf{Algorithm}} & \textbf{Disease-specific} & \multirow{2}{*}{\textbf{AUC}} & \multirow{2}{*}{\textbf{AUPR}}\\
      \hline
          \multirow{3}{*}{Tree-based} 
         & CART & $\checkmark$ &  $0.83$ & $0.81$  \\
         & RF & $\checkmark$ & $0.90$ & $0.89$ \\
         & BART & $\checkmark$ &$0.91$ & $0.89$\\
      \hline
    \multirow{3}{*}{Boosting models} 
         & XGBoost~\citep{zhang2021disease} & $\checkmark$ &  $0.87$ & $0.9$  \\
         & AdaBoost~\citep{zhang2021disease} & $\checkmark$ & $0.88$ & $0.9$ \\
         & M-CAP~\citep{jagadeesh2016m} & $\times$ &$0.79$ & $0.8$\\
          \hline
    \multirow{1}{*}{Ensemble models}  
         & REVEL~\citep{ioannidis2016revel} & $\times$ & $0.81$ & $0.79$ \\
         \hline
    \multirow{3}{*}{Language models}
         & ESM1b~\citep{brandes2023genome} &$\times$ &$0.82$ & $0.84$ \\
          & ESM2~\citep{lin2023evolutionary} &$\times$ &$0.74$ & $0.76$ \\
         & \textsc{dyna}  & $\checkmark$ & $\mathbf{0.88}$ &$\mathbf{0.91}$ \\
         \hline
    \end{tabular}
\end{table}

\begin{table}[!htbp]
\caption{ Performance for arrhythmias variant pathogenicity prediction.}
\label{tab:arm-baselines}
\centering
 \begin{tabular}{l |>{\centering\arraybackslash}p{3cm} >{\centering\arraybackslash}p{1.5cm} >{\centering\arraybackslash}p{1.5cm} >{\centering\arraybackslash}p{1.5cm} }
    \hline
      \multirow{2}{*}{\textbf{Category}}   & \multirow{2}{*}{\textbf{Algorithm}}  & \textbf{Disease-specific} & \multirow{2}{*}{\textbf{AUC}} & \multirow{2}{*}{\textbf{AUPR}}\\
        \hline
          \multirow{3}{*}{Tree-based} 
         & CART & $\checkmark$ &  $0.82$ & $0.86$  \\
         & RF & $\checkmark$ & $0.93$ & $0.92$ \\
         & BART & $\checkmark$ &$\textit{0.93}$ & $\textit{0.92}$\\
      \hline
    \multirow{3}{*}{Boosting models} 
         & XGBoost~\citep{zhang2021disease} & $\checkmark$& $0.90$ & $0.88$  \\
         & AdaBoost~\citep{zhang2021disease}& $\checkmark$ & $0.90$ & $0.90$ \\
         & M-CAP~\citep{jagadeesh2016m} & $\times$ & $0.85$ & $0.81$\\
          \hline
    \multirow{1}{*}{Ensemble models}  
         & REVEL~\citep{ioannidis2016revel} & $\times$ & $0.81$ & $0.79$ \\
         \hline
    \multirow{2}{*}{Language models}
         & ESM1b~\citep{brandes2023genome} & $\times$& $0.90$ & $0.89$ \\
           & ESM2~\citep{lin2023evolutionary} &$\times$ &$0.84$ & $0.88$ \\
         & \textsc{dyna}  &$\checkmark$ & $\mathbf{0.94}$ & $\mathbf{0.95}$ \\
         \hline
    \end{tabular}
\end{table}

\begin{figure}[!htbp]
    \centering
    
      \begin{subfigure}[b]{0.48\linewidth}
            \caption{}\label{fig:auc_cm_error}
        \includegraphics[width=\linewidth]{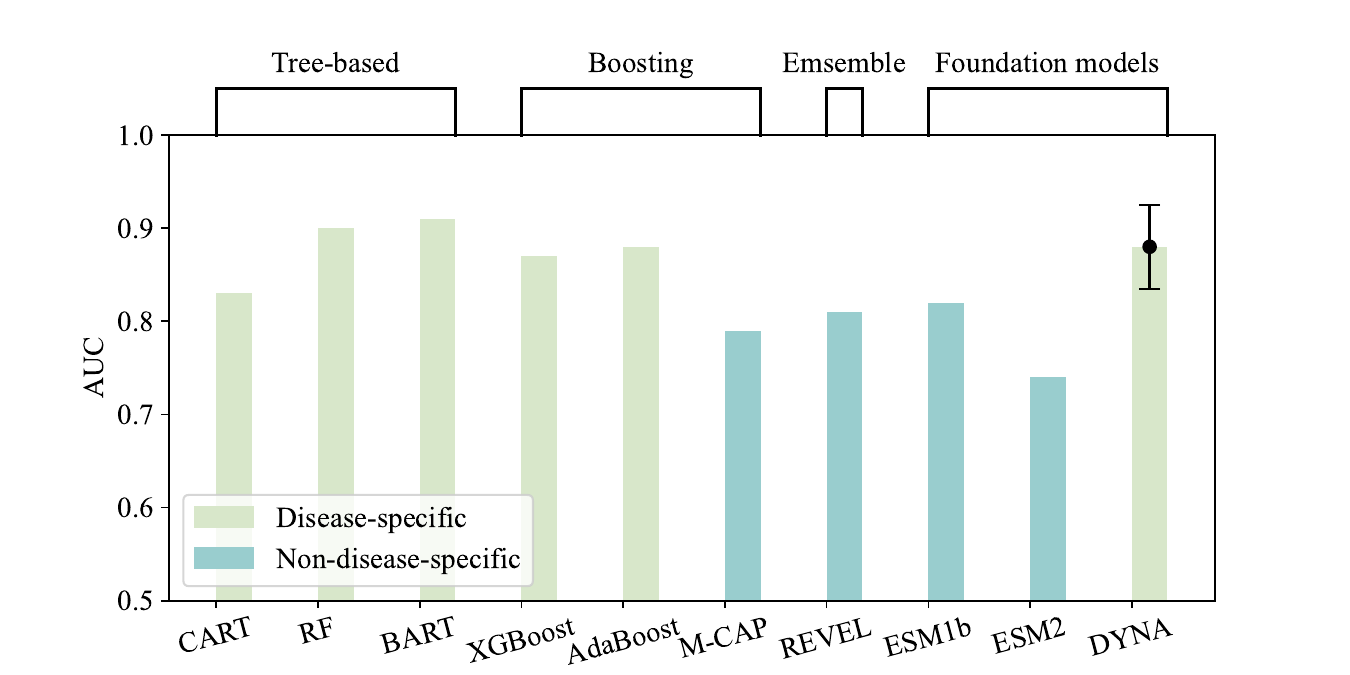}
    \end{subfigure}
    \hfill
        \begin{subfigure}[b]{0.48\linewidth}
            \caption{}\label{fig:auc_arm_error}
        \includegraphics[width=\linewidth]{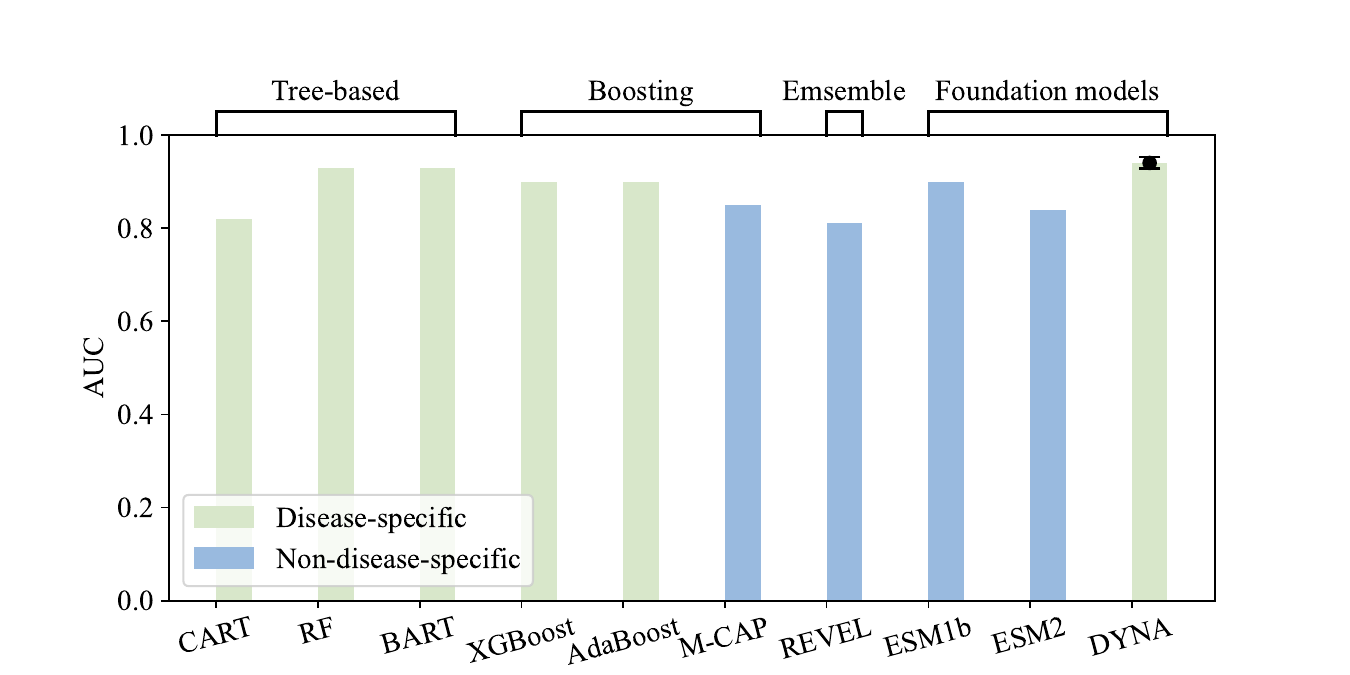}
    \end{subfigure}
       \begin{subfigure}[b]{0.48\linewidth}
            \caption{}\label{fig:aupr_cm_error}
        \includegraphics[width=\linewidth]{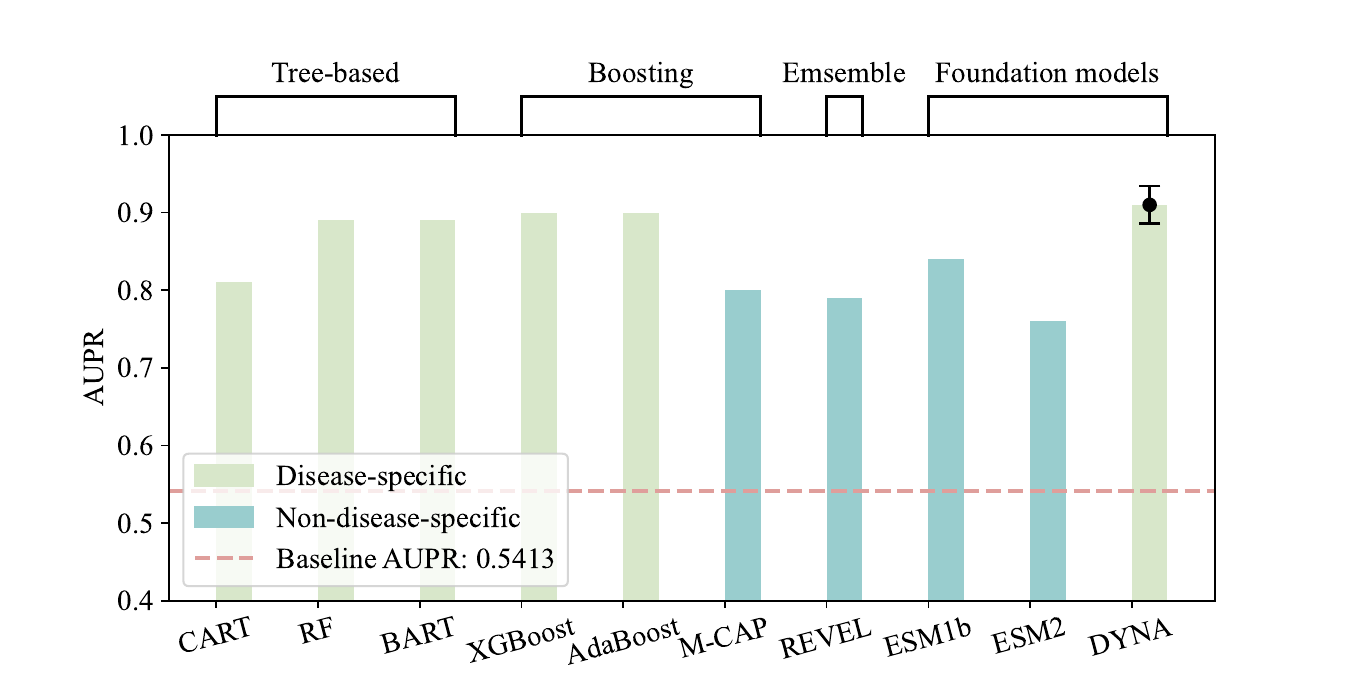}
    \end{subfigure}
    \hfill
        \begin{subfigure}[b]{0.48\linewidth}
            \caption{}\label{fig:aupr_arm_error}
        \includegraphics[width=\linewidth]{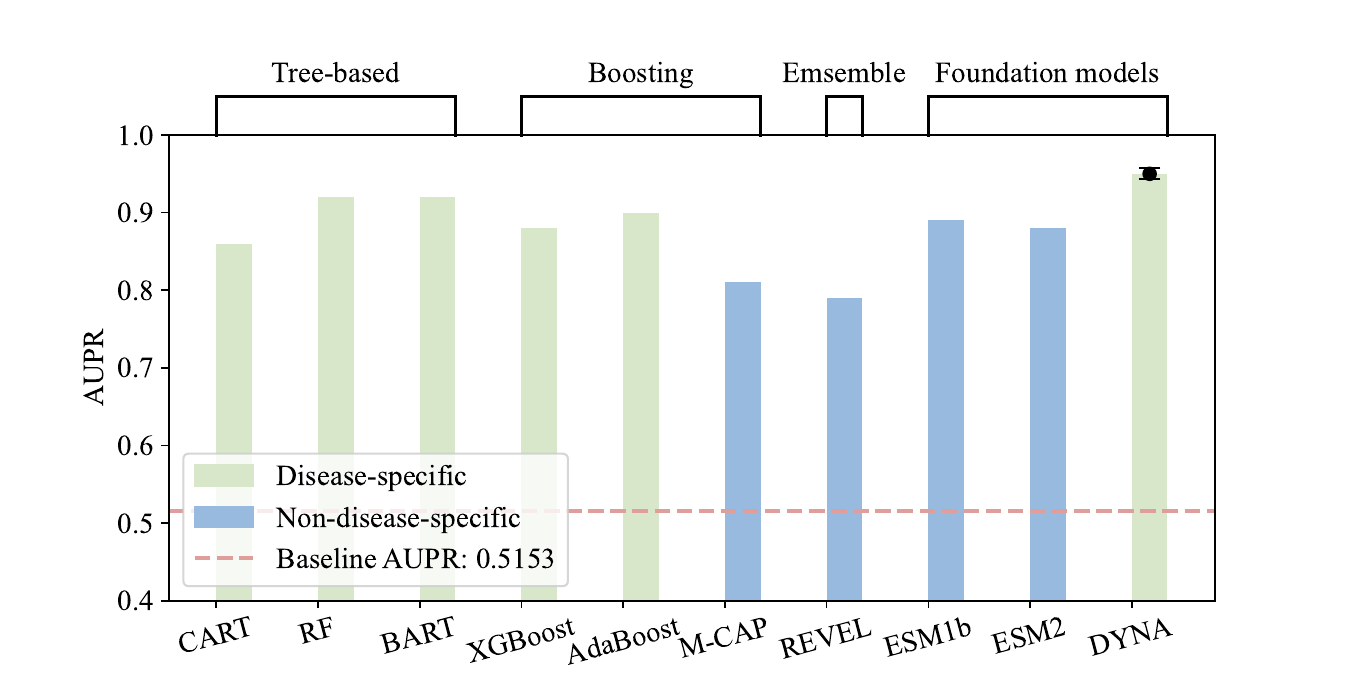}
    \end{subfigure}
\caption{ AUC and AUPR performances on CM and ARM for \textsc{dyna} and baselines.
}
\end{figure} 

\begin{figure}[!htbp]
    \centering


            \begin{subfigure}[b]{0.24\linewidth}
            \caption{}\label{fig:cm_distribution}
        \includegraphics[width=\linewidth]{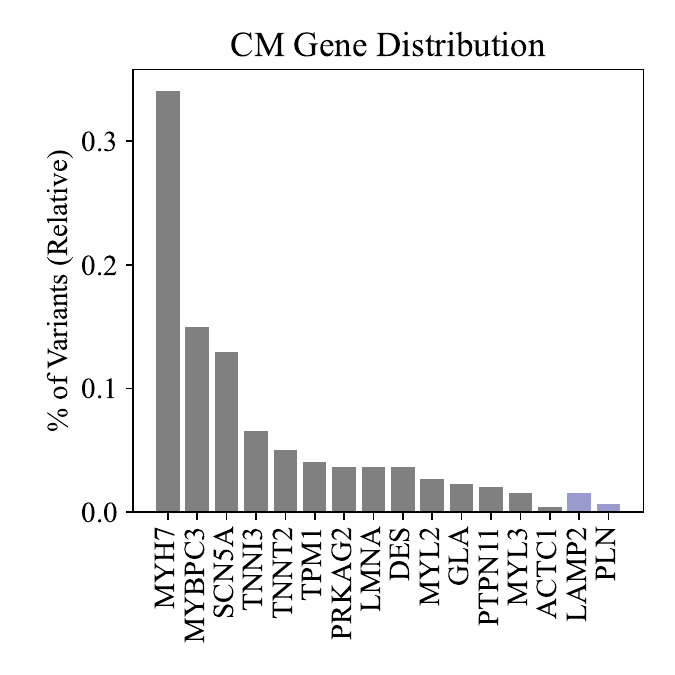}
    \end{subfigure}
    \begin{subfigure}[b]{0.24\linewidth}
            \caption{}\label{fig:arm_distribution}
        \includegraphics[width=\linewidth]{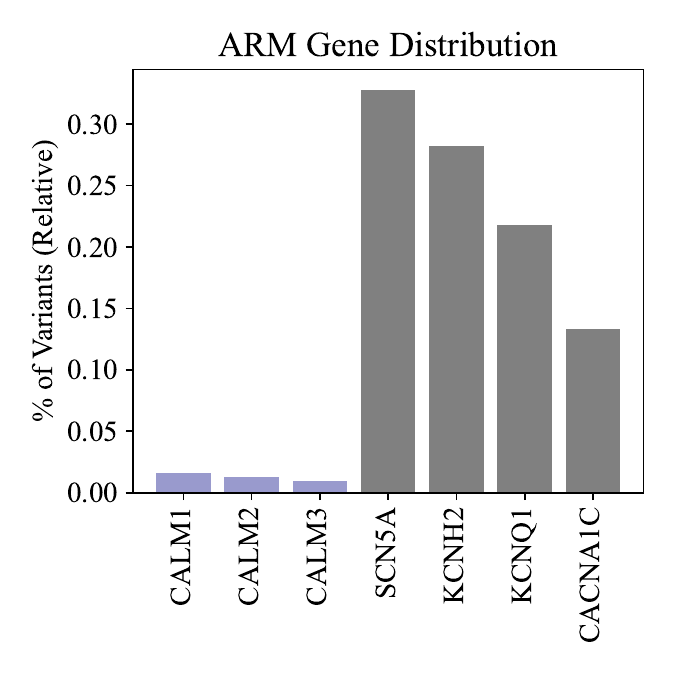}
    \end{subfigure}
    \caption{CM and ARM sample distribution based on genes.}
\end{figure}

Supplementary Figure~\ref{fig:cm_distribution} and~\ref{fig:arm_distribution} provide the sample distributions of CM and ARM based on genes, respectively. Grey bars indicate overlapping CM and ARM genes with ClinVar CM/ARM but differing mutation positions. Similarly, purple bars represent CM/ARM genes that differ from those in ClinVar CM/ARM in both genes and positions.

\section{Distributions of CM and ARM based on Genes}



\section{AUC and Weighted AUC for ClinVar CM and ARM}\label{seq-auc-clinvar}

\begin{figure}[!htbp]
    \centering
    
    \begin{subfigure}[b]{0.24\linewidth}
            \caption{}\label{fig:clinvar_arm_distribution}
        \includegraphics[width=\linewidth]{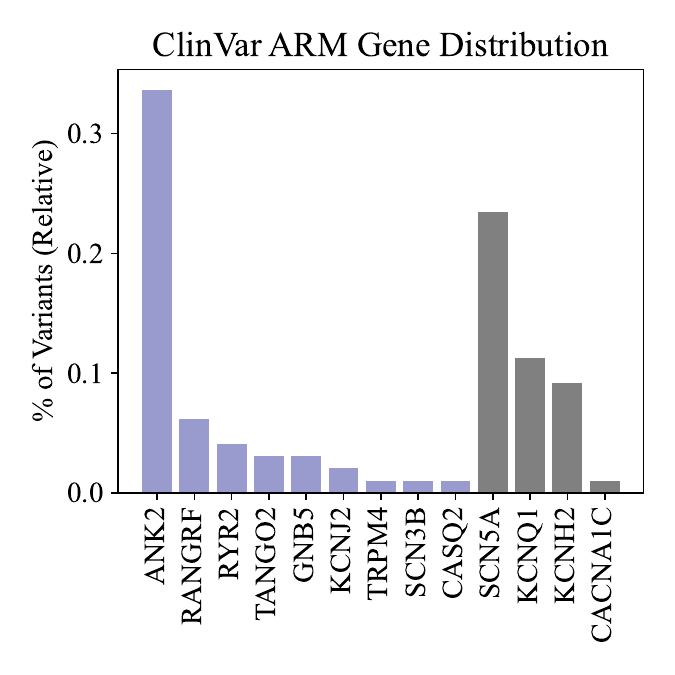}
    \end{subfigure}
    \begin{subfigure}[b]{\linewidth}
            \caption{}\label{fig:clinvar_cm_distribution}
        \includegraphics[width=\linewidth]
        {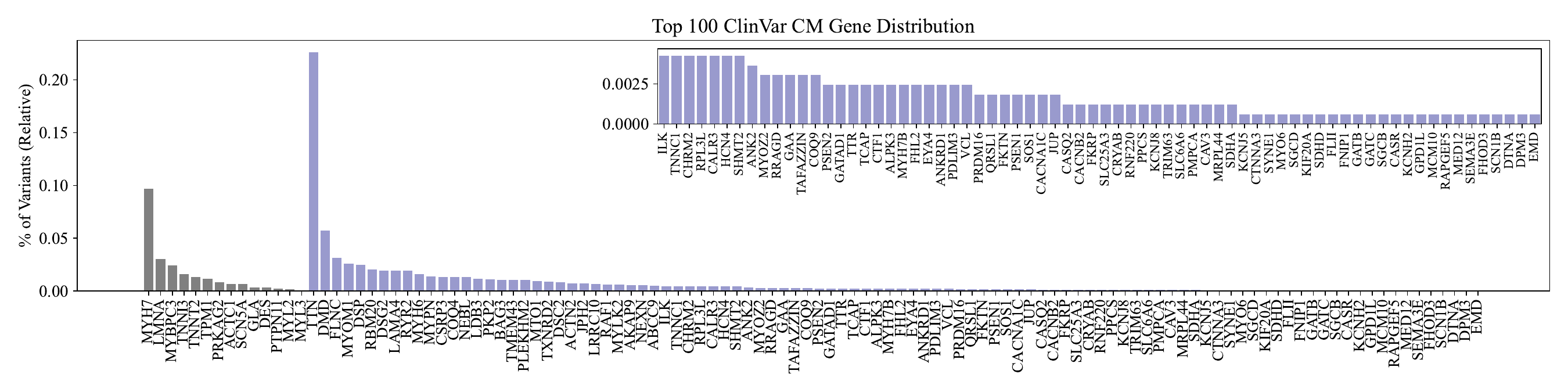}
    \end{subfigure}
\caption{
\textbf{\large{a}} Relative distribution of the top 100 ClinVar CM genes, highlighting the relative distributions of mutations within categories. Purple bars represent ClinVar CM genes sampled from diverse genes and positions, constituting 78.9\%, while grey bars depict CM genes with identical genes but varying mutation positions. \textbf{\large{b}} Relative distribution of mutations in ClinVar ARM genes, highlighting the relative distributions within categories. Purple bars represent ClinVar ARM genes sampled from diverse genes and positions, while grey bars depict ARM genes with identical genes but different mutation positions.
}
\end{figure}

Figure~\ref{fig:clinvar_cm_distribution} provides a detailed view of the top 100 ClinVar CM genes' distribution, demonstrating relative distributions within categories. Purple bars represent ClinVar CM genes sampled from diverse genes and positions, constituting $78.9\%$, while grey bars depict CM genes with identical genes but varying mutation positions.
In Figure~\ref{fig:clinvar_arm_distribution}, purple bars represent $9$ ClinVar ARM genes sampled from diverse genes and positions, while grey bars show $3$ ARM genes with identical genes but different mutation positions. This underscores ClinVar ARM's pronounced divergence from ARM distribution. 
Supplementary Figure~\ref{fig:cm_distribution} and~\ref{fig:arm_distribution} provide the sample distributions of CM and ARM based on genes, respectively. Grey bars indicate overlapping CM and ARM genes with ClinVar CM/ARM but differing mutation positions, while purple bars represent CM/ARM genes that differ from those in ClinVar CM/ARM in both genes and positions.

\begin{figure}[!htbp]
\centering
       \begin{subfigure}[b]{0.24\linewidth}
            \caption{}\label{fig:tpr_cm}
        \includegraphics[width=\linewidth]{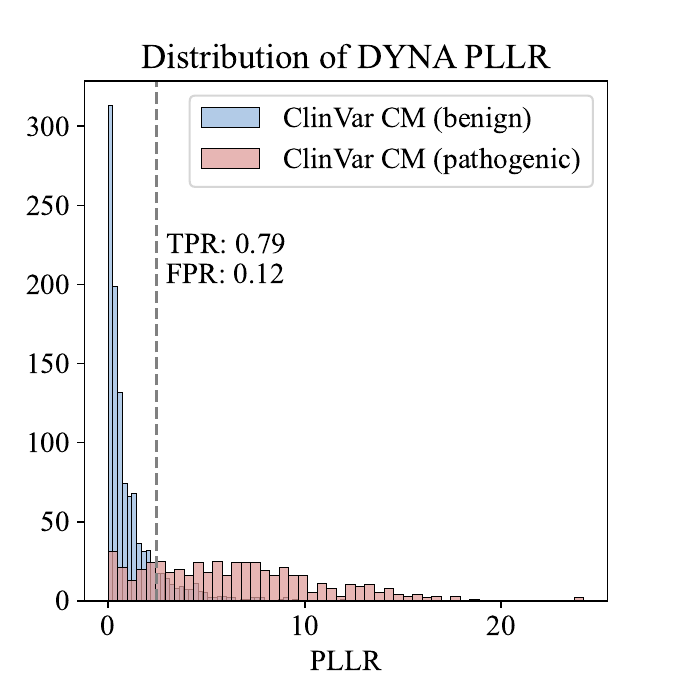}
    \end{subfigure}
        \begin{subfigure}[b]{0.24\linewidth}
            \caption{}\label{fig:tpr_arm}
        \includegraphics[width=\linewidth]{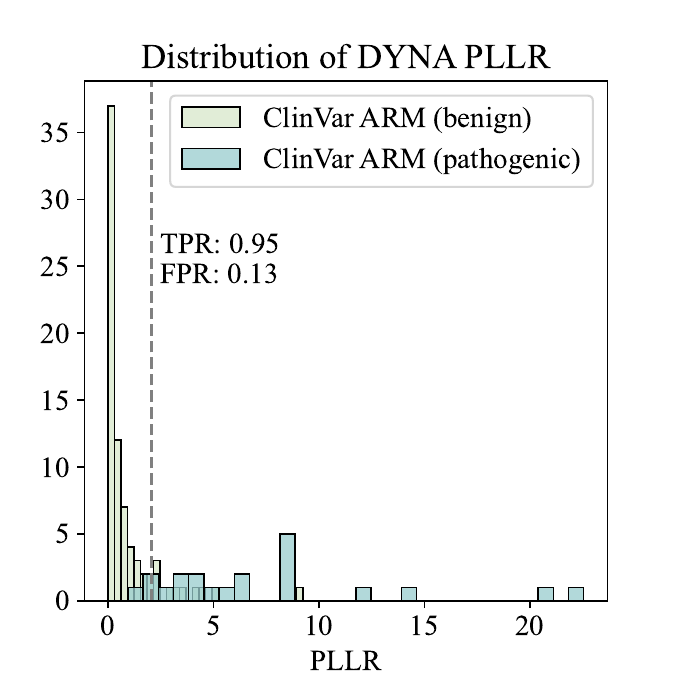}
    \end{subfigure}
           \caption{
\textbf{\large{a}} Distribution of ClinVar CM benign and pathogenic sequences. \textbf{\large{b}} Distribution of ClinVar ARM benign and pathogenic sequences. }
\end{figure}

\begin{figure}[!htbp]
    \centering

    \begin{subfigure}[b]{0.24\linewidth}
            \caption{}\label{fig:cm_benign_box}
        \includegraphics[width=\linewidth]{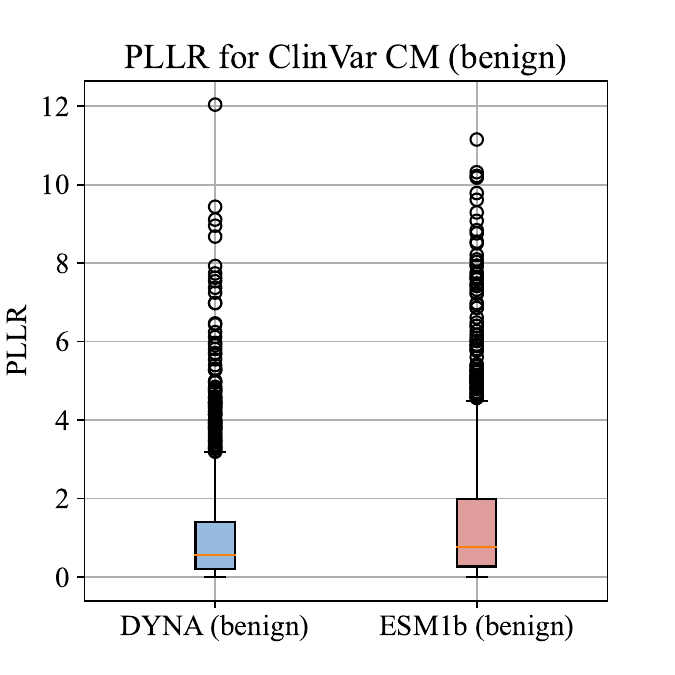}
    \end{subfigure}
    \begin{subfigure}[b]{0.24\linewidth}
            \caption{}\label{fig:cm_pathogenic_box}
        \includegraphics[width=\linewidth]{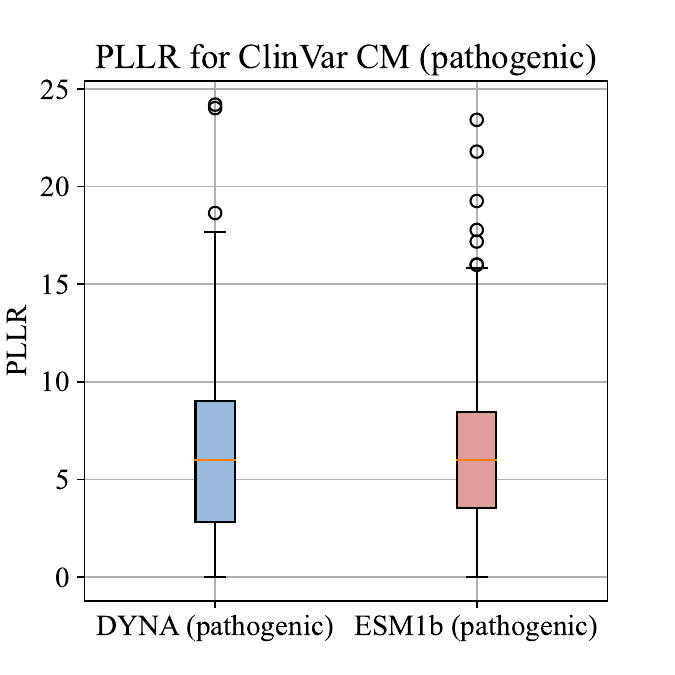}
    \end{subfigure}

        \begin{subfigure}[b]{0.24\linewidth}
            \caption{}\label{fig:arm_benign_box}
        \includegraphics[width=\linewidth]{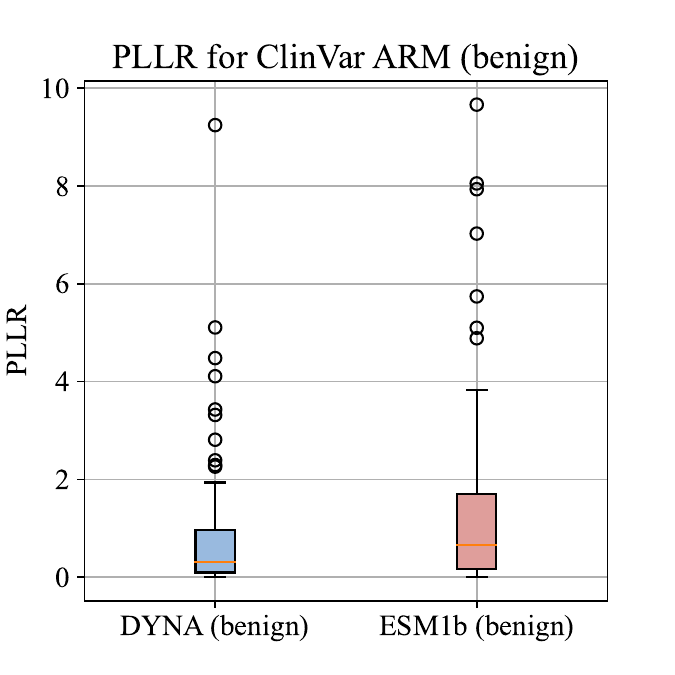}
    \end{subfigure}
    \begin{subfigure}[b]{0.24\linewidth}
            \caption{}\label{fig:arm_pathogenic_box}
        \includegraphics[width=\linewidth]{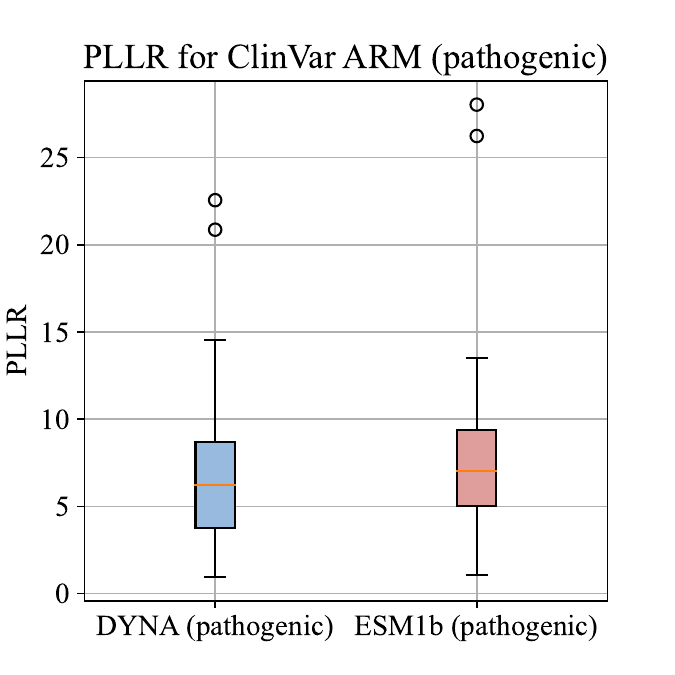}
    \end{subfigure}

           \caption{ 
           \textbf{\large{a}} presents the box plot of the distribution of PLLR for benign ClinVar CM sequences. \textbf{\large{b}} illustrates the box plot of the distribution of PLLR for pathogenic ClinVar CM sequences.
\textbf{\large{c}} presents the box plot of the distribution of PLLR for benign ClinVar ARM sequences. \textbf{\large{d}} illustrates the box plot of the distribution of PLLR for pathogenic ClinVar ARM sequences.}
\end{figure}

\begin{figure}[!htbp]
    \centering

    \begin{subfigure}[b]{0.48\linewidth}
            \caption{}\label{fig:cm_violin_p_value}
        \includegraphics[width=\linewidth]{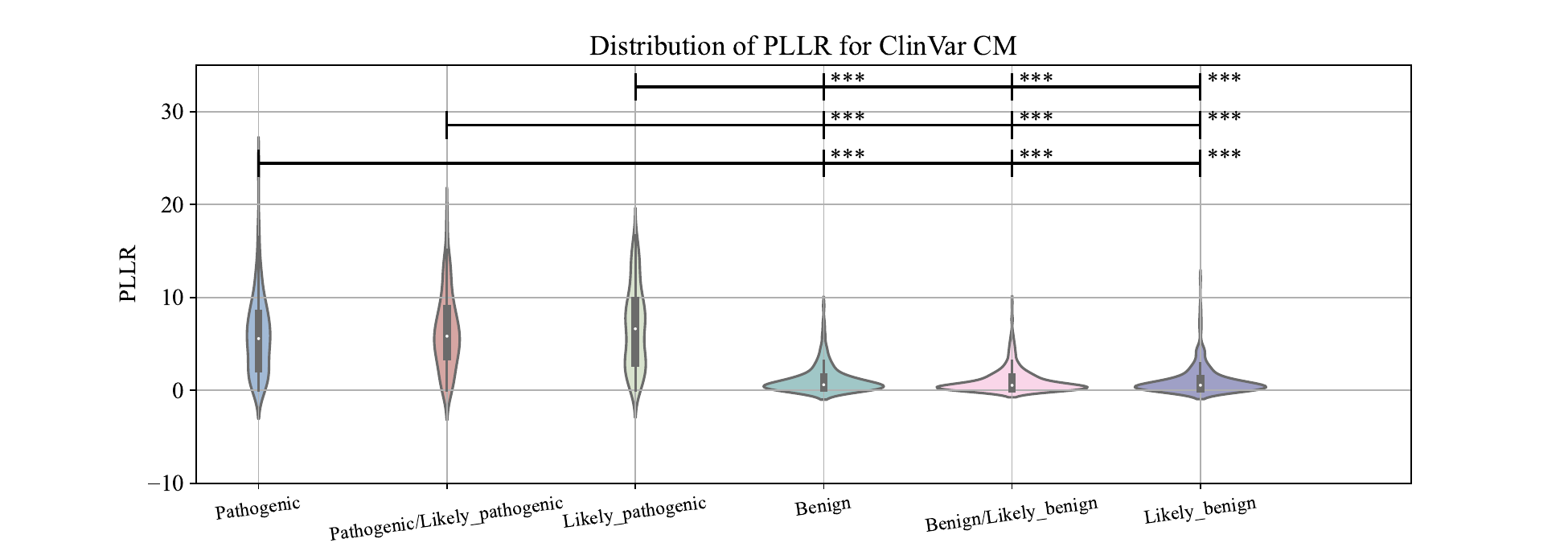}
    \end{subfigure}
    \hfill
        \begin{subfigure}[b]{0.48\linewidth}
            \caption{}\label{fig:arm_violin_p_value}
        \includegraphics[width=\linewidth]{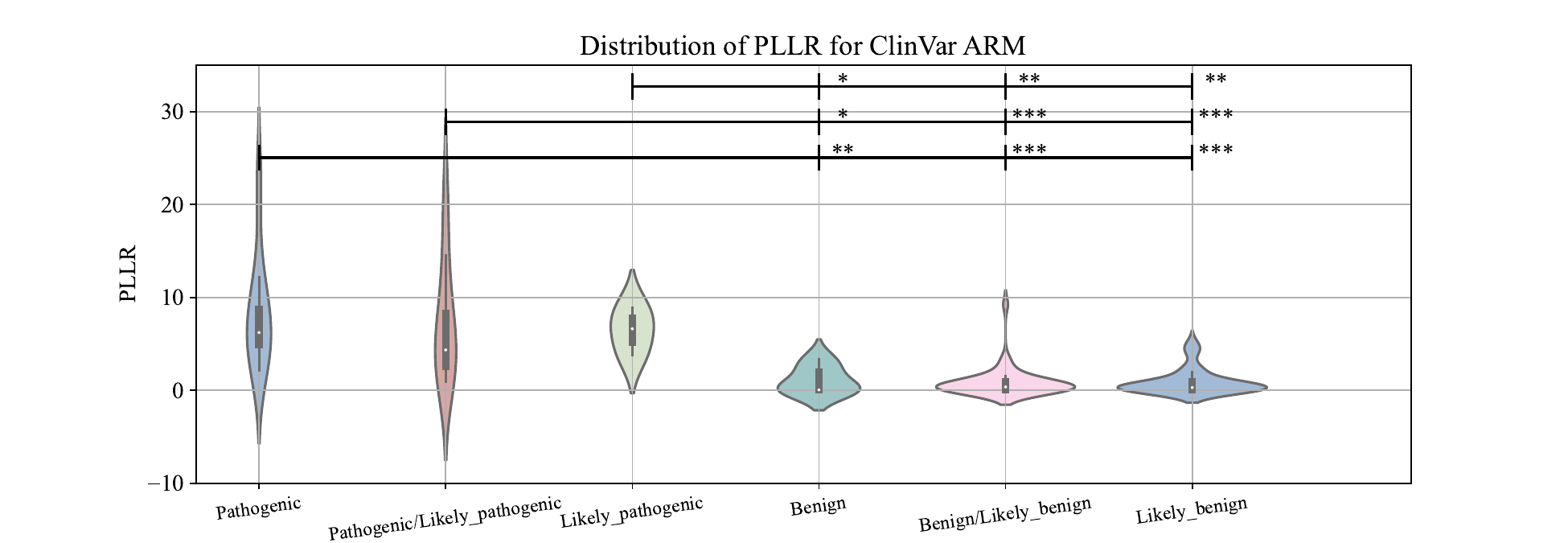}
    \end{subfigure}
\caption{
\textbf{\large{a}} Distribution of PLLR for ClinVar CM on common class labels.
\textbf{\large{b}} Distribution of PLLR for ClinVar ARM on common class labels.
}
\end{figure}

\begin{figure}[!htbp]
    \centering

    \begin{subfigure}[b]{0.24\linewidth}
            \caption{}
        \includegraphics[width=\linewidth]{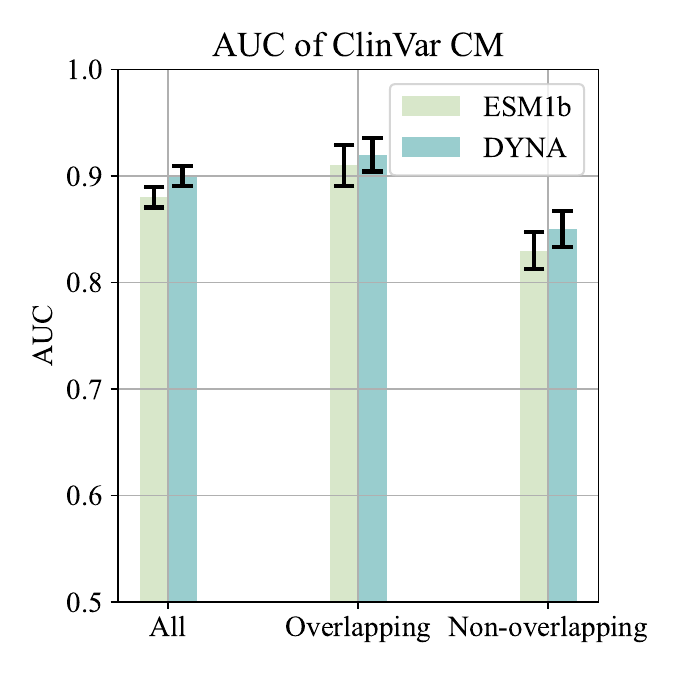}
    \end{subfigure}
                \hfill
    \begin{subfigure}[b]{0.24\linewidth}
            \caption{}
        \includegraphics[width=\linewidth]{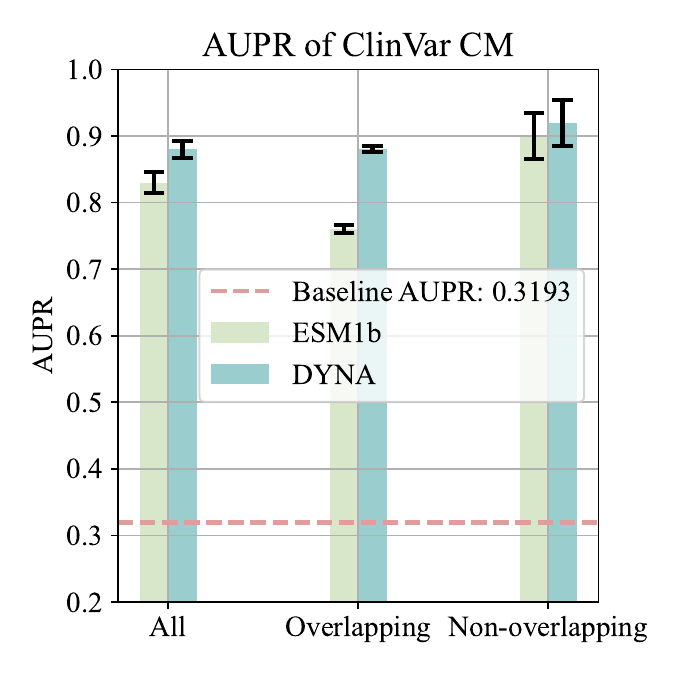} 
    \end{subfigure}
        \hfill
    \begin{subfigure}[b]{0.24\linewidth}
            \caption{}
        \includegraphics[width=\linewidth]{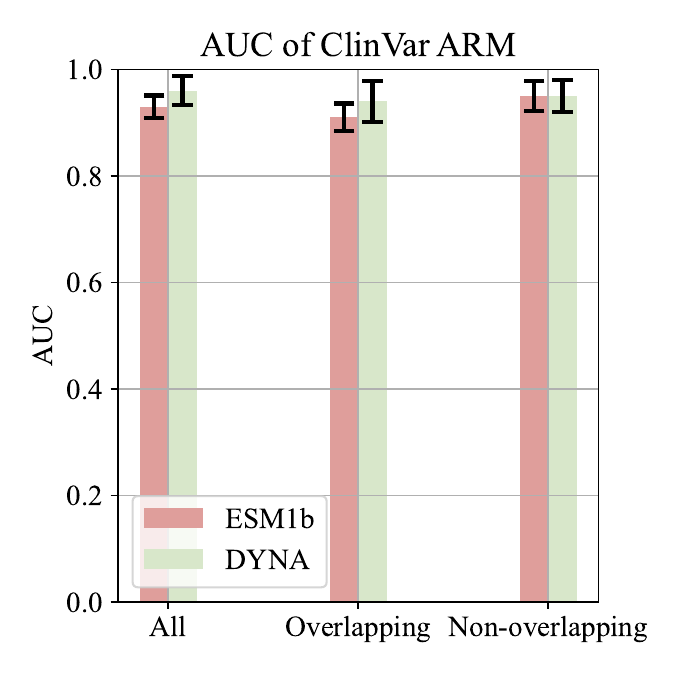}
    \end{subfigure}
                \hfill
    \begin{subfigure}[b]{0.24\linewidth}
            \caption{}\label{fig:apr_clinvar_arm}
        \includegraphics[width=\linewidth]{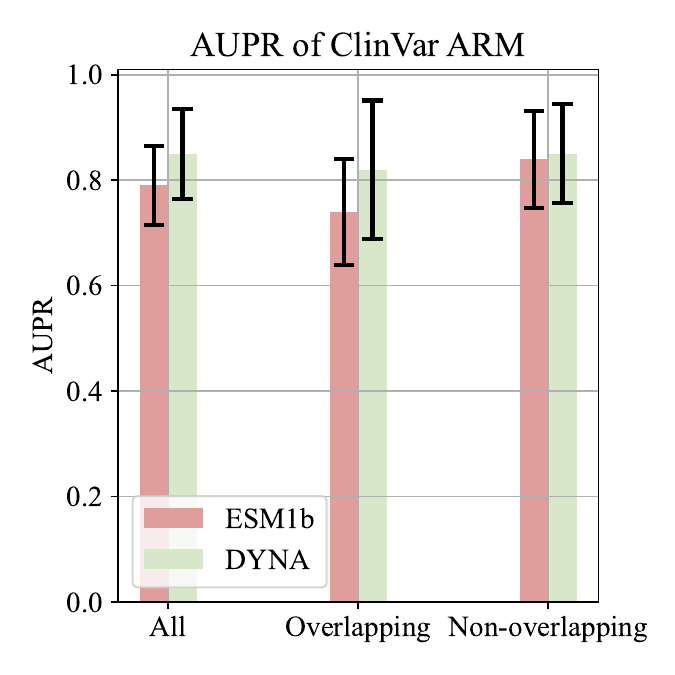}
        \end{subfigure}
        \caption{
Graphical representation of AUC and AUPR scores for \textsc{dyna} compared to ESM1b across the ClinVar CM and ARM datasets. These figures detail the performance metrics for non-weighted PLLR evaluations, demonstrating \textsc{dyna}'s superior generalization capabilities across all groups. Enhanced performance is particularly noted in non-overlapping genes, validating the model's strength in generalization to unseen disease-relevant genes compared to intra-gene generalization.
}\label{fig:auc_aupr_non_weighted}
\end{figure}

\begin{figure}[!htbp]
    \centering
    
    \begin{subfigure}[b]{0.24\linewidth}
            \caption{}
        \includegraphics[width=\linewidth]{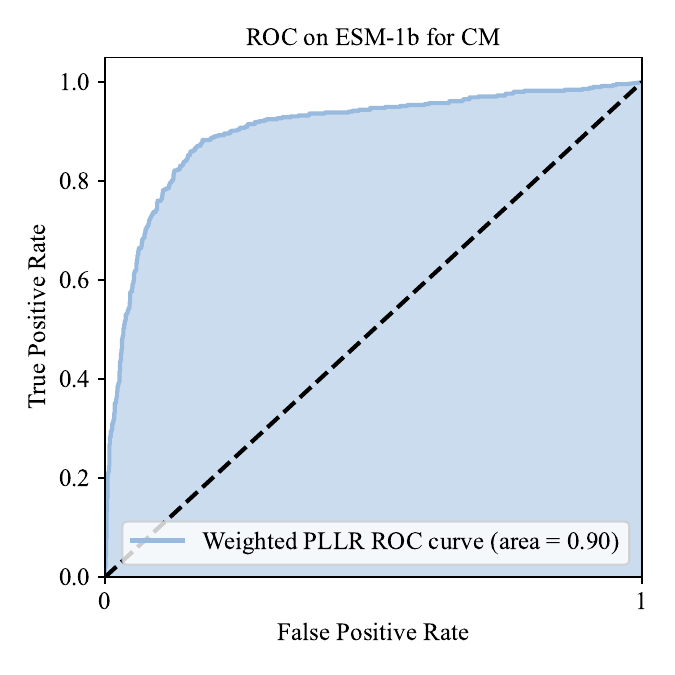}
    \end{subfigure}
    \hfill
    \begin{subfigure}[b]{0.24\linewidth}
            \caption{}
        \includegraphics[width=\linewidth]{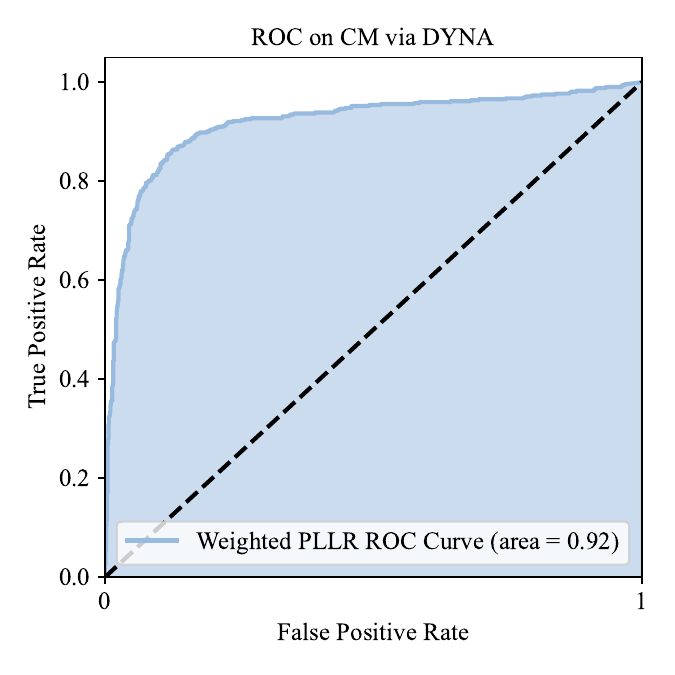}
    \end{subfigure}
    \hfill
    \begin{subfigure}[b]{0.24\linewidth}
            \caption{}
        \includegraphics[width=\linewidth]{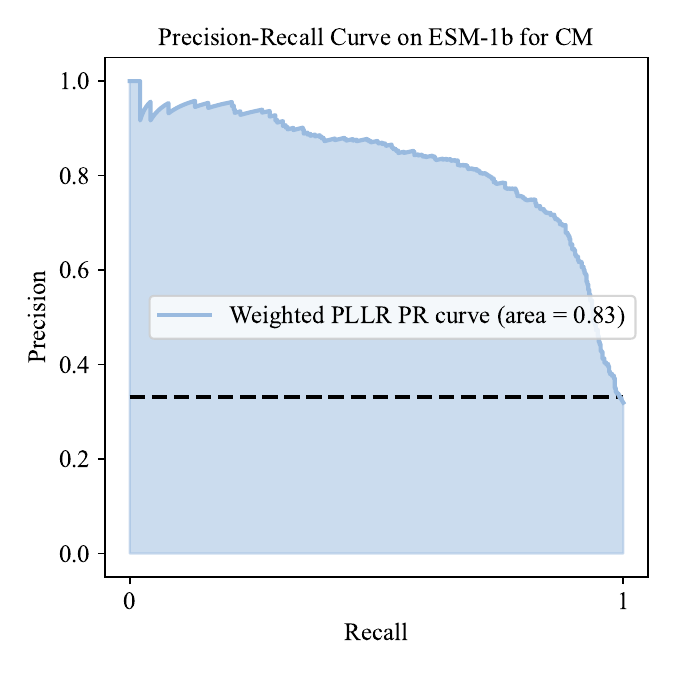}
    \end{subfigure}
    \hfill
    \begin{subfigure}[b]{0.24\linewidth}
            \caption{}
        \includegraphics[width=\linewidth]{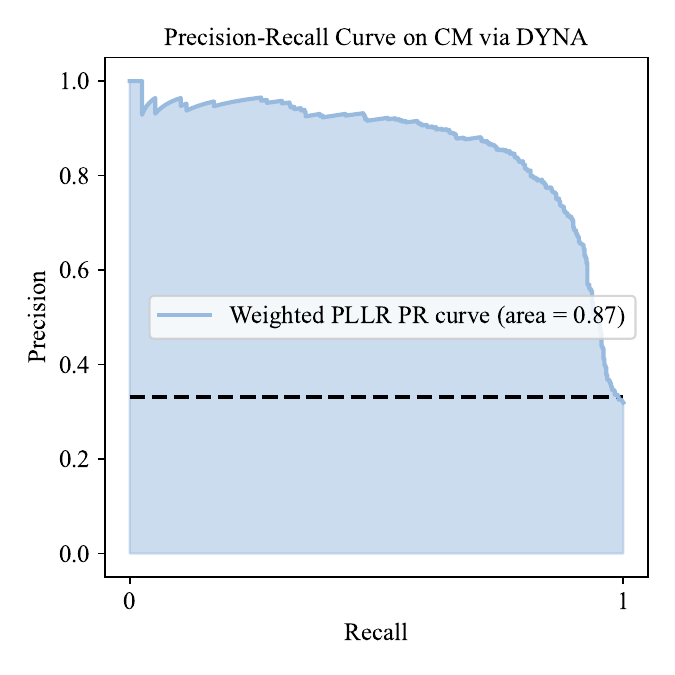}
    \end{subfigure}
    
    \begin{subfigure}[b]{0.24\linewidth}
            \caption{}
        \includegraphics[width=\linewidth]{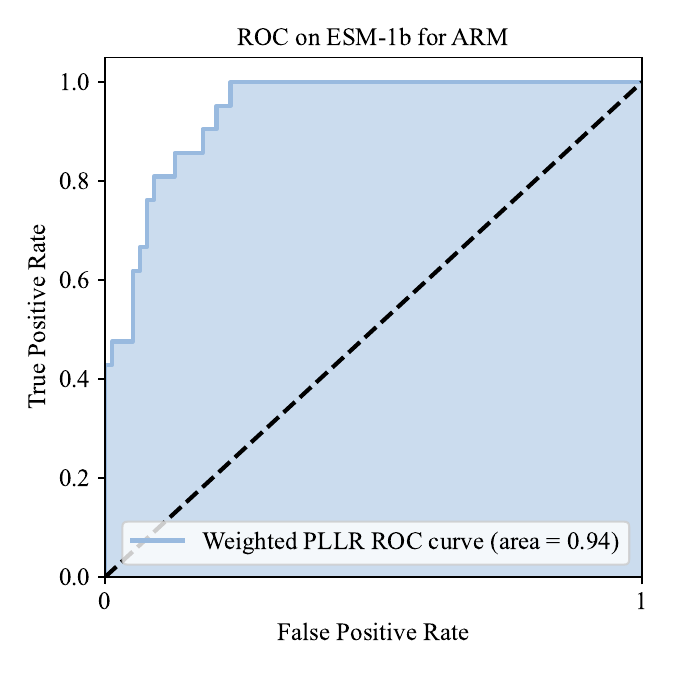}
    \end{subfigure}
    \hfill
    \begin{subfigure}[b]{0.24\linewidth}
            \caption{}
        \includegraphics[width=\linewidth]{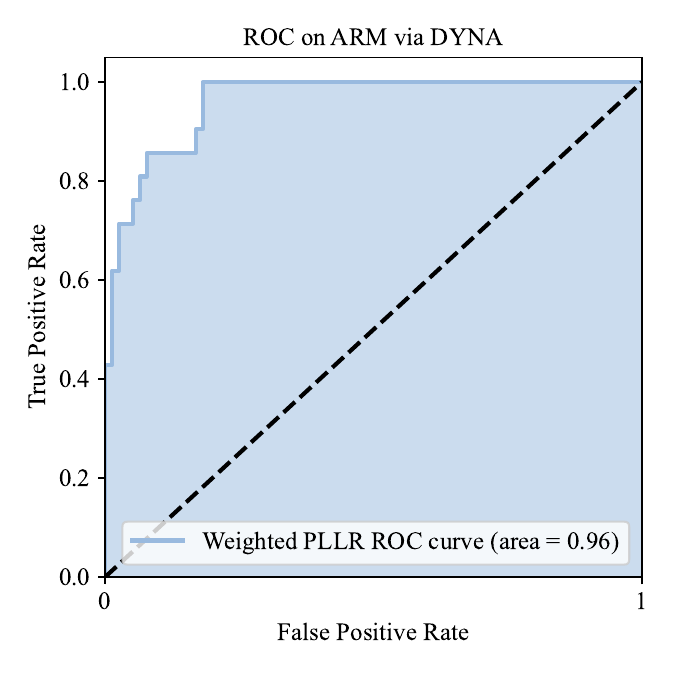}
    \end{subfigure}
    \hfill
    \begin{subfigure}[b]{0.24\linewidth}
            \caption{}
        \includegraphics[width=\linewidth]{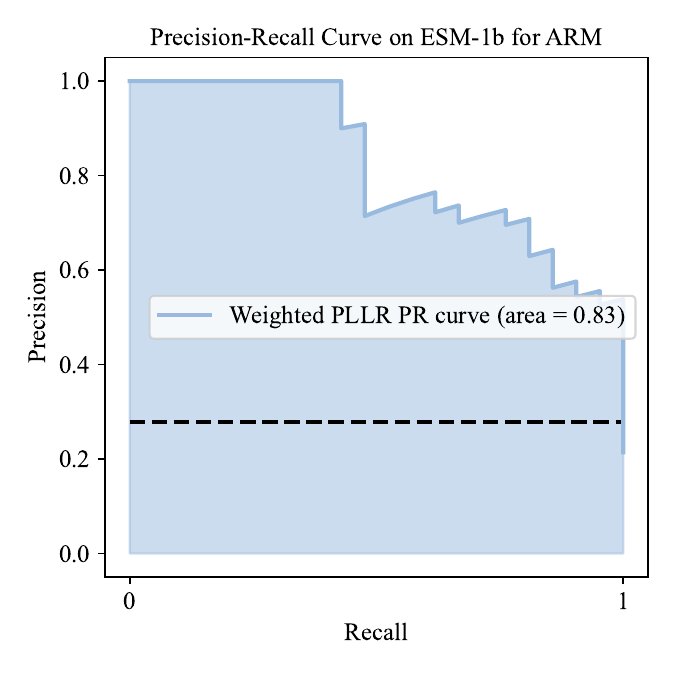}
    \end{subfigure}
    \hfill
    \begin{subfigure}[b]{0.24\linewidth}
            \caption{}
        \includegraphics[width=\linewidth]{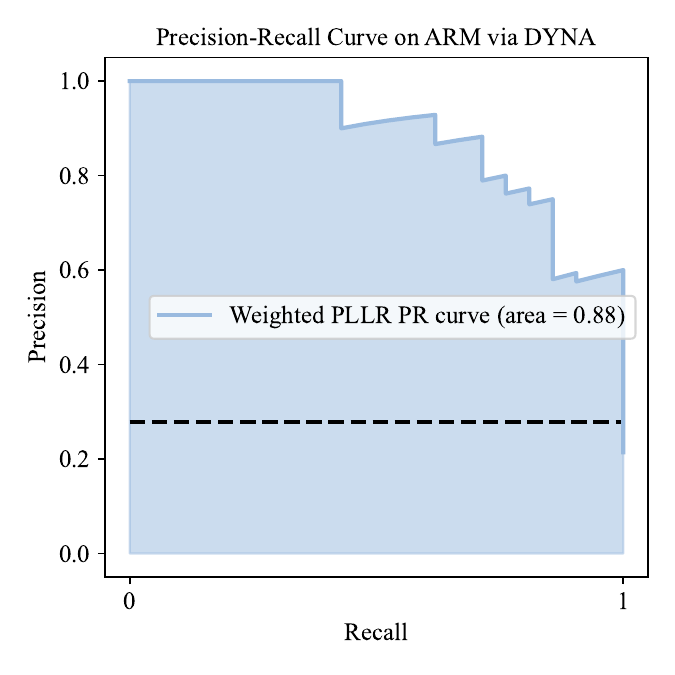}
    \end{subfigure}
    
    \caption{\textsc{dyna} ROC performances on ClinVar CM and ClinVar ARM.}\label{fig:clinvar_auc_esm1b}
\end{figure}

\begin{figure}[!htbp]
    \centering
    
    \begin{subfigure}[b]{0.24\linewidth}
            \caption{}
        \includegraphics[width=\linewidth]{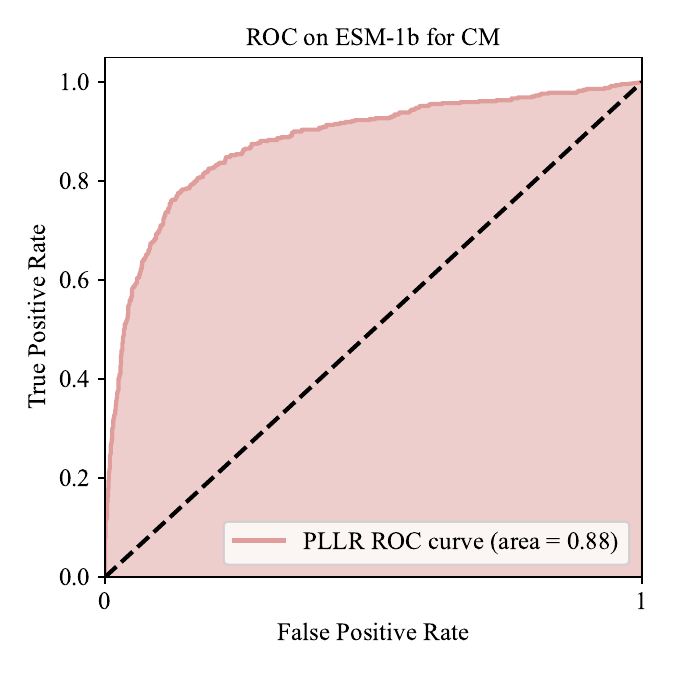}
    \end{subfigure}
    \hfill
    \begin{subfigure}[b]{0.24\linewidth}
            \caption{}
        \includegraphics[width=\linewidth]{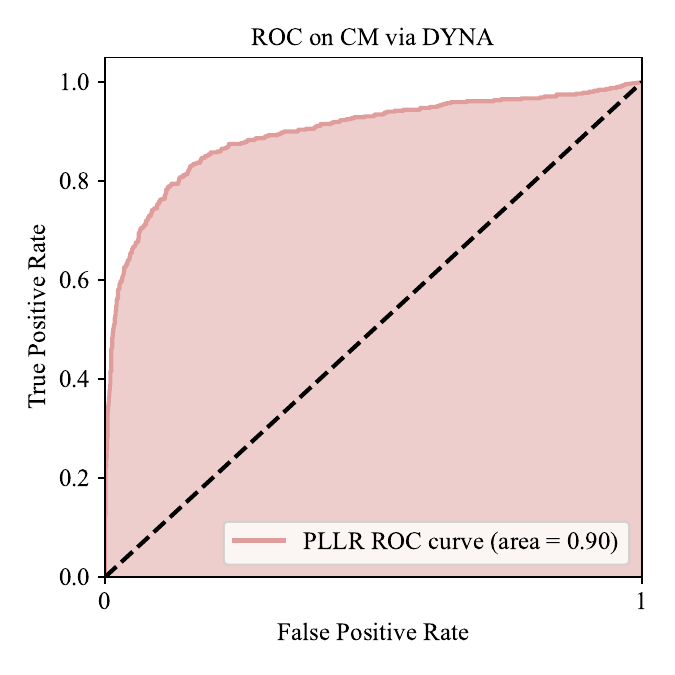}
    \end{subfigure}
    \hfill
    \begin{subfigure}[b]{0.24\linewidth}
            \caption{}
        \includegraphics[width=\linewidth]{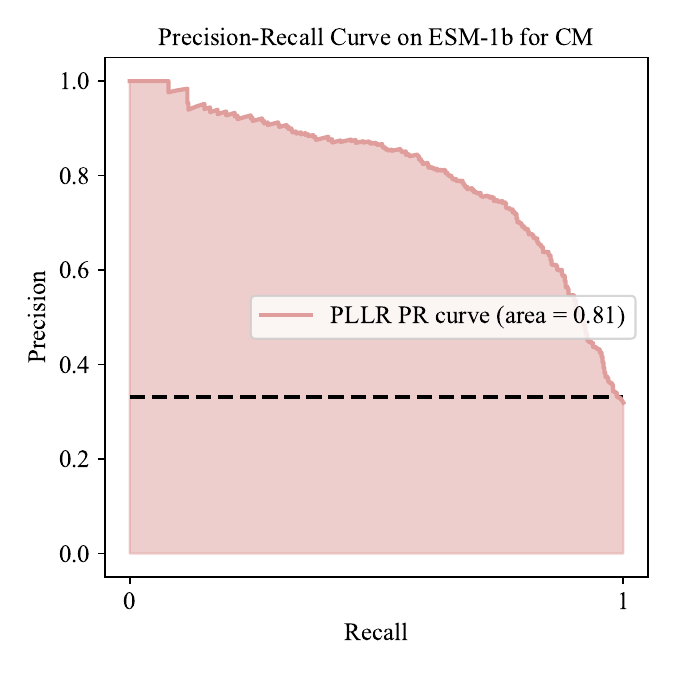}
    \end{subfigure}
    \hfill
    \begin{subfigure}[b]{0.24\linewidth}
            \caption{}
        \includegraphics[width=\linewidth]{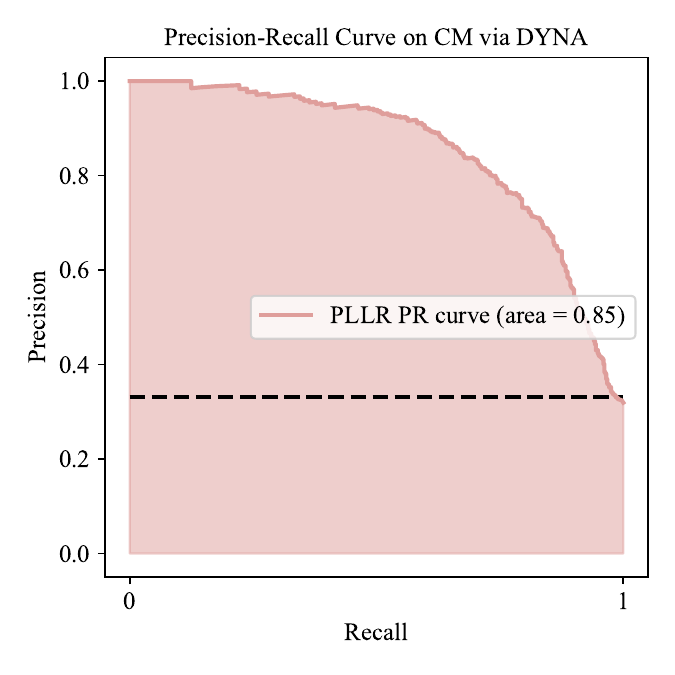}
    \end{subfigure}
    
    \begin{subfigure}[b]{0.24\linewidth}
            \caption{}
        \includegraphics[width=\linewidth]{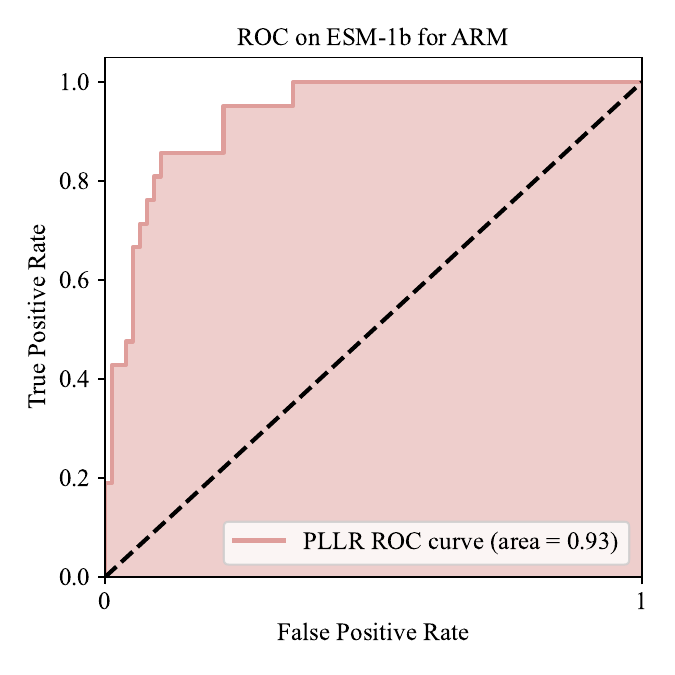}
    \end{subfigure}
    \hfill
    \begin{subfigure}[b]{0.24\linewidth}
            \caption{}
        \includegraphics[width=\linewidth]{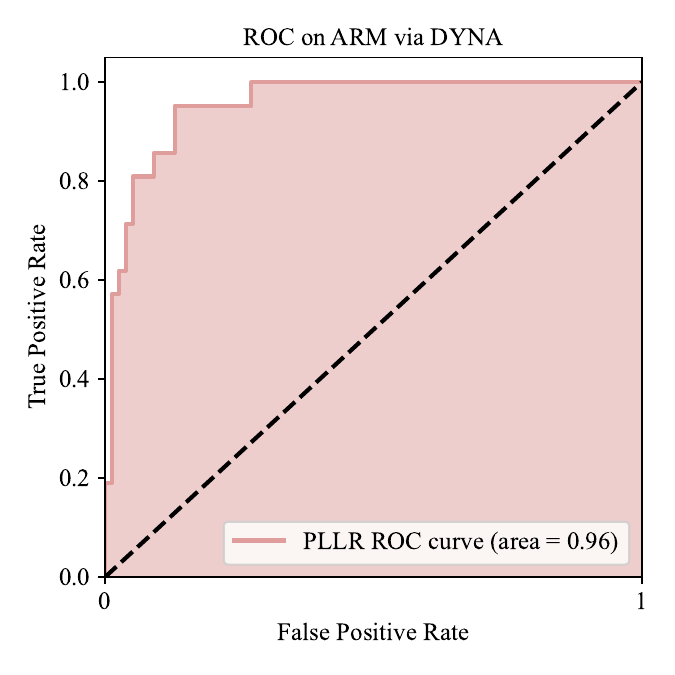}
    \end{subfigure}
    \hfill
    \begin{subfigure}[b]{0.24\linewidth}
            \caption{}
        \includegraphics[width=\linewidth]{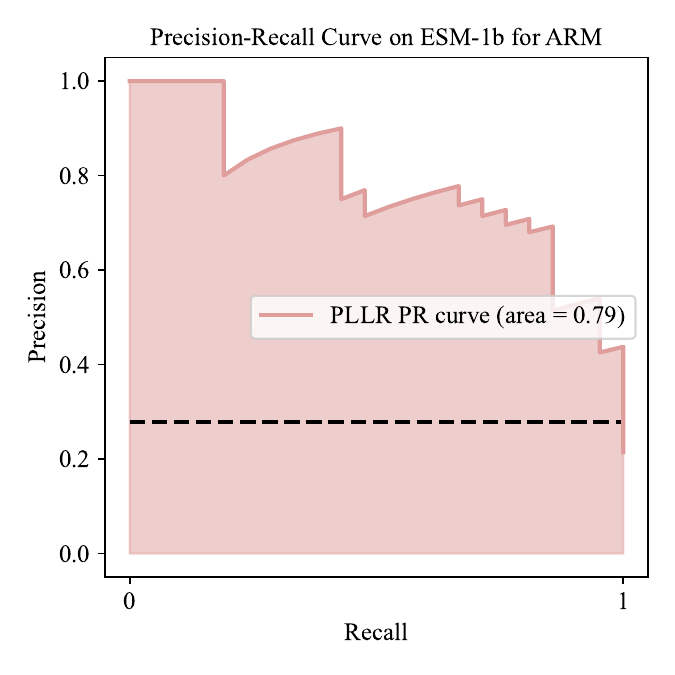}
    \end{subfigure}
    \hfill
    \begin{subfigure}[b]{0.24\linewidth}
            \caption{}
        \includegraphics[width=\linewidth]{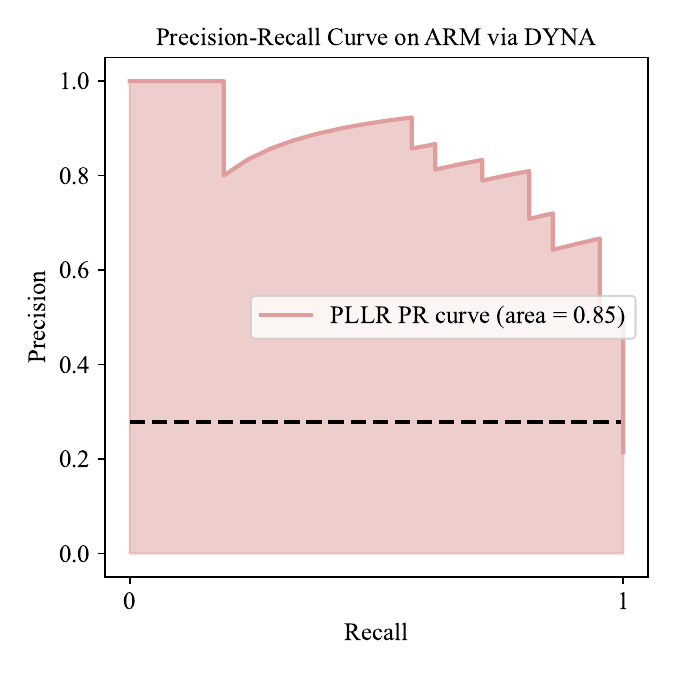}
    \end{subfigure}
    
    \caption{\textsc{dyna} AUPR performances on ClinVar CM and ClinVar ARM.}\label{fig:clinvar_aupr_esm1b}
\end{figure}

\begin{figure}[!htbp]
    \begin{subfigure}[b]{\linewidth}
    \caption{}\label{fig:top-50}
    \includegraphics[width=\linewidth]{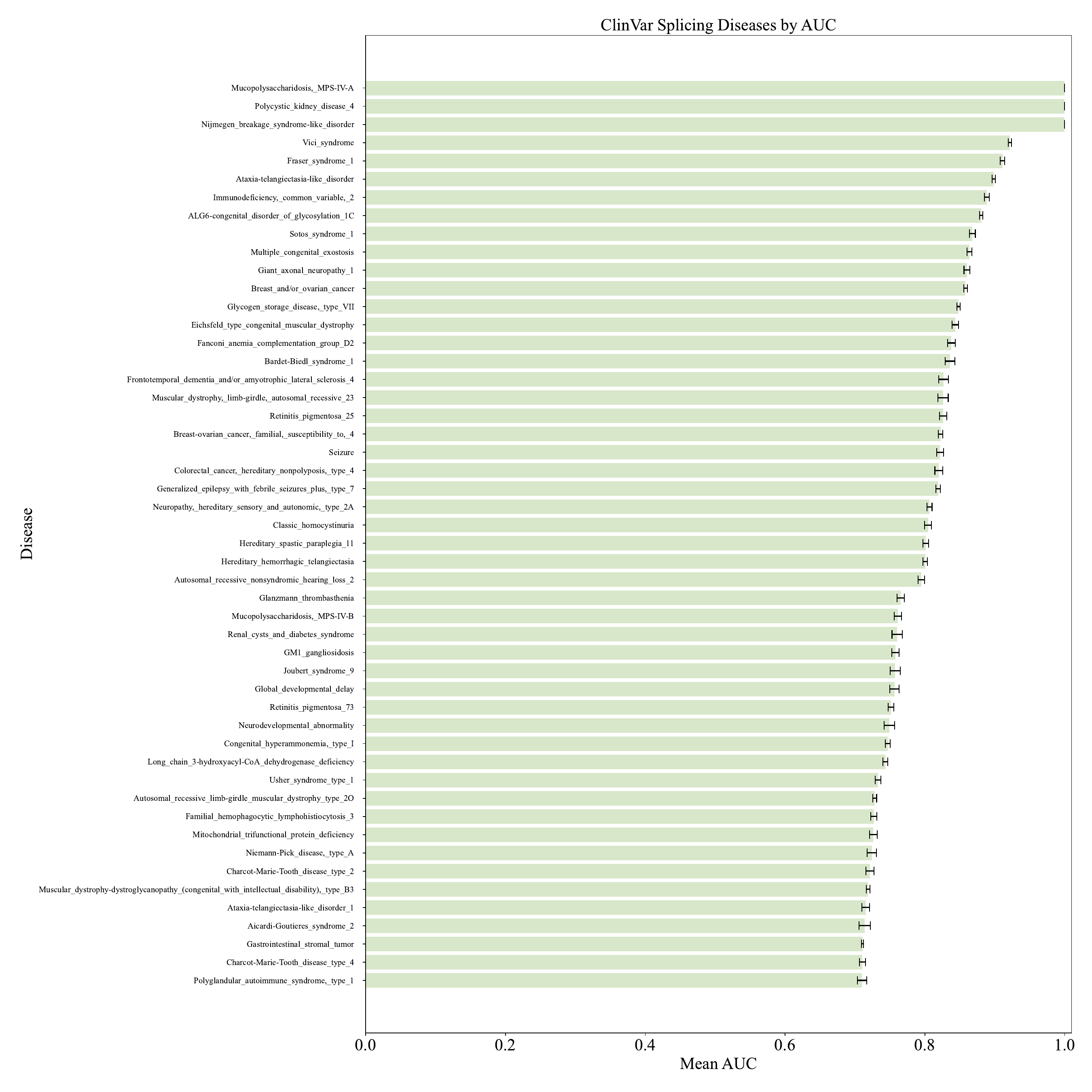}
    \end{subfigure}
    
        \begin{subfigure}[b]{\linewidth}
        \caption{}\label{fig:robust_splicing}
    \includegraphics[width=\linewidth]{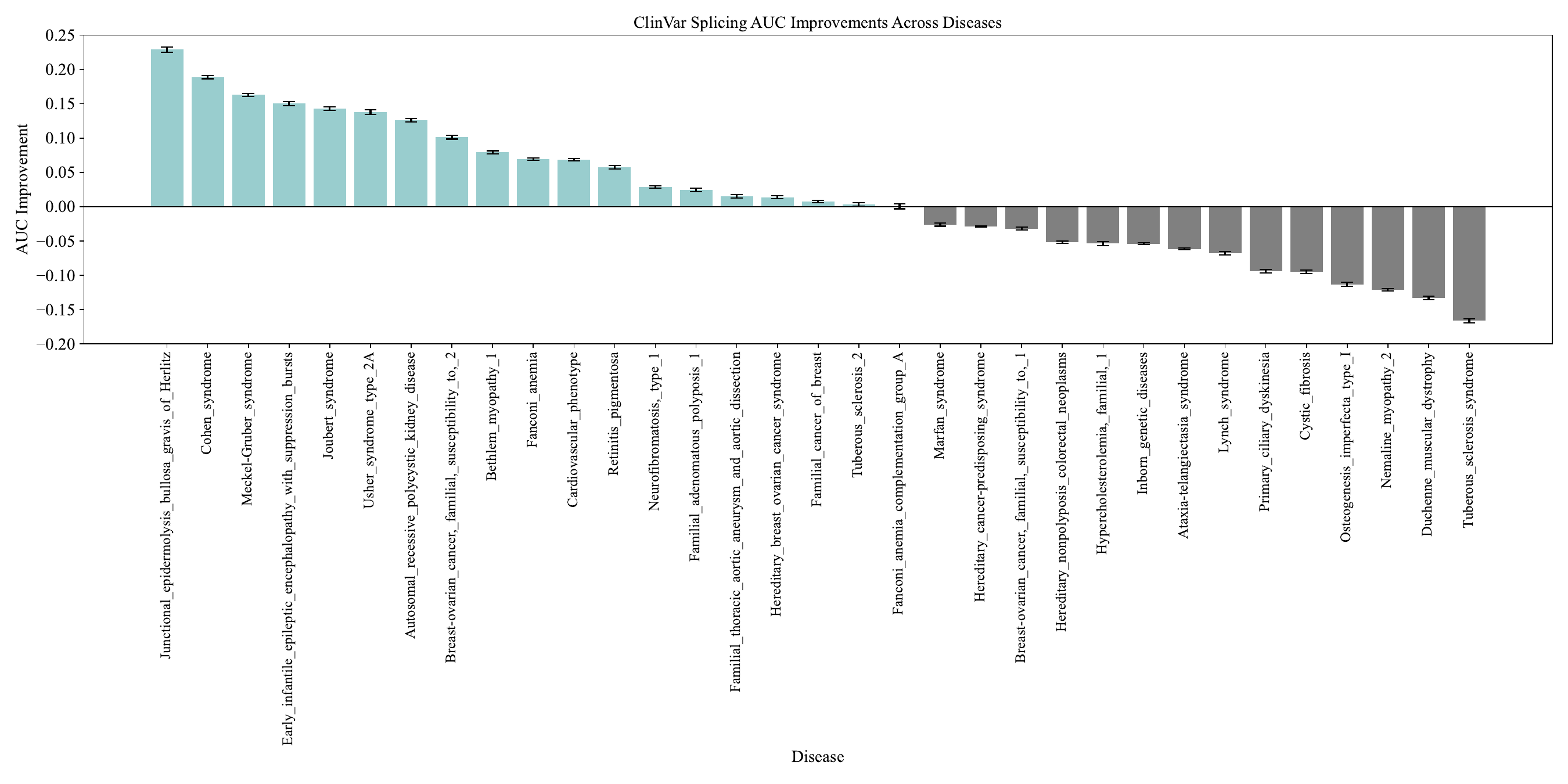}
    \end{subfigure}
    \caption{The top-50 AUC values per disease, along with the observed AUC improvement for \textsc{dyna} fine-tuned using the GPN model versus the baseline GPN model, underscore significant performance enhancements on each disease of the ClinVar Splicing dataset.}
    \label{fig:splicing_each_disease}
\end{figure}

\begin{figure}[!htbp]
    \centering

                    \begin{subfigure}[b]{0.48\linewidth}
                    \caption{}\label{fig:clinvar_splicing_aupr}
            \includegraphics[width=\linewidth]{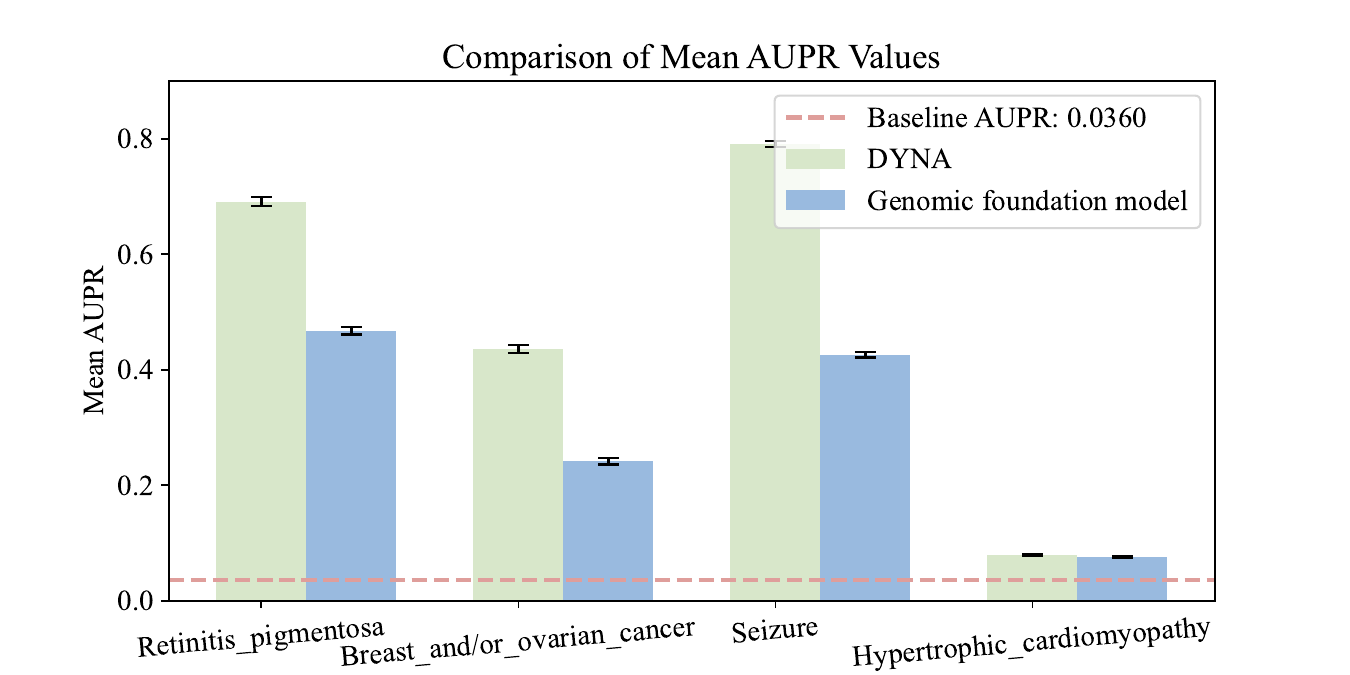}
    \end{subfigure} 
    

   
        \caption{AUPR results per disease within the ClinVar Splicing dataset. This figure displays the AUC four specific disease, each with more than five positive and negative variants, highlighting the performance improvements achieved by \textsc{dyna}, fine-tuned using the GPN model, over the baseline GPN model. Retinitis Pigmentosa, Breast and/or Ovarian Cancer, Seizure Disorders, and Hypertrophic Cardiomyopathy are shown due to their known splicing-related pathologies.}
        \end{figure}

\end{appendices}
     \clearpage


\bibliography{sn-bibliography}

\end{document}